\documentclass[twocolumn,floatfix,times,trackchanges]{aastex63}

\newcommand{\editbf}{}

\newcommand{\blue}{\color{blue}}

\newcommand{\Spitzer}{\textit{Spitzer}}
\newcommand{\Herschel}{\textit{Herschel}}
\newcommand{\frest}{$f_\mathrm{rest}$}

\newcommand{\Eu}{$E_\mathrm{up}$}

\newcommand{\Lbol}{$L_\mathrm{bol}$}
\newcommand{\Tbol}{$T_\mathrm{bol}$}
\newcommand{\Ltot}{$L_\mathrm{tot}$}
\newcommand{\tstar}{$t_\mathrm{\star}$}
\newcommand{\Mstar}{$M_\mathrm{\star}$}
\newcommand{\Rstar}{$R_\mathrm{\star}$}
\newcommand{\Tstar}{$T_\mathrm{\star}$}
\newcommand{\Mdisk}{$M_\mathrm{disk}$}
\newcommand{\RMAXdisk}{$R^\mathrm{outer}_\mathrm{disk}$}
\newcommand{\RMAXenv}{$R^\mathrm{outer}_\mathrm{env}$}
\newcommand{\RHOamb}{$\rho_\mathrm{amb}$}
\newcommand{\RHOcav}{$\rho_\mathrm{cav}$}
\newcommand{\Menv}{$M_\mathrm{env}$}
\newcommand{\Msun}{$M_\odot$}
\newcommand{\Msunyr}{$M_\odot\,\mathrm{yr}^{-1}$}
\newcommand{\Lsun}{$L_\odot$}
\newcommand{\Rsun}{$R_\odot$}

\newcommand{\Rc}{$R_C$}
\newcommand{\MDOTdisk}{$\dot{M}_\mathrm{disk}$}
\newcommand{\MDOTenv}{$\dot{M}_\mathrm{env}$}
\newcommand{\RMIN}{$R^\mathrm{inner}$}
\newcommand{\ZFAC}{$z^\mathrm{scale}_\mathrm{disk}$}
\newcommand{\incl}{$\cos\varphi$}

\newcommand{\RHOenvC}{$\rho_{\mathrm{env}.C}$}

\newcommand{\THETAcav}{$\theta_\mathrm{cav}$}

\newcommand{\Bdisk}{$B_\mathrm{disk}$}

\newcommand{\cmiii}{cm$^{-3}$}
\newcommand{\LED}{``warm--envelope'' scenario}
\newcommand{\FWHM}{$\overline{\mathrm{FWHM}}$}

\usepackage{CJKutf8}
\usepackage{graphicx}
\usepackage{longtable} \LTcapwidth=\textwidth
\usepackage{multirow}
\usepackage{subfigure}
\usepackage{ulem}
\usepackage{soul}
\usepackage{lipsum, babel}
\usepackage{enumitem}
\usepackage{amsmath}
\usepackage{euscript}
\usepackage{verbatim}

\setstcolor{red}

\accepted{\today}
\submitjournal{ApJ}

\shorttitle{The Warm--Envelope Origin of Hot Corinos}
\shortauthors{Hsu et al.}
\graphicspath{{./}{figures/}}

\begin{document}

\begin{CJK*}{UTF8}{bsmi}
\title{ALMA Survey of Orion Planck Galactic Cold Clumps (ALMASOP):\\ 
The Warm--Envelope Origin of Hot Corinos
}

\author[0000-0002-1369-1563]{Shih-Ying Hsu}
\email{seansyhsu@gmail.com}
\affiliation{National Taiwan University (NTU), No. 1, Section 4, Roosevelt Rd, Taipei 10617, Taiwan (R.O.C.)}
\affiliation{Institute of Astronomy and Astrophysics, Academia Sinica, No.1, Sec. 4, Roosevelt Rd, Taipei 10617, Taiwan (R.O.C.)}

\author[0000-0012-3245-1234]{Sheng-Yuan Liu}
\email{syliu@asiaa.sinica.edu.tw}
\affiliation{Institute of Astronomy and Astrophysics, Academia Sinica, No.1, Sec. 4, Roosevelt Rd, Taipei 10617, Taiwan (R.O.C.)}

\author[0000-0002-6773-459X]{Doug Johnstone}
\affiliation{NRC Herzberg Astronomy and Astrophysics, 5071 West Saanich Rd, Victoria, BC, V9E 2E7, Canada}
\affiliation{Department of Physics and Astronomy, University of Victoria, Victoria, BC, V8P 5C2, Canada}

\author[0000-0002-5286-2564]{Tie Liu}
\affiliation{Key Laboratory for Research in Galaxies and Cosmology, Shanghai Astronomical Observatory, Chinese Academy of Sciences, 80 Nandan Road, Shanghai 200030, People’s Republic of China}

\author[0000-0002-9574-8454]{Leonardo Bronfman}
\affiliation{Departamento de Astronom\'{i}a, Universidad de Chile, Casilla 36-D, Santiago, Chile}

\author[0000-0002-9774-1846]{Huei-Ru Vivien Chen}
\affiliation{Department of Physics and Institute of Astronomy, National Tsing Hua University, Hsinchu, 30013, Taiwan}

\author[0000-0002-2338-4583]{Somnath Dutta}
\affiliation{Institute of Astronomy and Astrophysics, Academia Sinica, No.1, Sec. 4, Roosevelt Rd, Taipei 10617, Taiwan (R.O.C.)}

\author[0000-0002-5881-3229]{David J. Eden}
\affiliation{Armagh Observatory and Planetarium, College Hill, Armagh, BT61 9DB, UK}

\author[0000-0001-5175-1777]{Neal J. Evans II}
\affiliation{Department of Astronomy, The University of Texas at Austin, 2515 Speedway, Stop C1400, Austin, Texas 78712-1205, USA}

\author[0000-0001-9304-7884]{Naomi Hirano}
\affiliation{Institute of Astronomy and Astrophysics, Academia Sinica, No.1, Sec. 4, Roosevelt Rd, Taipei 10617, Taiwan (R.O.C.)}

\author[0000-0002-5809-4834]{Mika Juvela}
\affiliation{Department of Physics, P.O.Box 64, FI-00014, University of Helsinki, Finland}

\author[0000-0002-4336-0730]{Yi-Jehng Kuan}
\affiliation{Department of Earth Sciences, National Taiwan Normal University, Taipei, Taiwan (R.O.C.)}
\affiliation{Institute of Astronomy and Astrophysics, Academia Sinica, No.1, Sec. 4, Roosevelt Rd, Taipei 10617, Taiwan (R.O.C.)}

\author[0000-0003-4022-4132]{Woojin Kwon}
\affiliation{Department of Earth Science Education, Seoul National University, 1 Gwanak-ro, Gwanak-gu, Seoul 08826, Republic of Korea}
\affiliation{SNU Astronomy Research Center, Seoul National University, 1 Gwanak-ro, Gwanak-gu, Seoul 08826, Republic of Korea}

\author[0000-0002-3024-5864]{Chin-Fei Lee}
\affiliation{Institute of Astronomy and Astrophysics, Academia Sinica, No.1, Sec. 4, Roosevelt Rd, Taipei 10617, Taiwan (R.O.C.)}
	
\author[0000-0002-3179-6334]{Chang Won Lee}
\affiliation{Korea Astronomy and Space Science Institute (KASI), 776 Daedeokdae-ro, Yuseong-gu, Daejeon 34055, Republic of Korea}
\affiliation{University of Science and Technology, Korea (UST), 217 Gajeong-ro, Yuseong-gu, Daejeon 34113, Republic of Korea}

\author[0000-0003-3119-2087]{Jeong-Eun Lee}
\affiliation{Department of Physics and Astronomy, Seoul National University, 1 Gwanak-ro, Gwanak-gu, Seoul 08826, Korea}

\author[0000-0003-1275-5251]{Shanghuo Li}
\affiliation{Max Planck Institute for Astronomy, Königstuhl 17, D-69117 Heidelberg, Germany}

\author[0000-0002-1624-6545]{Chun-Fan Liu}
\affiliation{Institute of Astronomy and Astrophysics, Academia Sinica, No.1, Sec. 4, Roosevelt Rd, Taipei 10617, Taiwan (R.O.C.)}

\author[0000-0001-8315-4248]{Xunchuan Liu}
\affiliation{Shanghai Astronomical Observatory, Chinese Academy of Sciences, Shanghai 200030, PR China}

\author[0000-0003-4506-3171]{Qiuyi Luo}
\affiliation{Shanghai Astronomical Observatory, Chinese Academy  of Sciences, Shanghai 200030, People’s Republic of China}
\affiliation{School of Astronomy and Space Sciences, University of Chinese Academy of Sciences, No. 19A Yuquan Road, Beijing 100049, People’s Republic of China} 

\author[0000-0003-2302-0613]{Sheng-Li Qin}
\affiliation{Department of Astronomy, Yunnan University, and Key Laboratory of Astroparticle Physics of Yunnan Province, Kunming, 650091, People's Republic of China}

\author[0000-0002-6529-202X]{Mark G. Rawlings}
\affiliation{Gemini Observatory/NSF’s NOIRLab, 670 N. A’ohoku Place, Hilo, Hawai’i, 96720, USA}

\author[0000-0002-4393-3463]{Dipen Sahu}
\affiliation{Physical Research laboratory, Navrangpura, Ahmedabad, Gujarat 380009, India}
\affiliation{Academia Sinica Institute of Astronomy and Astrophysics, 11F of AS/NTU Astronomy-Mathematics Building, No.1, Sec. 4, Roosevelt Rd, Taipei 10617, Taiwan, R.O.C.}

\author[0000-0002-7125-7685]{Patricio Sanhueza}
\affiliation{National Astronomical Observatory of Japan, National Institutes of Natural Sciences, 2-21-1 Osawa, Mitaka, Tokyo 181-8588, Japan}
\affiliation{Department of Astronomical Science, SOKENDAI (The Graduate University for Advanced Studies), 2-21-1 Osawa, Mitaka, Tokyo 181-8588, Japan}

\author[0000-0001-8385-9838]{Hsien Shang (尚賢)}
\affiliation{Institute of Astronomy and Astrophysics, Academia Sinica,  Taipei 10617, Taiwan}

\author[0000-0002-8149-8546]{Kenichi Tatematsu}
\affiliation{Nobeyama Radio Observatory, National Astronomical Observatory of Japan, National Institutes of Natural Sciences, 462-2 Nobeyama, Minamimaki, Minamisaku, Nagano 384-1305, Japan}
\affiliation{Department of Astronomical Science, The Graduate University for Advanced Studies, SOKENDAI,
2-21-1 Osawa, Mitaka, Tokyo 181-8588, Japan}

\author[0000-0001-8227-2816]{Yao-Lun Yang}
\affiliation{RIKEN Cluster for Pioneering Research, Wako-shi, Saitama, 351-0198, Japan}

\begin{abstract} 
Hot corinos are of great interest due to their richness in interstellar complex organic molecules (COMs) and the consequent potential prebiotic connection to solar--like planetary systems. 
Recent surveys have reported an increasing number of hot corino detections in Class 0/I protostars; however, the relationships between their physical properties and the hot--corino signatures remain elusive. 
In this study, our objective is to establish a general picture of the detectability of the hot corinos by identifying the origin of the hot--corino signatures in the sample of young stellar objects (YSOs) obtained from the Atacama Large Millimeter/submillimeter Array Survey of Orion Planck Galactic Cold Clumps (ALMASOP) project.  
We apply spectral energy distribution (SED) modeling to our sample and identify the physical parameters of the modeled YSOs directly, linking the detection of hot--corino signatures to the envelope properties of the YSOs. 
Imaging simulations of the methanol emission further support this scenario.
We, therefore, posit that the observed COM emission originates from the warm inner envelopes of the sample YSOs, based on both the warm region size and the envelope density profile. 
The former is governed by the source luminosity and is additionally affected by the disk and cavity properties, while the latter is related to the evolutionary stages. 
This scenario provides a framework for detecting hot--corino signatures toward luminous Class~0 YSOs, with fewer detections observed toward similarly luminous Class~I sources.
\end{abstract}


\section{Introduction}
\label{sec:2023:Intro}


The presence of interstellar complex organic molecules (iCOMs or COMs) in low--mass protostellar cores is of great interest.
These COMs, organic species consisting of six or more atoms \editbf{\citep{2009Herbst_COM_review}}, such as methanol (CH$_3$OH),
may play a role in habitability in planetary systems. 
Abundant (relative to molecular hydrogen $X>10^{-8}$) saturated COMs have been found, since 2004, in the localized, warm ($\sim$ 100~K) and compact ($\sim$ 100~au) zones surrounding low-- or intermediate--mass young stars, which are known as ``hot corinos" \citep{2004Ceccarelli_HotCorino}.

While earlier studies of hot corinos were more focused toward individual, often well--known and bright YSOs, such as IRAS 16293--2422 \citep{2003Cazaux_IRAS16293-2422} and HH--212 \citep{2016Codella_HH212, 2017Lee_HH212}, in recent years several chemical surveys toward low--mass protostellar objects have been conducted, leading to a greatly increasing number of the hot corino detection.
\citet{2019Bergner_Ser-emb_COM}, for example, found that three out of five low--mass Class~0/I protostars toward the Serpens cluster had hot-corino signatures. 
\citet{2020Belloche_COM_CALYPSO} observed 26 Class~0/I protostars, under the ``Continuum And Lines in Young ProtoStellar Objects (CALYPSO) '' Program, and report 12 sources harboring methanol. 
\citet{2020vanGelder_COM} examined COM emission toward seven Class~0 protostellar cores in the Perseus Barnard 1 cloud and Serpens Main region and found three COM--rich sources.
\citet{2021Yang_PEACHES_COM} surveyed COMs toward 50 protostars in the Perseus cloud, under the ``Perseus ALMA Chemistry Survey (PEACHES)'' project, and 28 out of 50 sources appear to harbor warm methanol. 
Under the ``ALMA Survey of Orion Planck Galactic Cold Clumps (ALMASOP)'' project, \citet{2022Hsu_ALMASOP} reported 11 sources having hot--corino signatures among 56 Class~0/I protostars.
Most recently, \citet{2022Bouvier_ORANGES} observed \editbf{19} protostars in the OMC-2/3 filament and detected \editbf{five} of them harboring hot corinos under the ORion ALMA New GEneration Survey (ORANGES).

Such sample studies, in contrast to the detailed investigations of individual objects, enabled the inspections of the nature of COM emission via statistical approaches.
{\citet{2020Belloche_COM_CALYPSO}, for example, found that all sources showing hot--corino signatures in their sample have luminosities higher than 4 \Lsun. 
The authors suspected that a lack of sensitivity might be producing the non--detections toward fainter sources, and therefore it remains unclear whether all low--mass protostars may go through a phase showing COM emission.
}
Using the ALMASOP sample, \citet{2022Hsu_ALMASOP} demonstrated a positive correlation between the YSO luminosity and both the emission extent and total amount of methanol molecules, which suggests that the detection of COM emission correlates with the size of the warm region in the protostellar system.  
Similarly, \citet{2022vanGelder_MCH3OH} compiled the methanol data in literature studies toward \editbf{184} low-- and high--mass protostars and found a positive correlation between the bolometric luminosities and the detected methanol (CH$_3$OH) mass. 
These characteristics are consistent with the contemporary paradigm of the hot corino, in which COMs formed on grain mantles and get thermally desorbed due to ice sublimation when the dust temperature reaches $\sim$ 100 K \citep[see, for example, ][]{2006Garrod_3phase, 2008Garrod_3phase, 2009Herbst_COM_review}. 
The detection of the hot--corino signature at a given sensitivity therefore depends on the total amount of gas--phase COMs, such as methanol, which is related to the warm region's size that is governed by the luminosity of the central YSO.

The luminosities, or correspondingly the resulting warm region sizes, of YSOs are likely not the only determining factor in detecting their associated hot--corino signatures.
In the ALMASOP sample, \citet{2022Hsu_ALMASOP} found no hot--corino signatures in the Class~I protostars, even though some of them have high bolometric luminosities comparable to those of the Class~0 protostars with COM emission. 
Indeed, the detection of hot corinos toward Class~I protostars in the literature to date appear significantly rarer than that of Class~0 ones (dozens of Class~0 to only about ten Class~I (\editbf{e.g. SVS13--A \citep{{2017Lefevre_SVS13A_HotCorino}} and L1551–-IRS5 \citep{2022Mercimek_ClassI_COMSurvey}}).
Similarly, \citet{2022vanGelder_MCH3OH} found that, at a given luminosity, in addition to YSOs with methanol masses following the empirical luminosity--mass correlation, there are also YSOs without methanol emission detection, suggesting additional factors other than luminosity at play.

The exact origin of the hot--corino signatures, which are diverse in the literature, may help investigate what physical parameters in addition to luminosity could govern the detectabilities of hot corinos. 
\citet{2019Jacobsen_L483_COM} suggests that the COMs reside in the ``innermost envelope'', at 40 -- 60~au of the embedded low--mass protostar L483 according to the extent of the COM emission and the non-detection of a Keplerian disk down to 15~au. 
Similarly, the Class~0 protostar B335 \citep{2016Imai_B335} appears to lack a Keplerian disk, at least down to 10~au \citep{2015Yen_B335_nodisk}. 
Furthermore, toward IRAS 16293-2422~A \citet{2016Oya_IRAS16293-2422A_envelope} find that methanol (CH$_3$OH) and methyl formate (HCOOCH$_3$) concentrate around the inner part of the infalling and rotating envelope. 
Alternatively, put forward by \citet{2017Lee_HH212} and \citet{2019Lee_HH212_COM_atm}, the COM emission arises from the ``disk atmosphere'', within a radius of $\sim$ 40~au according to the spatially resolved images and the position--velocity (PV) diagrams of the protostar HH--212. 

Do all solar--like protostars go through a phase showing a hot--corino signature?
If so, is it possible to build a general picture for the detectability of hot corinos?
In this study, we aim to answer these questions by identifying the origin of the hot--corino signatures in the ALMASOP sample. 

We briefly introduce the ALMASOP observations and the sample selection for this study in Sect.\ \ref{sec:2023:Obs}.  
In Sect.\ \ref{sec:2023:SED}, we present the YSO models for each source inferred from the Spectral Energy Distribution (SED) fitting and further show that the sources with hot--corino signature have relatively high envelope densities, high luminosities, and consequently, a large amount of warm envelope mass. 
In Sect.\ \ref{sec:2023:Sparx}, comparisons between simulated and observed methanol emission images imply that the envelope dominates the COM emission in the observed sources.
In Sect.\ \ref{sec:2023:Disc}, we review and highlight literature studies showing indicative signs of support for the warm envelope origin of COM emission and discuss the astrochemical implications of such a scenario.
We outline our conclusions in Sect.\ \ref{sec:2023:Conclusions}. 


\section{ALMASOP Project}
\label{sec:2023:Obs}

The ALMASOP survey selected 72 clumps in the Orion A, B, and $\lambda$ Orionis clouds as the observation targets, starting from a sample of Planck Galactic Cold Clumps \citep[PGCCs, ][]{2016Planck_PGCC} in the Orion Molecular Complex  \citep{2017Tatematsu_PGCC_N2Dp, 2018Liu_TOP-SCOPE, 2018Yi_PGCC_Orion, 2019Eden_SCOPE, 2020Tatematsu_ALMASOP, 2020Kim_ALMASOP_Nobeyama}.
The observations of the ALMASOP project were conducted with both the 12--m array (in two array configurations: C43--5 and C43--2 denoted, respectively, as TM1 and TM2) and the 7--m array (ACA/Morita Array) of the Atacama Large Millimeter/submillimeter Array (ALMA) as part of the Cycle 6 operations (\#2018.1.003.2.S, PI: Tie Liu). 
Our observations contain four spectral windows (SPWs) centering at 216.6, 218.9,  231.0 and 233.0 GHz with a uniform bandwidth of 1875 MHz and a resolution of $\sim$1.1 MHz.
The imaging was carried out using the \texttt{tclean} task in Common Astronomy Software Applications \citep[CASA, ][]{2007McMullin_CASA}. 
Under the ALMASOP observation, \citet{2020Dutta_ALMASOP} presents a sample of starless, prestellar, and protostellar cores. 
See \citet{2020Dutta_ALMASOP} for more details about the ALMASOP observations.

\section{YSO Model and SED Fitting Analysis}
\label{sec:2023:SED}

We aim to identify which YSO physical parameters might correlate with the (non-)detection of hot--corino signatures using our ALMASOP sample. 
We thus carry out an SED fitting analysis to derive the YSO models by a similar process to that introduced in Sect.\ 4 and Appendix C by \citet{2022Hsu_ALMASOP}. 
First, we select the sample for this study from the ALMASOP protostars based on their SED data points, including their fluxes and aperture radii at various wavelengths, as compiled by \citet{2020Dutta_ALMASOP}. 
Next, we employ a grid of YSO models and their corresponding SEDs at multiple wavelengths and aperture radii established by \citet{2006Robitaille_grid}. 
Finally, among the models in the grid, we identify the ones that best match with the observed SED data points.

\subsection{SED Modeling Process}

\subsubsection{Sample Selection}

We select the sample for this study from the ALMASOP protostars based on the corresponding SED data points tabulated in Table 6 of \citet{2020Dutta_ALMASOP}. 
These archival photometric data, covering wavelengths from 3.4 \micron\ to 870 \micron, 
were collected from
the Spitzer Space Telescope Survey of Orion A-B \citep{2012Megeath_MGM2012}, 
the Wide-field Infrared Survey Explorer \citep[WISE, ][]{2010Wright_SED_WISE}, 
the AKARI/IRC All--Sky Survey Point Source Catalogue \citep[AKARIPSC][]{2010Ishihara_SED_AKARI_IRC}, 
the Herschel Orion Protostellar Survey \citep[HOPS, ][]{2013Stutz_HOPS_APEX, 2015Tobin_SED_HOPS}, 
and the Atacama Pathfinder Experiment \citep[APEX, ][]{2013Stutz_HOPS_APEX}. 
The aperture radii for \Spitzer, \Herschel, and APEX measurements are quoted from \citet{2016Furlan_HOPS_SED} and the FWHMs of the beam for the WISE and AKARI measurements are based on \citet{2010Wright_SED_WISE} and \citet{2010Ishihara_SED_AKARI_IRC} , respectively. 
In addition, we append a SED data point at 1.3~mm using the integrated flux within twice the beam size in the continuum images made with the ALMASOP combined data (TM1+TM2+ACA).
At 1.3~mm the missing flux due to interferometric observations could help with the situation by filtering out the very extended cloud component and being sensitive to only the core scale structures most relevant to the SED fitting. 
See Table \ref{tab:2023:sed_info} for more detailed information.

We further down--selected 16 Class~0 and seven Class~I \editbf{protostars} based on the additional criteria below in order to have better-constrained the SED models: 
\begin{enumerate}
    \item 
    We require photometric measurements for each source at mid--IR (3.4, 3.4, 4.5, 4.6 and 5.8 \micron), far--IR (9, 12, 18, 22, 24, 70, 100, and 160 \micron), sub-mm (350 and 870 \micron), and mm (1,300 \micron) bands, as they are sensitive to different components and mechanisms in a protostellar system \citep[e.g., ][]{2003Whitney_SED,2006Robitaille_grid,2016Furlan_HOPS_SED}. 
    In particular, we further require the source to have a photometric measurement at 100\,\micron\ at which the SED of Class~0/I protostars typically peak \citep[see, for example, ][]{2016Furlan_HOPS_SED}.
    Consequently, the sources in this study are all protostellar objects observed by the Herschel Orion Protostar Survey,  \citep[HOPS,][]{2013Stutz_HOPS_APEX,2015Tobin_SED_HOPS} from which we collected the 100 \micron\ photometric data. 
    These sources are distributed across the Orion A and B clouds.
    
    \item The source is not blended with other sources in the [70 -- 160] \micron\ bands. 
    Since ALMASOP has a better angular resolution than the other photometric literature data, sometimes more than one protostar identified in ALMASOP is part of the same HOPS object. 
    To support the SED analysis and COMs investigation, we ensure that each selected YSO has only one--to--one corresponding HOPS counterpart cataloged by \citet{2016Furlan_HOPS_SED}. 
\end{enumerate}

Overall, based on \citet{2020Dutta_ALMASOP} catalogues, the 23 sources in our sample have a bolometric temperature range from 31$\pm$10~K to 381$\pm$60~K and a bolometric luminosity range from 0.4$\pm$0.2 \Lsun\ to 180$\pm$70 \Lsun.
Please refer to Table \ref{tab:2023:src_info} for more information about the source coordinates, properties, and corresponding HOPS indices. 

\begin{deluxetable*}{lccccrrlc}[tbp!]
\label{tab:2023:src_info}
\caption{Sample list in this study.}
\tablehead{
\colhead{Name} & \colhead{$\mathrm{\alpha_{J2000}}$} & \colhead{$\mathrm{\delta_{J2000}}$} & \colhead{Cloud} & \colhead{Stage} & \colhead{$T_\mathrm{bol}$} & \colhead{$L_\mathrm{bol}$} & \colhead{HOPS Index}  & \colhead{Hot Corino} \\ 
\colhead{} & \colhead{} & \colhead{} & \colhead{} & \colhead{} & \colhead{(K)} & \colhead{(\Lsun)} & \colhead{}  & \colhead{} 
}
\startdata
G205.46--14.56M1--A & 05h46m08.6s & -00d10m38.49s & Orion B & Class 0 & 47$\pm$12 & 4.8$\pm$2.1 & HOPS--317 & \\
G205.46--14.56S1--A & 05h46m07.3s & -00d13m30.23s & Orion B & Class 0 & 44$\pm$19 & 22$\pm$8 & HOPS--358 & $\bigcirc$ \\
G206.12--15.76 & 05h42m45.3s & -01d16m13.94s & Orion B & Class 0 & 35$\pm$9 & 3.0$\pm$1.4 & HOPS--400 & \\
G206.93--16.61W2 & 05h41m25.0s & -02d18m06.75s & Orion B & Class 0 & 31$\pm$10 & 6.3$\pm$3.0 & HOPS--399 & $\bigcirc$ \\
G208.68--19.20N1 & 05h35m23.4s & -05d01m30.60s & Orion A & Class 0 & 38$\pm$13 & 36.7$\pm$14.5 & HOPS--87 & $\bigstar$ \\
G209.55--19.68S1 & 05h35m13.4s & -05d57m57.89s & Orion A & Class 0 & 50$\pm$15 & 9.1$\pm$3.6 & HOPS--11 & $\bigcirc$ \\
G209.55--19.68S2 & 05h35m09.1s & -05d58m26.87s & Orion A & Class 0 & 48$\pm$11 & 3.4$\pm$1.4 & HOPS--10 & \\
G210.37--19.53S & 05h37m00.4s & -06d37m10.90s & Orion A & Class 0 & 39$\pm$10 & 0.6$\pm$0.3 & HOPS--164 & \\
G210.49--19.79W--A & 05h36m18.9s & -06d45m23.54s & Orion A & Class 0 & 51$\pm$20 & 60$\pm$24 & HOPS--168 & $\bigstar$ \\
G210.97--19.33S2--A & 05h38m45.5s & -07d01m02.02s & Orion A & Class 0 & 53$\pm$15 & 3.9$\pm$1.5 & HOPS--377 & \\
G211.01--19.54N & 05h37m57.0s & -07d06m56.23s & Orion A & Class 0 & 39$\pm$12 & 4.5$\pm$1.8 & HOPS--153 & \\
G211.01--19.54S & 05h37m58.8s & -07d07m25.72s & Orion A & Class 0 & 52$\pm$8 & 0.9$\pm$0.4 & HOPS--152 & \\
G211.16--19.33N2 & 05h39m05.8s & -07d10m39.29s & Orion A & Class 0 & 70$\pm$20 & 3.7$\pm$1.4 & HOPS--133 & \\
G211.47--19.27S & 05h39m56.0s & -07d30m27.61s & Orion A & Class 0 & 49$\pm$21 & 180$\pm$70 & HOPS--288 & $\bigstar$ \\
G212.10--19.15S & 05h41m26.2s & -07d56m51.93s & Orion A & Class 0 & 43$\pm$12 & 3.2$\pm$1.2 & HOPS--247 & \\
G212.84--19.45N & 05h41m32.1s & -08d40m09.77s & Orion A & Class 0 & 50$\pm$13 & 3.0$\pm$1.2 & HOPS--224 & \\
\hline
G205.46--14.56S2 & 05h46m04.8s & -00d14m16.67s & Orion B & Class I & 381$\pm$60 & 12.5$\pm$4.7 & HOPS--385 & \\
G205.46--14.56S3 & 05h46m03.6s & -00d14m49.57s & Orion B & Class I & 178$\pm$33 & 6.4$\pm$2.4 & HOPS--315 & \\
G210.82--19.47S--B & 05h38m03.4s & -06d58m15.89s & Orion A & Class I & 74$\pm$12 & 0.4$\pm$0.2 & HOPS--156 & \\
G210.97--19.33S2--B & 05h38m45.0s & -07d01m01.68s & Orion A & Class I & 82$\pm$24 & 4.1$\pm$1.6 & HOPS--144 & \\
G211.16--19.33N5 & 05h38m45.3s & -07d10m56.03s & Orion A & Class I & 112$\pm$16 & 1.3$\pm$0.5 & HOPS--135 & \\
G212.10--19.15N2--A & 05h41m23.79s & -07d53m46.74s & Orion A & Class I & 114$\pm$10 & 1.1$\pm$0.5 & HOPS--263 & \\
G212.10--19.15N2--B & 05h41m24.0s & -07d53m42.22s & Orion A & Class I & 160$\pm$30 & 1.1$\pm$0.5 & HOPS--262 & \\
\enddata
\tablecomments{
$\mathrm{\alpha_{J2000}}$ and $\mathrm{\delta_{J2000}}$ are the right ascension and declination, respectively, of the peak position in our combined 1.3~mm continuum observations.
\Tbol\ and \Lbol\ are respectively the bolometric temperature and bolometric luminosity adopted from \citet{2020Dutta_ALMASOP}. 
In the last column (Hot Corino), the symbols ``$\bigstar$'' and ``$\bigcirc$'' represent the hot corino sources discovered in the ACA-only data \citep{2020Hsu_ALMASOP} and the combined data \citep{2022Hsu_ALMASOP}, respectively. 
The horizontal line separates the sources at the Class 0 (top) and Class I (bottom) stages, split by \Tbol\ $=70$ K. 
\editbf{All the sources are found to be low--/intermediate--mass protostars based on the SED modeling analysis in this study.}
}
\tablerefs{
\Lbol\ and \Tbol: \citet{2020Dutta_ALMASOP}; 
HOPS Index:~\citet{2016Furlan_HOPS_SED}; 
Hot Corino: \citet{2020Hsu_ALMASOP} and \citet{2022Hsu_ALMASOP}
}
\end{deluxetable*}

\begin{figure*}[htb!]
\includegraphics[width=0.95\textwidth]{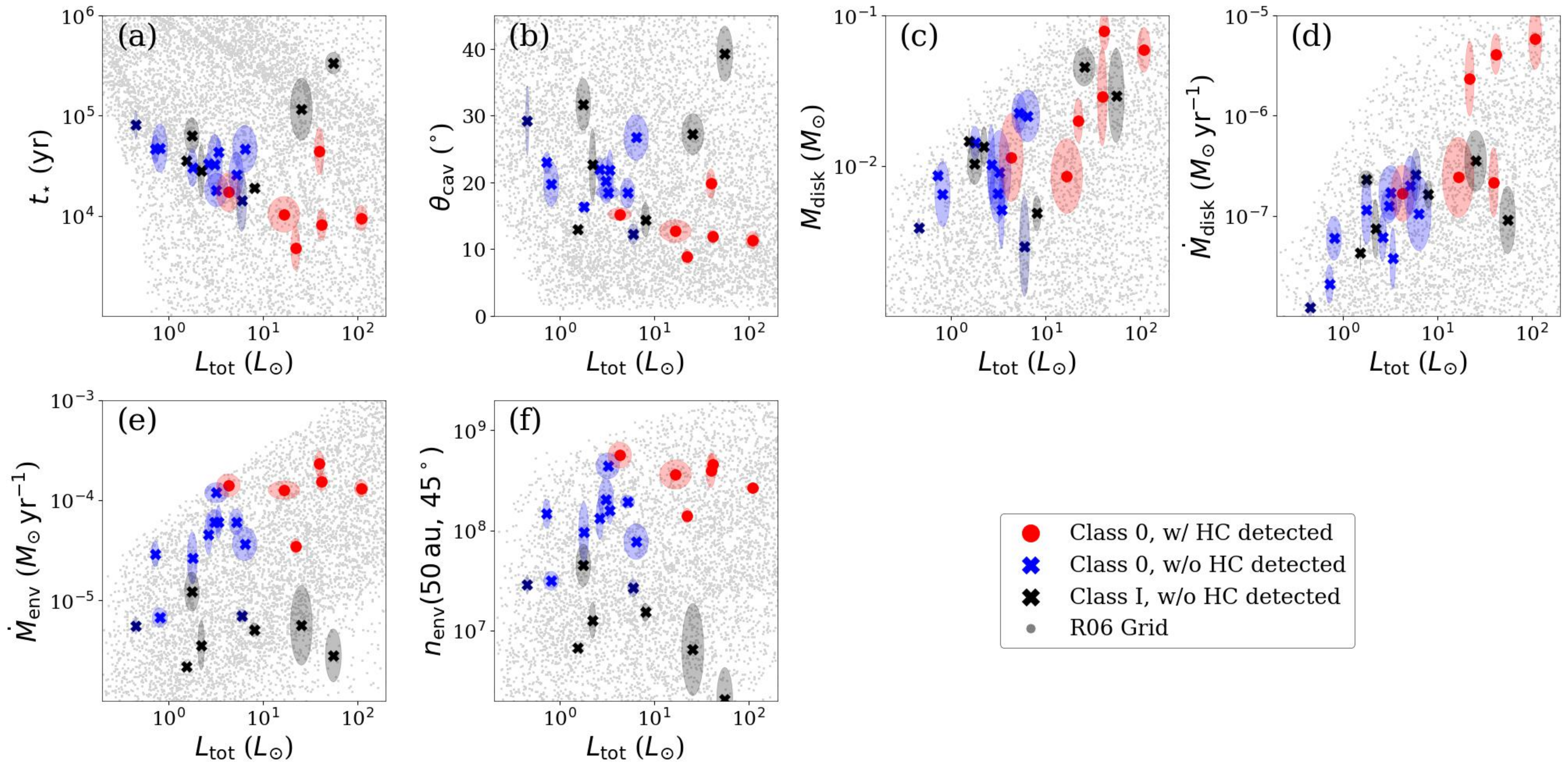}
\caption{\label{fig:2023:YSOPar_part} 
Weighted average and standard deviation of the following YSO model parameter versus total luminosity (\Ltot). 
The \tstar\ is the evolutionary age of the protostellar system. 
The \THETAcav\ is the opening angle of the cavity. 
The \Mdisk\ is the total mass of the disk.
The \MDOTdisk\ is the disk accretion rate.
The \MDOTenv\ is the envelope infall rate.
Panel (f) is the envelope number density at radius $r=50$~au and polar angle $\theta=45^\circ$. 
The red circles represent the sources with hot--corino signature \citep{2020Hsu_ALMASOP,2022Hsu_ALMASOP}. 
The blue and black crosses represent the Class~0 and Class~I protostars toward which hot corinos signatures are not detected, respectively. 
The grey dots represent all the YSO models in the R06 grid. 
Note that the last two YSO parameters are not the original R06 grid parameters. 
}
\end{figure*}

\subsubsection{YSO Model Grid}
For the physical parameters of the YSO models, we adopt the model grid published by \citet{2006Robitaille_grid} (hereafter, ``R06 grid''). 
In the R06 grid, the SEDs were produced by \texttt{HO-CHUNK}\footnote{https://gemelli.colorado.edu/\(\sim \)bwhitney/codes/codes.html}, a Monte Carlo radiation transfer code for protostellar geometry from Class~0 to Class~III sources \citep{2003Whitney_SED,2003Whitney_SEDII}.
In \texttt{HO-CHUNK}, the YSO physical model consists of a central protostar, an infalling and rotating envelope, a flared disk, and a bipolar cavity.
The envelope density model follows the ``CMU'' prescription \citep{1976Ulrich_CMU,1981Cassen_CMU}. 
The disk density model is a standard flared accretion disk model \citep[see, for example, ][]{1981Pringle_disk_model,2003Whitney_SED}. 
Thus, the full physical model is axisymmetric and described by 14 parameters: stellar mass (\Mstar), stellar radius (\Rstar), stellar temperature (\Tstar), envelope infall rate (\MDOTenv), envelope outer radius (\RMAXenv), cavity density (\RHOcav), cavity opening angle (\THETAcav), disk mass (\Mdisk), outer disk radius (\RMAXdisk), the inner radius of disk and envelope (\RMIN), disk accretion rate (\MDOTdisk), disk scale height factor (\ZFAC), disk flaring power component (\Bdisk), and ambient density (\RHOamb). 
The R06 grid has 200,000 physical YSO models, each with 10 SEDs evaluated at different inclination values (\incl).
More information of the R06 grid is available by \citet{2003Whitney_SED}, \citet{2006Robitaille_grid}, \citet{2007Robitaille_sedfitter}, and the instruction file of the \texttt{HO-CHUNK} package. 

\subsubsection{SED Fitting}
We employed the \texttt{SED Fitter} \footnote{https://sedfitter.readthedocs.io/en/stable/}\citep{2007Robitaille_sedfitter} for evaluating the YSO models best matching with the observation.
An additional extinction, $A_V$, was set to be in the range of [0, 30] mag based on \citet{2016Furlan_HOPS_SED}.
The source distance, $D$, was fixed to be 398 pc and 404 pc, respectively, for sources in the Orion A and B clouds \citep{2018Kounkel_Orion_distance}.
For each combination of SED grid model, foreground extinction ($A_V$), and distance ($D$), the \texttt{SED Fitter} evaluates the ``goodness of fit'' based on the $\chi^2$ value defined by: 

\begin{equation}
\label{eq:chi2}
    \chi^2\equiv
    \sum_{i=1}^{N_\mathrm{data}}\left [ 
    \frac{\left< \log_{10} F_\nu(\lambda_i) \right>-\log_{10} M_\nu(\lambda_i)}
         {\sigma\left ( \left< \log_{10} F_\nu(\lambda_i) \right> \right )} 
    \right ]^2
    \;\;\mathrm{,} 
\end{equation}
where $F_\nu(\lambda_i)$ and $M_\nu(\lambda_i)$ are the observed flux density and the modeled flux density at a given wavelength $\lambda_i$ within the corresponding observational aperture, respectively, and $\sigma$ is the uncertainty of the logarithm of the flux. 
In addition to the $\chi^2$ value, the continuum and molecular line images obtained with ALMASOP yield more constraints to the acceptable model. 
First, we set an upper limit of the disk outer radius (\RMAXdisk) at 90~au for those without clear disk signatures in the C$^{18}$O $J=2-1$ moment--0 images. 
Second, we limited the projected opening angle adapted from \citet{2020Dutta_ALMASOP} and \citet{2020Hsu_ALMASOP} for the sources showing clear bipolar outflow in their CO $J=2-1$ moment--0 images.
See \citet{2007Robitaille_sedfitter} and Appendix C of \citet{2022Hsu_ALMASOP} for more details. 


For statistical robustness, we select the nine best-fit models (i.e., the nine models having the lowest $\chi^2$ values) for each source. 
Table \ref{tab:2023:YSOModel_name} provides the model names from the R06 grid. 
For each source, we derive the weighted average and standard deviation of the inferred physical parameters based on the $\chi^{2}$ values to illustrate the distribution of the parameters and their statistical uncertainties (Table \ref{tab:2023:YSOModel_par}).
The weight for each model $w_j$ is defined as: 

\begin{equation}
    w_j\equiv\frac{1}{\chi_j^2}
    \;\;\mathrm{,} 
\end{equation}

\subsection{Best-fit YSO Parameters}

We show the distribution of YSO model parameters relevant to our discussion versus the total luminosity (\Ltot, i.e., the sum of stellar luminosity and accretion luminosity) in Figure~\ref{fig:2023:YSOPar_part} and attach the distributions for the full set of (14) parameters in Appendix \ref{appx:sed} (Figure~\ref{fig:2023:YSOPar_complete}) for completeness. 
For the ALMASOP sample, red circles and blue crosses represent the Class~0 protostars with and without hot--corino signatures \citep{2020Hsu_ALMASOP,2022Hsu_ALMASOP}, respectively. 
The black crosses represent Class~I protostars and there are no hot corinos detected within this group. 
\editbf{The majority of the sources in our sample have modeled stellar masses \Msun\ below $\sim$ 2~\Msun, with one exception having a modeled stellar mass \Mstar\ of $\sim$ 4~\Msun.
This suggests that our sample consists of low--/intermediate--mass YSOs.
}
At first glance, there appear to be some noticeable correlations between certain physical parameters and the total luminosity. 
Some of these correlations are in fact related to the underlying grid sampling of the parameters, which is illustrated by the background, light gray markers in each panel. 
For example, the clear positive correlation between the stellar mass (\Mstar) and the total luminosity (\Ltot) of our sample in Figure~\ref{fig:2023:YSOPar_complete}--(b) is intrinsically limited by the sampling. 

\subsection{Trends in parameters from Class~0 to Class~I}
\label{sec:2023:sed_trend}
The SED fitting approach admittedly has its limitations as it imposes very simplified YSO physical models. 
The lack of near--infrared (near--IR) photometric data also makes the stellar parameters such as stellar temperature and radius somewhat uncertain. 
However, the inferred total luminosity, which dictates the thermal structure, is mainly determined at 100~\micron\ and will not be severely affected.
By comparing the distributions of the inferred parameters of YSO sources at different stages, we find reasonable trends implying that valuable insight is offered by these parameters extracted from the SED fitting. 

\begin{itemize}
    \item Evolutionary age (\tstar): 
    As shown in Figure~\ref{fig:2023:YSOPar_part}--(a), Class~I protostars are more evolved than their Class~0 counterparts. 
    The evolutionary ages of Class~0 protostars mostly range from $2.0\times10^3$ to $3.6 \times 10^4$ yr, while those of Class~I protostars range from $2.5\times10^4$ to $3.2 \times 10^5$ yr.
    There seems to be a tentative transition band between Class~0 and Class~I protostars at \tstar\ $\sim2.5\times10^4$ yr, which is comparable to the estimation of 2 -- 6 $\times10^4$ yr suggested by \citet{2006Froebrich_Class0_age} . 
    
    \item Cavity opening angle (\THETAcav): 
    As shown in Figure~\ref{fig:2023:YSOPar_part} (b), 
    { Class~I protostars have larger cavity opening angles (\THETAcav) being around 15$^{\circ}$ and above while Class~0 protostars have smaller opening angles ranging from a few degrees to around 20$^{\circ}$.} 
    This is consistent with the results by \citet{2006Arce_cavity_angle} and  \citet{2008Seale_YSO_cavity}, that the outflow opening angle of a protostellar source becomes wider as the YSO evolves. 

    \item Disk mass (\Mdisk): 
    As shown in Figure~\ref{fig:2023:YSOPar_part}--(c), there is no clear division in the disk mass (\Mdisk)  between the Class~0 and I stage.
    This result appears to be consistent with \citet{2009Jorgensen_PROSAC_II_Class0toI} and \citet{2020Tobin_VANDAM-II} in which no obvious growth of disk masses from Class~0 to Class~I stage was found.

    \item Disk accretion rate (\MDOTdisk):
    As shown in Figure~\ref{fig:2023:YSOPar_part}--(d), the disk accretion rates for Class 0 protostars span mostly from  $10^{-7}$ to $10^{-5}$ \Msunyr. 
    For Class I protostars, the disk accretion rates range from $10^{-8}$ to $10^{-6}$ \Msunyr. These results are  consistent with \citet{2021Fiorellino_MDOTdisk_ClassI} and \citet{2023Fiorellino_MDOTdisk_ClassI}. 
    The declining mass accretion rate from the Class~0 to the Class~I stage is also hinted at by \citet{2017Yen_MDOTdisk_Class0_to_I} based on their observations as well as literature. 
\end{itemize}

We note that while G211.16--19.33N2 and G210.82--19.47S--B have bolometric temperatures (\Tbol\ of $70\pm20$ and $74\pm12$~K, respectively) at the borderline between the Class~0 and Class~I categories, the trends discussed above remain valid.
In addition, we examine and comment on the validity of our SED modelling for five fields with the presence of multiplicity reported by the VLA/ALMA Nascent Disk and Multiplicity (VANDAM) survey of Orion protostars \citep{2020Tobin_VANDAM-II},
In the cases of G205.46--14.56M1--A (HOPS--317) and G205.46--14.56S1--A (HOPS--358), the separations between the two components (HOPS–317–A/B and HOPS–358–A/B, respectively) exceed 2,000~au. 
Only the A components are registered as the HOPS objects and the SED measurements are also dominated by the A components. Our modelling is thus applicable to the A components in these two systems.
In the cases of G210.49--19.79W--A (HOPS--168) and G206.12--15.76 (HOPS--400), the separations between the two components (HOPS–168–A/B and HOPS–400–A/B, respectively) are $\sim$100~au. 
Their SED measurements are likely dominated by the A components, as they are 1.5--5 times brighter than their counterparts in the VANDAM survey.
Moreover, their CO outflows observed by ALMASOP are primarily associated with the brighter components.
Our modelling is therefore also applicable to the A components in these two systems.
Lastly, G211.47--19.27S (HOPS--288) was reported by the VANDAM survey as a triple system (HOPS–288–A–A, HOPS–288–A–B, and HOPS–288-B), with a very close separation of $\sim$50~au between HOPS–288--A--A/B and a intermediate distance of $\sim$200~au to HOPS–288-B. It is not fully obvious from which component the CO outflow(s) reported by ALMASOP is originated. Given the close proximity of the triple system, our SED fitting result is more likely a descriptive model for the envelope of the circumbinary (HOPS–288--A--A/B) or the circum-triple system.


\subsection{Hot Corino and Envelope Infall Rate, Density, and Mass}
\label{sec:2023:Envelope}

\begin{figure}[htb!]
\includegraphics[width=0.45\textwidth]{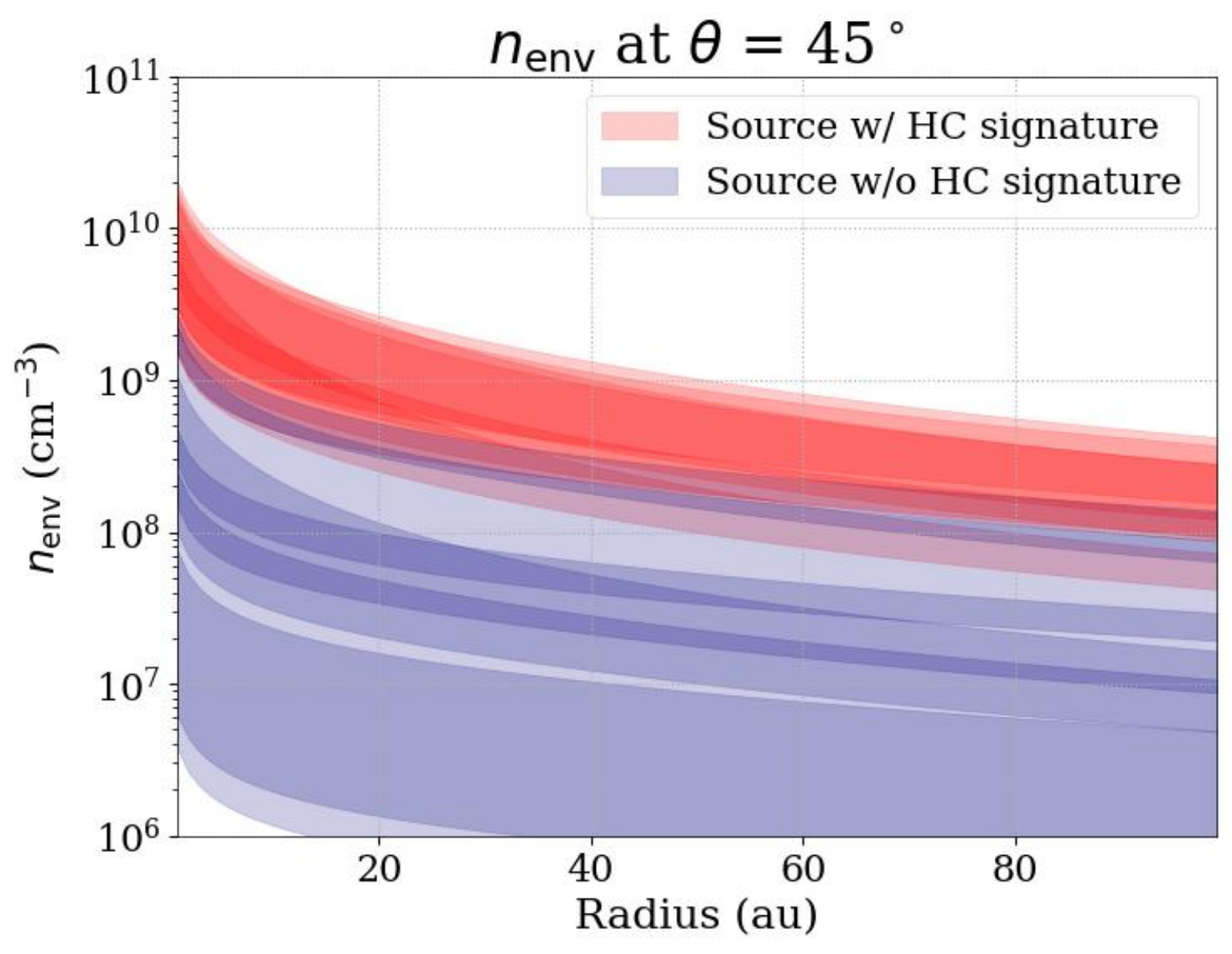}
\caption{\label{fig:2023:numDENSenv} 
The number density profile from the ``CMU'' envelope models. 
Each color band represents an individual source in our sample at modeled total luminosity \Ltot\ $> 5$ \Lsun. }
\end{figure}

\begin{figure*}[htb!]
\centering
\includegraphics[width=0.99\textwidth]{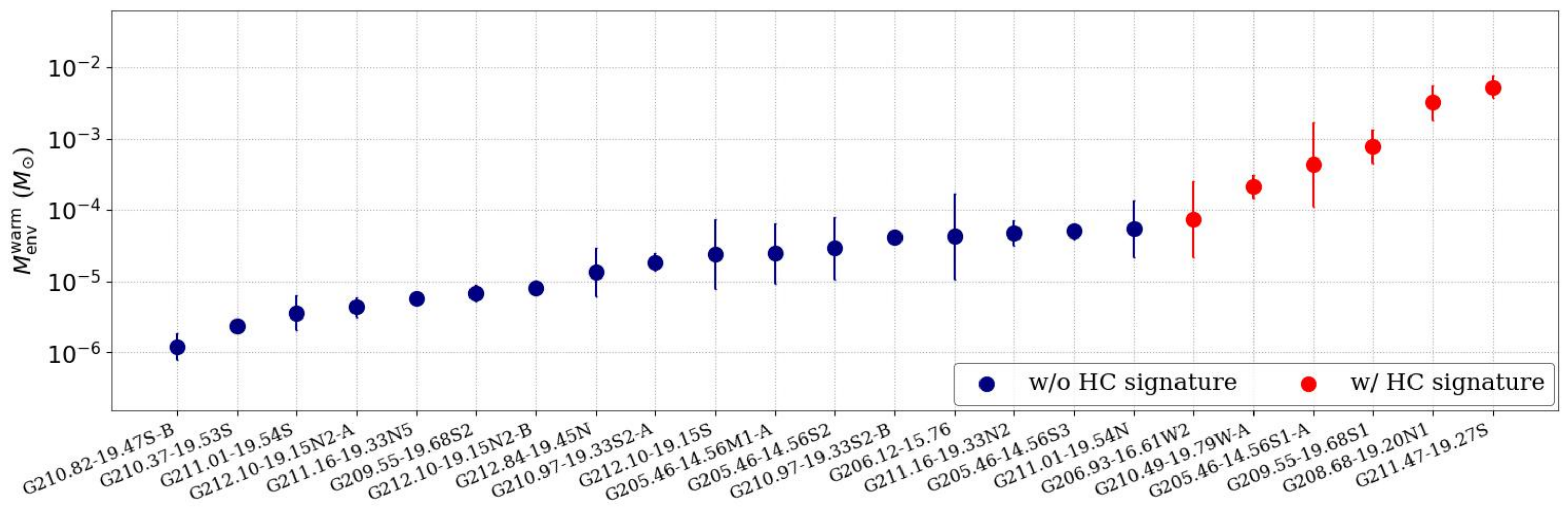}
\caption{\label{fig:2023:Menv_warm} 
The warm envelope mass inferred from the nine best-fit model physical parameters.  
The red and navy dots represent the sources with and without hot--corino signatures \citep{2020Hsu_ALMASOP,2022Hsu_ALMASOP}. 
}
\end{figure*}

At high luminosities\editbf{, namely modeled total luminosity \Ltot$>5$\Lsun}, we further find that the envelope infall rate (\MDOTenv) exhibits a trend between sources with and without hot--corino signatures. 
As shown in Figure~\ref{fig:2023:YSOPar_part}--(e), The envelope infall rates (\MDOTenv) for Class~0 and Class~I YSOs are clearly differentiated, with those of Class~0 YSOs ($\sim 5\times 10^{-5}$ \Msunyr) being significantly higher than those of Class~I YSOs ($\sim 5\times 10^{-6}$ \Msunyr). \editbf{These values are comparable to the median envelope infall rates obtained by \citet{2016Furlan_HOPS_SED} ($\sim 2.5\times 10^{-5}$ and $\sim 1.0\times 10^{-6}$ \Msunyr\ for Class~0 and Class~I YSOs, respectively).}
This differentiation presents a potential connection between envelope infall rate and COM emission intensity. 
The COM emission intensity presumably correlates with the total amount of gaseous COMs. 
The total amount of gaseous COMs should positively correlate with the warm envelope mass if COMs thermally desorb into the gas-phase within the envelope due to ice sublimation at $\sim 100$ K. 
Thus, the warm envelope mass is in turn related to the envelope infall rate. 

The envelope density model in \texttt{HO-CHUNK} is the ``CMU'' model \citep{1976Ulrich_CMU,1981Cassen_CMU},  in which the circumstellar material is freely falling with a slight rotation and the trajectories terminate at the disk midplane. 
The envelope density $\rho_\mathrm{env}$ at radius $r$ and polar angle $\theta$ in the ``CMU'' model is: 

\begin{equation}
\label{eq:RHOenv}
\begin{split}
    \rho_\mathrm{env}(r, \mu) & = 
    \rho_{\mathrm{env}.C} \left ( \frac{r}{R_C} \right )^{-3/2}\\
    & \times \left ( 1 + \frac{\mu}{\mu_0} \right )^{-1/2} 
    \left ( \frac{\mu}{\mu_0} + \frac{2\mu_0^2}{r/R_C} \right )^{-1}
    \;\;\mathrm{,} 
\end{split}
\end{equation}
where \Rc\ is centrifugal radius (same as the outer radius of the disk \RMAXdisk\ in the R06 grid), $\mu=\cos\theta$,  $\mu_0$ is the value of $\mu$ for the streamline of infalling particles as $r\rightarrow\infty$, satisfying:

\begin{equation}
\label{eq:Rc}
    \frac{r}{R_C}=\frac{1-\mu_0^2}{1-\mu/\mu_0}
    \;\;\mathrm{,} 
\end{equation}
and the characteristic density factor $\rho_{\mathrm{env}.C}$ is defined by: 

\begin{equation}
\label{eq:RHOenvC}
    \rho_{\mathrm{env}.C} 
    = \frac{M_\mathrm{env}}{4\pi R_C^3}
    = \frac{\dot{M}_\mathrm{env}}{4\pi (GM_\star R_C^3)^{1/2}}
    \;\;\mathrm{,} 
\end{equation}
where \Menv\ is envelope mass, \MDOTenv\ is envelope mass infall rate, and \Mstar\ is the stellar mass. 
Note that a singularity in $\rho_{\mathrm{env}.C}$ occurs in the midplane at the centrifugal radius (\Rc). 

We evaluate the nominal envelope number density ($n_\mathrm{env} = \rho_\mathrm{env}/m_\mathrm{H_2}$, where $m_\mathrm{H_2}$ is the mean molecular hydrogen mass) at radius $r=50$~au (half of the typical hot corino size) and polar angle $\theta=45^\circ$ (to avoid outflow cavity and singularity in the midplane) for all the R06 grid, which includes the sets of best-fit models for the ALMASOP sample. 
As shown in Figure~\ref{fig:2023:YSOPar_part}--(f),
the Class~I sources generally have lower envelope density than the Class~0 sources. 
This differentiation is particularly evident at high luminosities, where Class~0 YSOs show hot--corino signatures but Class~I YSOs do not. 
In Figure~\ref{fig:2023:numDENSenv}, we compare the envelope molecular gas number density profile at polar angle $\theta=45^{\circ}$ (to avoid outflow cavity and density singularity in the midplane) for sources with modeled total luminosity \Ltot\ $>$ 5 \Lsun {(the lower limit for our sources with hot corino detected)}. 
The differentiation between the sources with and without hot--corino signatures in our sample is generally valid. 
We note again that this envelope density value is an indicator of the envelope density level. 




To take into account the overall physical structure and compute the warm envelope mass, we run the \texttt{HO-CHUNK} simulations for our sample and then integrate the density within the following region: (1) having a temperature higher than 100 K (ice sublimation temperature) and (2) being outside the disk-dominant region (within three times of the disk scale height) or the cavity interior.
As shown in Figure~\ref{fig:2023:Menv_warm}, the sources with hot--corino signature indeed have high warm envelope mass. 


\section{Methanol Emission in Warm Envelopes}
\label{sec:2023:Sparx}
Our SED analysis favors the hypothesis that COM emission leading to the hot--corino signature originates from a warm envelope. 
To verify whether the existence of COMs in an envelope can produce emission consistent with actual observations, we carried out imaging simulations of CH$_3$OH emission based on the previous SED model parameters. 
\editbf{
We select the three most luminous sources (in terms of either their observed bolometric luminosity, the averaged total luminosity from the nine best-fit models, or their total luminosity from the best-fit model)} in which hot corinos have been detected in the ALMASOP project \citep{2020Hsu_ALMASOP, 2022Hsu_ALMASOP}, namely  G208.68-19.20N1 (HOPS--87), G210.49-19.79W--A (HOPS--168), and G211.47-19.27S (HOPS--288), for the simulation, hereafter G208N1, G210WA, and G211S, respectively. 

\subsection{Physical Parameters}
We first use \texttt{HO-CHUNK} to calculate the thermal structure, namely, the temperature profile based on the best-fit YSO model exported by the \texttt{SED Fitter}.
Refer to Table \ref{tab:2023:YSOModel_sparx} for the parameters used in the \texttt{HO-CHUNK} simulations. 

\begin{figure*}[htb!]
    \gridline{\fig{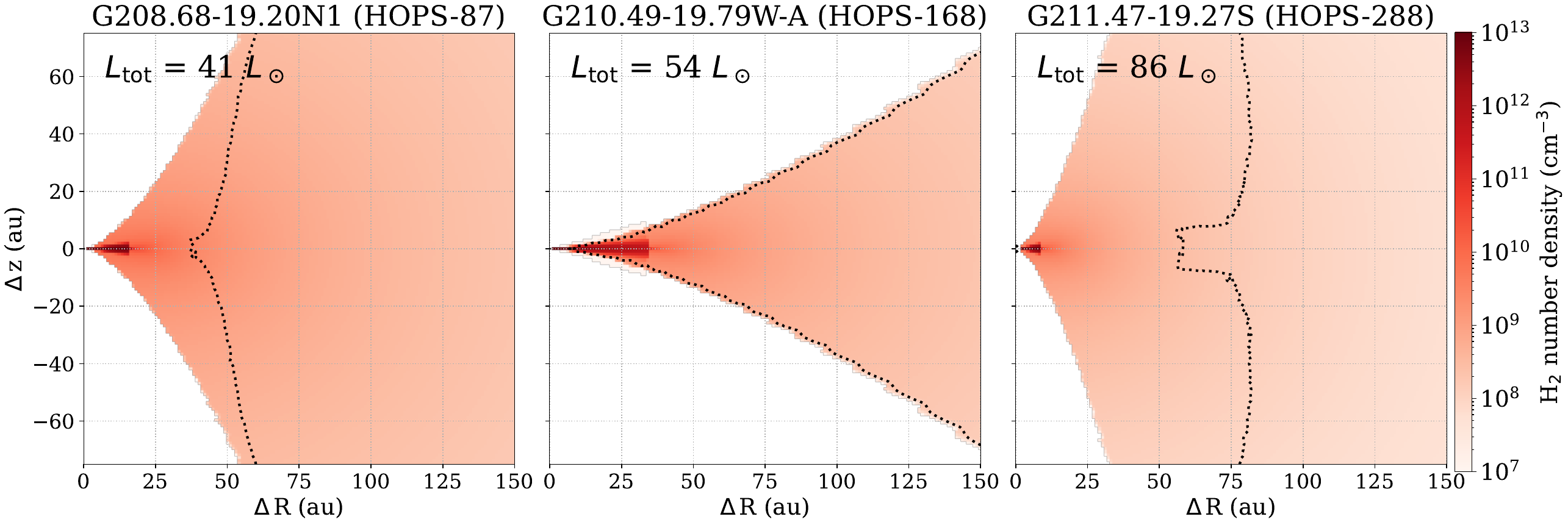}{0.99\textwidth}{(a) H$_2$ number density}}
    \gridline{\fig{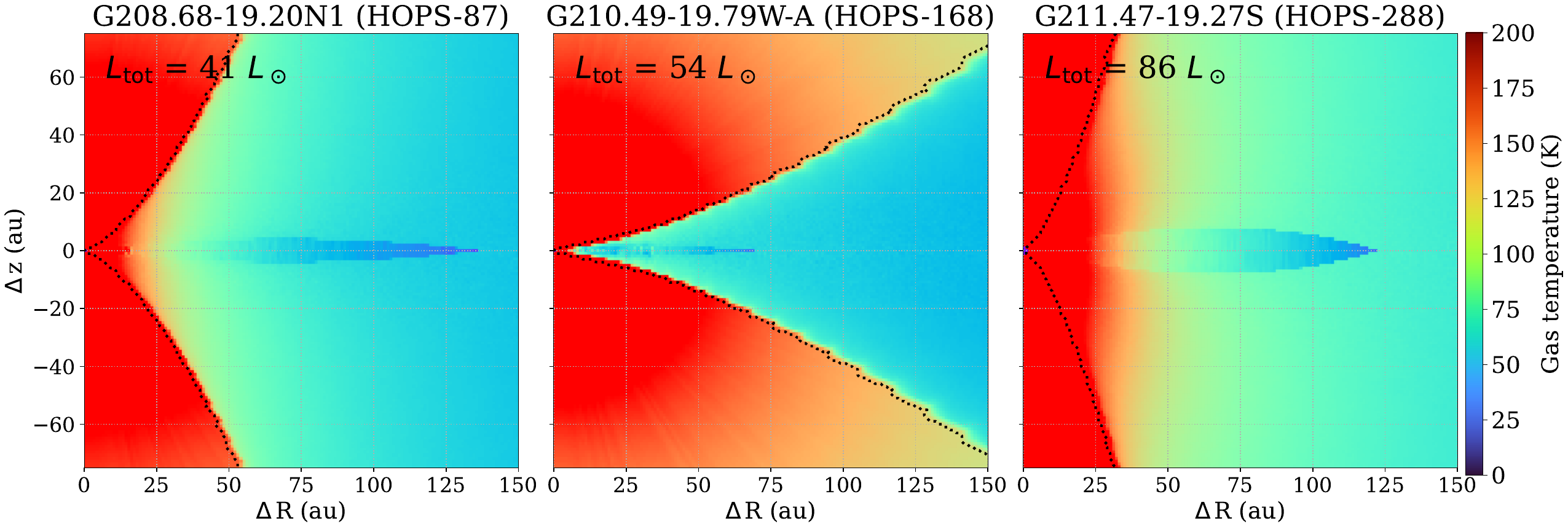}{0.99\textwidth}{(b) Gas temperature}}
    \gridline{\fig{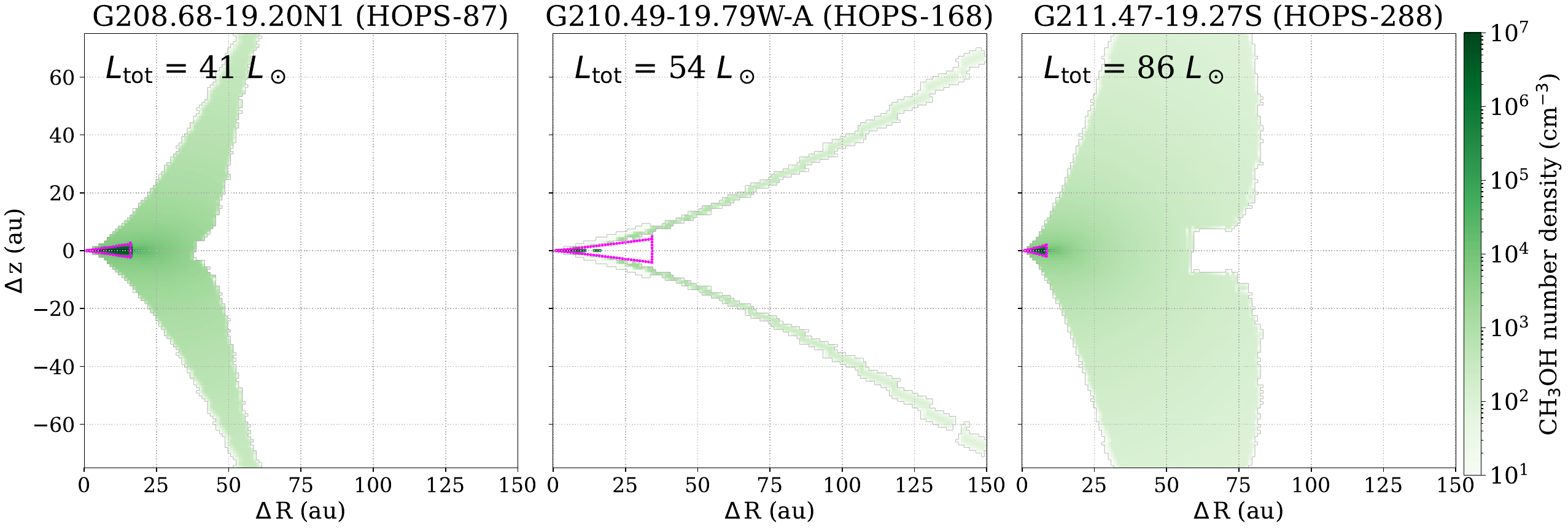}{0.99\textwidth}{(c) CH$_3$OH number density}}
\caption{\label{fig:2023:map_RHO_Tgas} 
The molecular hydrogen number density (a), gas temperature (b), and methanol number density (c) maps for the three sources where hot corinos were ACA-detected. 
The text in the upper left corner of each panel shows the total luminosity of the best-fit model exported by the \texttt{SED Fitter}. 
The dotted contours in the top panels (a) show the location where the gas temperature attains 100 K.  
The dotted contours in the middle panels (b) illustrate the edge of the outflow cavities. 
The shadowed regions in the bottom panels are the areas between the dashed contours in (a) and (b), while the color scale represents the number density of methanol. 
The magenta dotted lines in the bottom panels (c) illustrate the disk--dominant boundaries, set at three times of the disk scale height. 
At the disk outer radius (\RMAXdisk), the disk--dominant boundaries are of 2.4, 4.2, and 2.1~au, respectively, in their vertical heights.  
}
\end{figure*}

Figure~\ref{fig:2023:map_RHO_Tgas}--(a) and --(b) show the model-derived gas number density and the temperature distribution, respectively, for the three sources.
The flaring disks dominate the inner densest regions near the midplane, and the warm bipolar regions are due to the outflow cavities. 
As found by \citet{2003Whitney_SED}, the outer midplane is relatively cooler since the optically thick disk blocks much of the stellar radiation.
In contrast, the bipolar cavity is relatively warmer due to its direct radiation exposure.
Based on these thermal structures, we next determine the distribution of gas-phase methanol.
We assume that the entire warm (i.e., $T > 100$ K , the ice sublimation temperature) region, excluding the cavity interior, harbors gaseous CH$_3$OH, the most commonly-detected COM.

\subsection{Image Simulation of Methanol Emission}
We use \texttt{SPARX}\footnote{https://sparx.tiara.sinica.edu.tw/} (Simulation Package for Astronomical Radiative Xfer), a radiative transfer code for calculating molecular line and dust continuum radiation, to produce synthetic CH$_3$OH image cubes with the number density and temperature profiles exported by \texttt{HO-CHUNK} (Sect. \ref{sec:2023:SED}). 
Following the approach detailed in \texttt{HO-CHUNK} \citep{2003Whitney_SED} and used by \citet{2010Keto_SPARX_velocity}, we adopt the gas velocity in the infalling envelope via the ``CMU'' \citep{1976Ulrich_CMU,1981Cassen_CMU,2004Mendoza_YSO_velocity} model, and assume that the gas velocity in the disk follows a Keplerian velocity. 
We set a fixed fractional abundance (i.e., the relative number density with respect to H$_2$) of gas-phase CH$_3$OH of $X=10^{-6}$ for all sources. This abundance is comparable to that of $\sim10^{-7}$--$10^{-6}$ determined by \citet{2022Hsu_ALMASOP} and to the CH$_3$OH abundance found in ices \citep{2015Boogert_ice_review, 2022Brunken_CH3OCH3}.
The methanol line that we chose for the demonstration is the E-CH$_3$OH $4(2,3)-3(1,2)$ transition, which is the strongest methanol transition in the ALMASOP observations having upper energy \Eu\ $=46$~K and rest frequency \frest\ $=218440.0$~MHz. 
The level populations are assumed to be under local thermodynamic equilibrium (LTE), as the modeled H$_2$ number density is well--above the critical density at $\sim 10^6$ cm$^{-3}$ of this transition.

\subsubsection{Envelope+disk case}
As noted previously, the YSO physical model consists of a central protostar, an infalling and rotating envelope, a flared disk, and a bipolar cavity. 
In the ``envelope+disk'' case, the gas-phase COM emission will arise from the (inner) warm disk and the envelope surface below the outflow cavity, forming a bipolar cone-shaped shell, as shown in Figure~\ref{fig:2023:map_RHO_Tgas}--(c).

To compare the resulting synthetic CH$_3$OH image cubes with the ALMASOP observational maps, we further made simulated maps by subtracting the synthetic (line-free) continuum image from the image cubes, convolving the continuum-subtracted cubes with the ALMASOP elliptical beam (the geometric mean $\overline{\mathrm{HPBW}}$ is of $\sim$0\farcs{45}, or $\sim$ 180~au at a distance $d$ = 400~pc. 
See Table \ref{tab:2023:YSOModel_par}). 
We made the moment--0 maps within the velocity intervals based on the line width reported by \citet{2022Hsu_ALMASOP} around the line center and imposed a simulated noise level consistent with the real observations. 
Given that the CH$_3$OH emission is compact in all cases, we expected negligible missing flux due to lack of short baselines, and therefore carry out imaging simulations without employing the "\texttt{simobserve}" task in Common Astronomy Software Applications package (CASA).

\begin{figure*}[htb!]
    \gridline{\fig{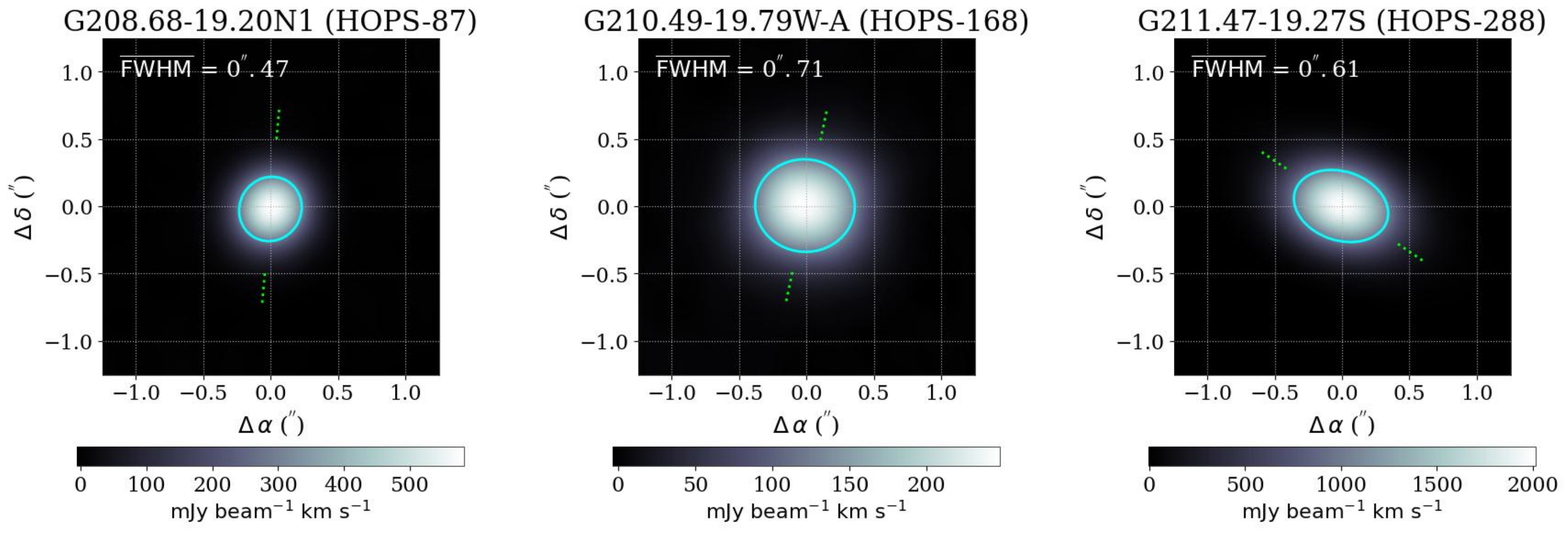}{0.99\textwidth}{(a) ``Envelope+disk'' case}}
    \gridline{\fig{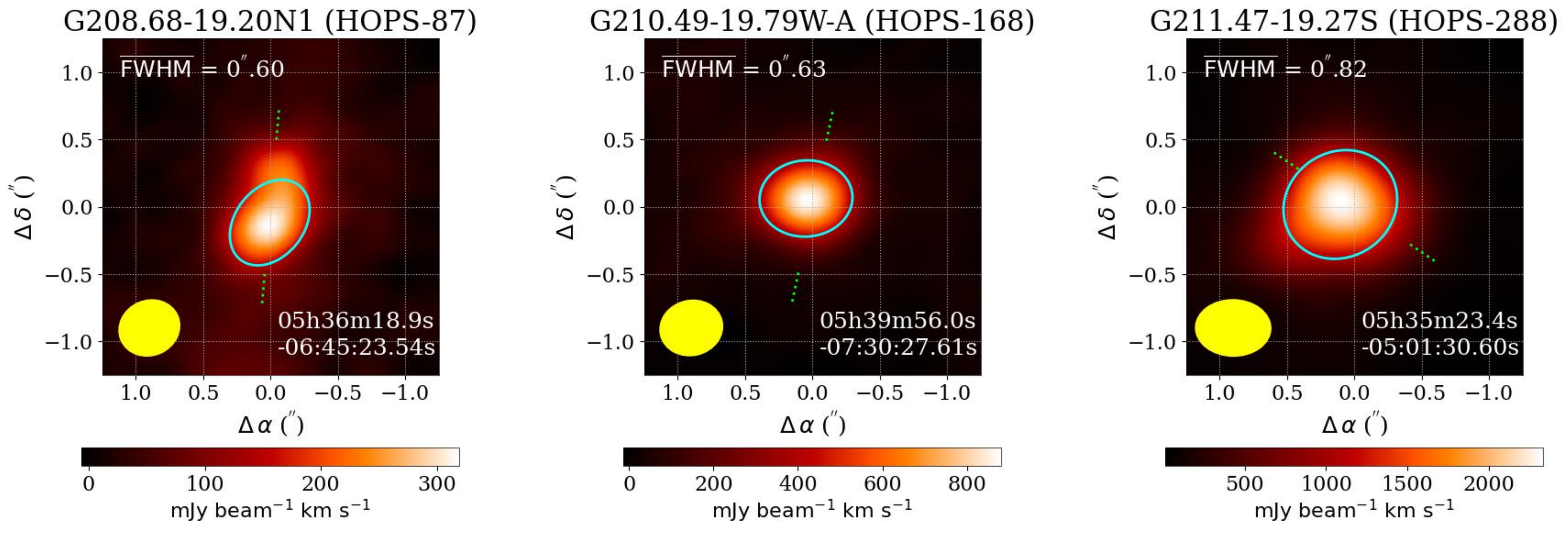}{0.99\textwidth}{(b) ALMASOP observation}}
    \gridline{\fig{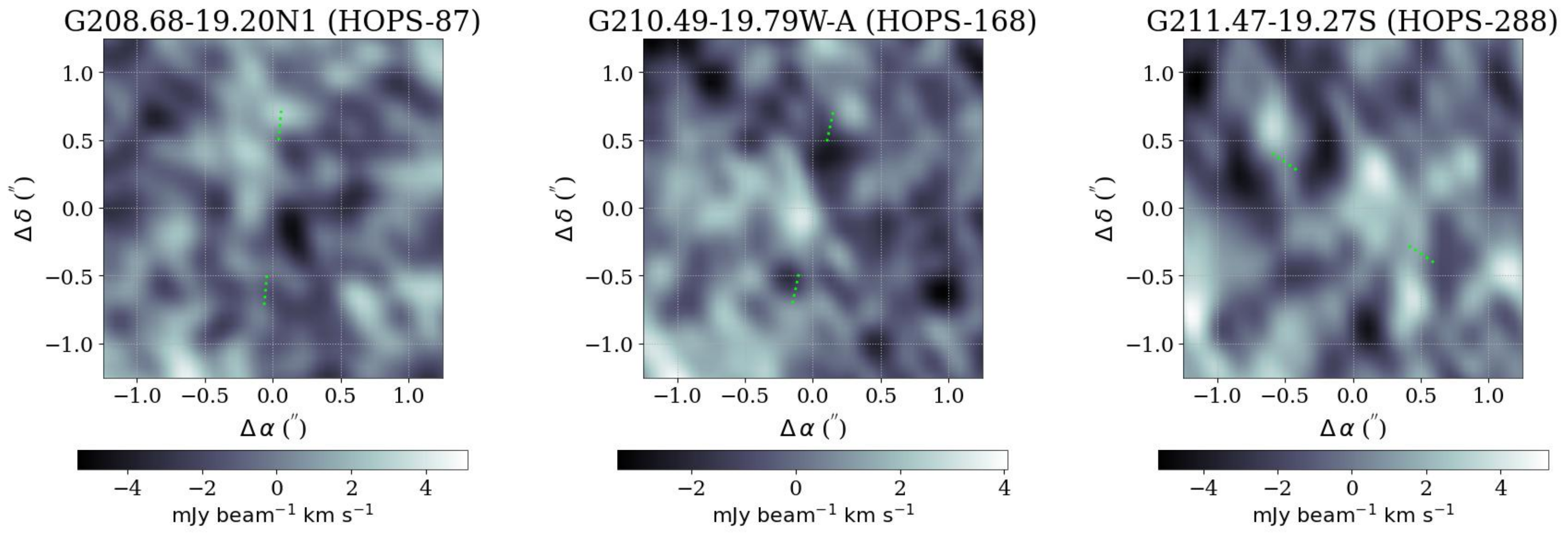}{0.99\textwidth}{(c) Disk case} }
\caption{\label{fig:2023:map_CH3OH46} 
The top (a) and the bottom (c) panels are the \texttt{SPARX} simulation results for the ``envelope+disk'' case and the ``disk'' case. 
The middle (b) panels are the observed data from ALMASOP, and the center coordinates are labeled at the bottom right corners. 
The x-axes and the y-axes are the relative right ascension ($\Delta\,\alpha$) and the relative declination ($\Delta\,\delta$), respectively. 
The aqua ellipse illustrates the FWHM from 2D Gaussian fitting, and the geometric mean of the FWHM is shown at the top of each panel. 
Please refer to Table \ref{tab:2023:sparx} for more details of the 2D Gaussian fitting result. 
The lime dashed lines show the cavity axis \citep{2020Dutta_ALMASOP,2022Hsu_ALMASOP}. 
The two rightmost panels in the bottom row have no 2D Gaussian fitting result since their integrated intensities are too weak. 
\hfill\break
\hfill\break
\hfill\break
}
\end{figure*}

Figure~\ref{fig:2023:map_CH3OH46}--(a) shows the simulated moment--0 maps of the CH$_3$OH emission for the ``envelope+disk'' case, and Figure~\ref{fig:2023:map_CH3OH46}--(b) shows the ALMASOP observed spectral images for the same sources based on their best--fit YSO models.
We apply 2D Gaussian fitting to the integrated CH$_3$OH emission images in both sets of maps.
In each panel of Figure~\ref{fig:2023:map_CH3OH46}--(a) and --(b) , the aqua ellipse illustrates the full--width--half--maximum (FWHM) of the image and the label shows geometric mean of the FWHMs. 
Moreover, we consider all nine best--fit YSO models for the three sources and calculate the weighted average and standard deviation of the simulated methanol emission peak intensity and size.
As summarized in Table \ref{tab:2023:sparx}, with a fixed CH$_3$OH fractional abundance $X=10^{-6}$, the differences between the \texttt{SPARX} imaging simulations and the ALMASOP observation are in general less than a factor of two, both in terms of the integrated intensity and the emission extent.
The only major difference appears in the integrated intensity for the case of G210.49--19.79W--A, which is a factor of six. 
\editbf{ 
Such a discrepancy might be caused by a relatively higher fractional abundance of methanol $X$ in the source, which is fixed to $10^{-6}$ in the imaging simulation.
}


\begin{deluxetable}{lrrrr}
\tabletypesize{\small}
\setlength{\tabcolsep}{4pt}
\tablecaption{\label{tab:2023:sparx} Comparison between \texttt{SPARX} simulation and ALMASOP observation}
\tablehead{
\colhead{Source} & \multicolumn{2}{c}{\FWHM} & \multicolumn{2}{c}{Peak} \\
\colhead{} & \multicolumn{2}{c}{(${''}$)} & \multicolumn{2}{c}{(mJy beam$^{-1}$ km s$^{-1}$)} \\
\hline
\colhead{} & \colhead{\texttt{SPARX}} & \colhead{ALMASOP} & \colhead{\texttt{SPARX}} & \colhead{ALMASOP}
}
\startdata
G208N1 & $0\farcs50\pm0\farcs02$ & $0\farcs41 \pm 0\farcs01$ & $623\pm227$ & $321 \pm 4$ \\
G210WA & $0\farcs72\pm0\farcs05$  & $0\farcs45 \pm 0\farcs03$ & $145\pm70$ & $834 \pm 36$ \\
G211S & $0\farcs62\pm0\farcs04$ & $0\farcs67 \pm 0\farcs03$ & $1706\pm356$  & $2340 \pm 54$ \\
\enddata
\tablecomments{
The transition is the CH$_3$OH-46 transition. 
The methanol abundance model in \texttt{SPARX} is the ``envelope+disk'' model. 
The result for \texttt{SPARX} are the weighted average and standard deviation for the nine best--fit YSO models. 
}
\end{deluxetable}


\subsubsection{Disk case}
The above "envelope+disk" cases reproduce well the observed CH$_3$OH emission.
To examine whether the CH$_3$OH emission is predominantly from the disk or from the envelope, we further generated simulated maps for ``disk''-only cases in which gas-phase methanol molecules just reside in disk regions.
In other words, different from the ``envelope+disk'' case, we only consider methanol within three times the scale height above and below the  disk midplane, as illustrated by the magenta dashed curves in Figure~\ref{fig:2023:map_RHO_Tgas}--(c).

As shown in Figure~\ref{fig:2023:map_CH3OH46}--(c), the simulated CH$_3$OH integrated intensity images for these ``disk'' cases appear quite different from the observed moment-0 maps.
The simulated methanol integrated emission is even below the noise level of the observations.
For the ``disk" cases in general, despite gas phase CH$_3$OH residing in the disks, the modeled CH$_3$OH spectral emission is either too compact and weak or getting veiled by the very dense and consequently optically-thick dust continuum.
 

In summary, the lack of correlation between the disk parameters and the presence of hot--corino signatures, as indicated in Sect.\ \ref{sec:2023:SED}, implies that the COM emission does not originate from only disks in these objects. 
Moreover, the ``envelope+disk'' cases reproduce the ALMASOP CH$_3$OH results much better, which implies that CH$_3$OH in the warm envelope is likely responsible for the observed COM emission.


\section{Discussion}
\label{sec:2023:Disc}

\subsection{Warm-envelope Origin of Hot Corino}
Our findings, including the correlation between the corino detection and the warm envelope mass and the detailed imaging simulation strongly favors the hypothesis that the warm COM emission seen in the ALMASOP sample originates within the YSO warm envelope.
This \LED\ follows the thermal desorption paradigm, in which COMs thermally desorb from dust grains when the temperature exceeds the ice sublimation temperature at $\sim 100$ K \citep[e.g., ][]{2006Garrod_3phase,2008Garrod_3phase, 2009Herbst_COM_review}. 
The central protostar is crucial in heating the surrounding environments, including the circumstellar envelope and the Keplerian disk. 
However, the disk-shadowing effect, as depicted in \ref{fig:2023:map_RHO_Tgas}--(a) and --(b) leads to lower temperatures in the regions near the midplane, resulting in a reduced warm region within the envelope and particularly the disk. 
Such disk-shadowing effect, for example, has been observed by \citet{2015Murillo_disk-shadowing} toward the Class~0 protostar VLA~1623A and demonstrated by \citet{2022Nazari_CH3OH_YSOmodel} through the comparison between their ``envelope-only'' models and  ``envelope-plus-disk'' models. 
In contrast, as shown in Figure~\ref{fig:2023:map_RHO_Tgas}--(a) and --(b), regions in the envelope further away from the midplane are less affected by the disk--shadowing effect. 
Additionally, as indicated by Figure~\ref{fig:2023:map_RHO_Tgas}--(a) and --(b), photons emitted by the central protostar can penetrate the outflow cavities and heat up the envelope. 
Without an outflow cavity, the warm region size in the envelope may be significantly reduced. 
As a result, the envelope component predominates the warm region in a protostellar system and leads to the dominance of the observed hot--corino signature.

In the \LED, the detectability of gaseous COMs is predominately dictated by both the luminosity and the envelope density profile. 
A higher YSO luminosity leads to a broader region harboring gas-phase COMs, as demonstrated by \citet{2022Hsu_ALMASOP}.
Furthermore, the higher envelope density (with a fixed fractional abundance) results in a greater amount of COMs.
At a fixed observing sensitivity, a Class~0/I protostar having a more massive warm envelope will give rise to more intense COM emission, making the hot corino in this protostar more easily detectable. 


The common warm-envelope origin found in the ALMASOP sample implies that any Class~0/I protostars could harbor a hot corino, provided that their inner envelope reaches a sufficient temperature, \editbf{as was initially suggested by \citet{2004Ceccarelli_HotCorino}}. 
Consequently, with \editbf{appropriate observations, such as a high--sensitivity census of CH$_3$OH toward protostars with yet--detected hot corinos}, it is likely that the majority of Class~0/I protostars will exhibit hot--corino signatures. 
The warm--envelope origin can be verified by not only spatially resolved imaging but also accurate kinematics allowing precise position-velocity (PV) diagrams.

Before such observations are achieved, it is worthwhile to examine whether the above warm--envelope scenario discovered in ALMASOP could be generalized to the hot corino detection in other studies. 
In the following, we discuss results in the literature showing indicative signs of support to the scenario.

\subsubsection{The Luminosity and the Radius of COM Extent}
For a sample of protostars having comparable envelope densities (e.g., all at Class~0 stage), the COM emission extent is expected to be positively correlated with source luminosity, as is indeed suggested by \citet{2022Hsu_ALMASOP}.  
Results reported in the literature well--constraining the COM extent for hot--corino sources, L483 and B335, also suggest similar trends. 
Both L483 and B335 are nearly edge--on (i.e., inclination angle $\varphi\geq80^{\circ}$) Class~0 protostellar cores lacking detected Keplerian disk. 
L483, located in the Serpens–Aquila Rift with a bolometric luminosity, \Lbol$ =10.5$ \Lsun\, was found to exhibit COM emission at a scale of 40 -- 60~au.  \citep{2000Shirley_L483_Tbol,2018Oya_L483_incl,2019Jacobsen_L483_COM}. 
\citet{2019Jacobsen_L483_COM} attributed this emission to the innermost warm envelope.
B335, in Bok globule B335, has a relatively low bolometric luminosity (\Lbol$=1.36$ \Lsun), and the COM distribution remains unresolved at a scale of $\sim$ 45~au \citep[or HPBW$=$0\farcs{55} at a distance $d\sim165$ pc,][]{1988Hirano_B335_incl,2010Yen_B335_incl,2016Imai_B335,2020Watson_B335_distance,2022Evans_B335_modeling}. 
There exist luminosity variations in B335 \citep{2022Evans_B335_modeling}, which have admittedly led to variations in the warm region size.

\subsubsection{The envelope density of Class~I Protostars with Hot Corino Detected}
A Class~I protostar generally has a lower envelope density than a Class~0 protostar, which based on our hypothesis, would make the detection of COM emission toward a Class~I protostar more difficult. 
In our ALMASOP sample, for example, the modeled envelope number densities at our nominal reference location, $r=50$~au and $\theta=45^\circ$, of all Class~I protostars are below $3\times 10^8$ \cmiii, as shown in Figure~\ref{fig:2023:YSOPar_part}--(f). 
Even though some of them have high luminosities and narrow cavity angles (e.g., \Ltot$\sim30$\Lsun\ and \THETAcav$\sim$27\degr\ for G205.46--14.56S3), no hot--corino signatures were detected toward them.  
Besides, \citet{2014Lindberg_IRS7B_disk} and \citet{2018Artur_IRS67_disk} interpreted the non--detection of CH$_3$OH emission toward protostellar cores R Cra--IRS 7B (a borderline Class~0/I protostar at 4.6 \Lsun) and Oph--IRS 67 (a Class~I binary protostellar system at 4.0 \Lsun) 
\editbf{as a result of the low density in their warm inner envelopes because of the presence of a disk and the consequent flat envelope density profiles.}

However, Class~I protostars with high envelope densities may change such a situation. 
For example, both Ser--emb 17 and L1551--IRS5, having their average cavity opening angles also less than 45\degr\ \citep{2020Bergner_Ser-emb-17_chem, 2009Wu_L1551-IRS5_outflow}, are Class~I protostellar cores exhibiting hot--corino signatures.
Ser--emb 17 has an envelope mass (\Menv) of $3.6$ \Msun\ \citep{2011Enoch_Ser-emb_bol,2019Bergner_Ser-emb_COM}. 
Assuming a centrifugal radius \Rc\ $\sim$ 100~au, at the nominal reference location, the estimated envelope density would be $2.2\times 10^{10}$ \cmiii\ (Equation \ref{eq:RHOenv}), which is even higher than those of the ALMASOP sources showing hot--corino signatures in Figure~\ref{fig:2023:YSOPar_part}--(g).
Similarly, for L1551--IRS5, \citet{2003Osorio_L1551IRS5_SEDFitting} presented a protostellar model consisting of a flattened infalling envelope surrounding a binary system with individual circumstellar disks and a circumbinary disk through SED fitting. 
The reported envelope mass and the centrifugal radius are of 4 \Msun\ and 300~au, respectively.
By adopting these two parameters, we estimate that the corresponding envelope number density at the reference location is $1.7\times 10^{9}$ \cmiii\ (Equation \ref{eq:RHOenv}), comparable to those of sources with hot--corino signatures in ALMASOP, as indicated in Figure~\ref{fig:2023:YSOPar_part}--(g).

\subsubsection{The sensitivity and the presence of hot--corino signature}
The hot--corino signatures of YSOs with higher luminosities are more likely to be detected at a given observational sensitivity if their envelope densities are comparable. 
This implication results from the fact that the sources with hot--corino signature in ALMASOP are mostly, if not all, Class~0 YSOs at the highest luminosities among the whole sample of Class~0 protostars. 
This picture also appears applicable to archival Class~I hot corino surveys. 
\citet{2019Bergner_Ser-emb_COM}, for example, observed two Class~I protostars; among the two, Ser--emb 17 with the detection of COM emission has a significantly higher luminosity (of 3.8 \Lsun) than the other (of 0.4 \Lsun) with no COM emission detected. 
\citet{2022Mercimek_ClassI_COMSurvey} conducted a chemical survey of four Class~I protostars distributed in the Taurus and Perseus clouds. 
Of the four Class~I protostars, only L1551–IRS5, having the highest bolometric luminosity (30 -- 40 \Lsun\ for L1551--IRS5 and 3.5 -- 5.0 \Lsun\ for others), appeared to be COM-rich.
Moreover, \citet{2020Bianchi_L1551-IRS5} conducted a chemical survey of the L1551--IRS5 binary system and found that the brighter (1.3 mm) component (N) dominates the COM emission. 
For the ALMASOP YSO sample, observations with improved sensitivities will verify the existing but undiscovered hot--corino signatures in the YSOs with low envelope masses (marked the \editbf{navy dots} in Figure~\ref{fig:2023:Menv_warm}).

\subsection{Complications}
While this study suggests that the hot--corino signature commonly originates from the warm envelope, there are complications to this scenario regarding the presence of COMs and their detection in protostellar objects.
\editbf{
For example, there may be temporal variations in the total luminosity and consequently the level of heating due to episodic accretion events, such as those seen in FU Ori objects. 
The extent of the warm region and equivalently the location of the snowline will therefore move outwards as a result of the outburst, as have been shown by literature such as \citet{2018vantHoff_snowline} and \citet{2019Lee_snowline_outburst}.}
In addition, the radiation emitted from the central protostar may not be the only heating source in a protostellar system. 
\citet{2021Tabone_HH212_cavity} speculated that a different heating mechanism due to the accretion shocks might explain the presence of disk--origin COM emission near the centrifugal barrier of the Class~0 protostar HH--212 \citep{2017Lee_HH212,2019Lee_HH212_COM_atm}.
A similar mechanism was later suggested by \citet{2020Oya_IRAS16293-2422-A_few_au} based on the radial profile of the rotational temperature in H$_2$CS in IRAS~16293--2422~A. 
Recently, this localized heating mechanism was further supported by the ring--like elevated warm structure in the Class~0 protostar B335 \citep{2022Okoda_B335_few_au}.

Furthermore, other mechanisms may also contribute to the presence of gaseous COMs in protostellar systems. 
These mechanisms include outflow shocks, where energetic outflows result in shock--induced desorption of COMs \citep{2022Vastel_BHB2007-11_COM_outflow}, 
and reactive desorption, where the penetration of photons into the cavity leads to the photodissociation of icy COMs and the subsequent recombination of their photoproducts \citep{2015Drozdovskaya_cavity_COMs}. 
These mechanisms play a significant role in shaping the chemistry in protostellar systems and should be considered when studying the distribution and composition of COMs in YSO environments.

We also note that, in addition to the limitation due to observational sensitivity, other factors may affect the detectability of COM emission. 
Dust continuum opacity, for instance, could be one possible reason for a non--detection.
\citet{2019Sahu_IRAS4A1_hot_corino_atmosphere} observed the protostellar binary system IRAS~4A1 and 4A2 and suggested that the absence of COM emission in IRAS 4A1 could be due to an optically thick circumstellar disk and/or different layers of a temperature-stratified and dense envelope, which veils the COM emission and in fact results in COM absorption signatures.
This hypothesis was subsequently supported by \citet{2020DeSimone_dust_opacity} through observations at longer, centimeter wavelengths at which the dust continuum opacity becomes optically thin.
\editbf{ 
Similar effects may occur in SVS13--A binary system, as suggested by \citep{2022Bianchi_SVS13A_COM_binary}. 
}

\editbf{
Finally, the non--detection of gaseous COMs may simply be due to a lack of COM emission in protostar cores.
\citet{2019Oya_Elias29_sulfur} found that Elias 29, a Class~I protostar, is (likely) poor in COMs as the desorbed COMs might have been destroyed by the proton transfer reactions. 
Alternatively, there may be chemical diversity between protostellar cores by nature.
For example, \citet{2008Sakai_L1527_WCCC} showed that the protostellar core in the low--mass star--forming region L1527 is rich in carbon--chain molecules but shows a lack of saturated COMs typically found in hot corinos.
The authors introduced the concept of ``warm--carbon--chain chemistry (WCCC)'' to explain such chemical composition in the protostellar system.
However, there are sources, \citep[e.g., L483, ][]{2017Oya_L483_HCC_WCCC} exhibiting signatures of both hot--corino and WCCC simultaneously.
It thus remains uncertain whether the sources subjected to WCCC indeed lack in gaseous saturated COMs. 
}

\subsection{Implications for Astrochemistry in Cores and Disks}
\label{sec:2023:Chem_c2d}

Complex organic species, such as CH$_3$OH, previously residing on icy grain mantles, desorb into the gas phase as gas and grains migrate from the outer colder region into the warmer inner region of the infalling envelope. 
This process can be most naturally ubiquitous in embedded YSOs\editbf{, as proposed by \citet{2004Ceccarelli_HotCorino}}. 
The detection of hot--corino signatures, as entailed by our analysis, is dictated by the total amount of COMs, which is related to the size and density of the inner warm region in the infalling envelope and, equivalently, the luminosity and evolution stage of the YSO.
The gaseous COMs (CH$_3$OH specifically in this study) seen in the warm inner infalling envelope may reflect the ``fresh" ice composition in grain mantles, perhaps more closely than seen in massive hot molecular cores where the chemistry suffers additional alteration due to harsher UV radiation from the \editbf{central} massive YSOs \citep[e.g., ][]{2022Olguin_G335MM1ALMA1_CH3OH}. 
The connection of the hot--corino signature to the chemistry in the later developed protoplanetary disks may depend on whether the COMs will get dispersed \citep{2006Ferreira_diskwind}, destroyed in the gas phase \citep{1992Charnley_hot_core_orion,1998Millar_model_core,2009Nomura_disk_accretion,2016Yoneda_chemistry_disk}, or eventually settle within the disk.
The inheritance nature of COMs in protoplanetary disks, as hinted by 
\citet{2016Yoneda_chemistry_disk}, may depend on the location within the disk and the COM species. 
COMs seen in later stages of protoplanetary disks are sometimes interpreted as of thermal desorption origin \citep{2020vantHoff_temperature_disk_Taurus, 2021Booth_CH3OH_HD100546, 2021vanderMarel_CH3OH_OphIRS48}. 
In case their abundances are indicative of being inherited from the prestellar stage, the icy mantles are likely preserved with the dust grains accreted onto the outer and colder disk, as also discussed by \citet{2013Hincelin_chemical_ISM_disk}. 
The ice mantles then sublimated when dust grains migrate into the inner disk inside the snowline, or the snowline gets pushed outwards from a luminosity burst event \citep{2019Lee_snowline_outburst}. 
Minor bodies in the solar system exhibiting inheritance of ISM chemistry may have a similar accretion and thermal history.


\section{Conclusions}
\label{sec:2023:Conclusions}

\begin{enumerate}
  \item We carry out SED fitting analysis for the Class~0/I protostars previously observed by the ALMASOP project and report the distributions of the modeled YSO physical parameters. 
  The differentiation in the YSO model parameters among the sources with and without hot--corino signatures implies a link between the detection of the hot--corino signature and the envelope density profile for YSOs with similar luminosity.
  Our modeling results suggest that the sources with hot--corino signatures have high luminosities, envelope densities, and consequently high warm envelope mass. 
  
  \item We further carry out simulations of the methanol moment--0 integrated intensity images and compared them with the ALMASOP observations. 
  The ``envelope+disk'' models produce the synthetic CH$_3$OH images matching with the ALMASOP observations significantly better than the ``disk'' models.
  



  \item Our study favors the hypothesis that the origin of the detected hot--corino signature is commonly the warm region within the protostellar envelope.
  Under this scenario, the detectability of a hot corino depends on the warm envelope mass, determined by the warm region size and the envelope density profile. 
  The former is governed by the source luminosity and is additionally affected by the whole protostellar structure, such as the disk and cavity.
  The latter is related to the YSO evolutionary stage. 
  
  \item 
  The presence of gaseous COMs in the warm inner envelopes can be most naturally ubiquitous in embedded YSOs.
  The gaseous COMs seen in the warm inner envelope in sources with hot--corino signature may reflect the ``fresh'' ice composition in grain mantles. 
  

\end{enumerate}


\acknowledgments
This paper makes use of the following ALMA data: ADS/JAO.ALMA\#2018.1.00302.S. ALMA is a partnership of ESO (representing its member states), NSF (USA), and NINS (Japan), together with NRC (Canada), MOST and ASIAA (Taiwan), and KASI (Republic of Korea), in cooperation with the Republic of Chile. The Joint ALMA Observatory is operated by ESO, AUI/NRAO, and NAOJ.
SYH and SYL acknowledge support from the Ministry of Science and Technology (MoST) with grants 110-2112-M-001-056- and 111-2112-M-001-042-.
D.S. acknowledges the support from  Ramanujan Fellowship (SERB)  and PRL, India.
This work has been supported by the National Key R\&D Program of China (No. 2022YFA1603100). 
Tie Liu acknowledges the supports by National Natural Science Foundation of China (NSFC) through grants No.12122307 and No.12073061, the international partnership program of Chinese Academy of Sciences through grant No.114231KYSB20200009, Shanghai Pujiang Program 20PJ1415500 and the science research grants from the China Manned Space Project with no. CMS-CSST-2021-B06
MJ acknowledges support from the Academy of Finland grant No. 348342.
The work of MGR is supported by NOIRLab, which is managed by the Association of Universities for Research in Astronomy (AURA) under a cooperative agreement with the National Science Foundation.
LB gratefully acknowledges support by the ANID BASAL project FB210003. 
Y.-L.Y. acknowledges support from Grant-in-Aid from the Ministry of Education, Culture, Sports, Science, and Technology of Japan (JP 22K20389, JP 20H05845, JP 20H05844), and a pioneering project in RIKEN (Evolution of Matter in the Universe).
This research made use of Astropy,\footnote{http://www.astropy.org} a community-developed core Python package for Astronomy \citep{astropy:2013, astropy:2018}. 
K.T. was supported by JSPS KAKENHI (Grant Number JP20H05645). 
DJ is supported by NRC Canada and by an NSERC Discovery Grant.
PS was partially supported by a Grant-in-Aid for Scientific Research (KAKENHI Number JP22H01271 and JP23H01221) of JSPS. 


\software{
astropy \citep{astropy:2013, astropy:2018},
\texttt{CASA} \citep{2007McMullin_CASA},
\texttt{HO-CHUNK} \citep{2003Whitney_SED}
\texttt{SED Fitter} \citep{2006Robitaille_grid, 2007Robitaille_sedfitter}
\texttt{SPARX}
}

\clearpage
\appendix
\section{SED Analysis\label{appx:sed}}
\resetapptablenumbers

All the photometric data points we use for the SED fitting can be found in Table 6 of \citet{2020Dutta_ALMASOP}. 
Also, Table C1 in \citet{2022Hsu_ALMASOP} shows the wavelength, the aperture radius, the instrument, and the observatory for each SED photometric data point. 
The aperture is the field of observation for extracting the flux. 
The references include
UKIRT/UKIDSS \citep{2007Lawrence_SED_UKIDSS}, 
WISE \citep{2010Wright_SED_WISE}, 
\Spitzer \citep{2004Werner_Spitzer}, 
IRAC \citep{2004Fazio_IRAC}, 
MIPS \citep{2004Rieke_MIPS}, 
AKARI PSC \citep{2010Ishihara_SED_AKARI_IRC}, 
AKARI BSC \citep{2010Yamamura_SED_AKARI_FIS}, 
\Herschel \citep{2010Pilbratt_Herschel}, 
PACS \citep{2010Poglitsch_PACS}, 
APEX \citep{2006Gusten_APEX}, 
SABOCA \citep{2010Siringo_APEX350_SABOCA}, 
LABOCA \citep{2009Siringo_APEX870_LABOCA}, and 
JCMT/SCUBA2 \citep{2008Francesco_JCMTS_submm}. 

\begin{figure*}[htb!]
\includegraphics[width=\textwidth]{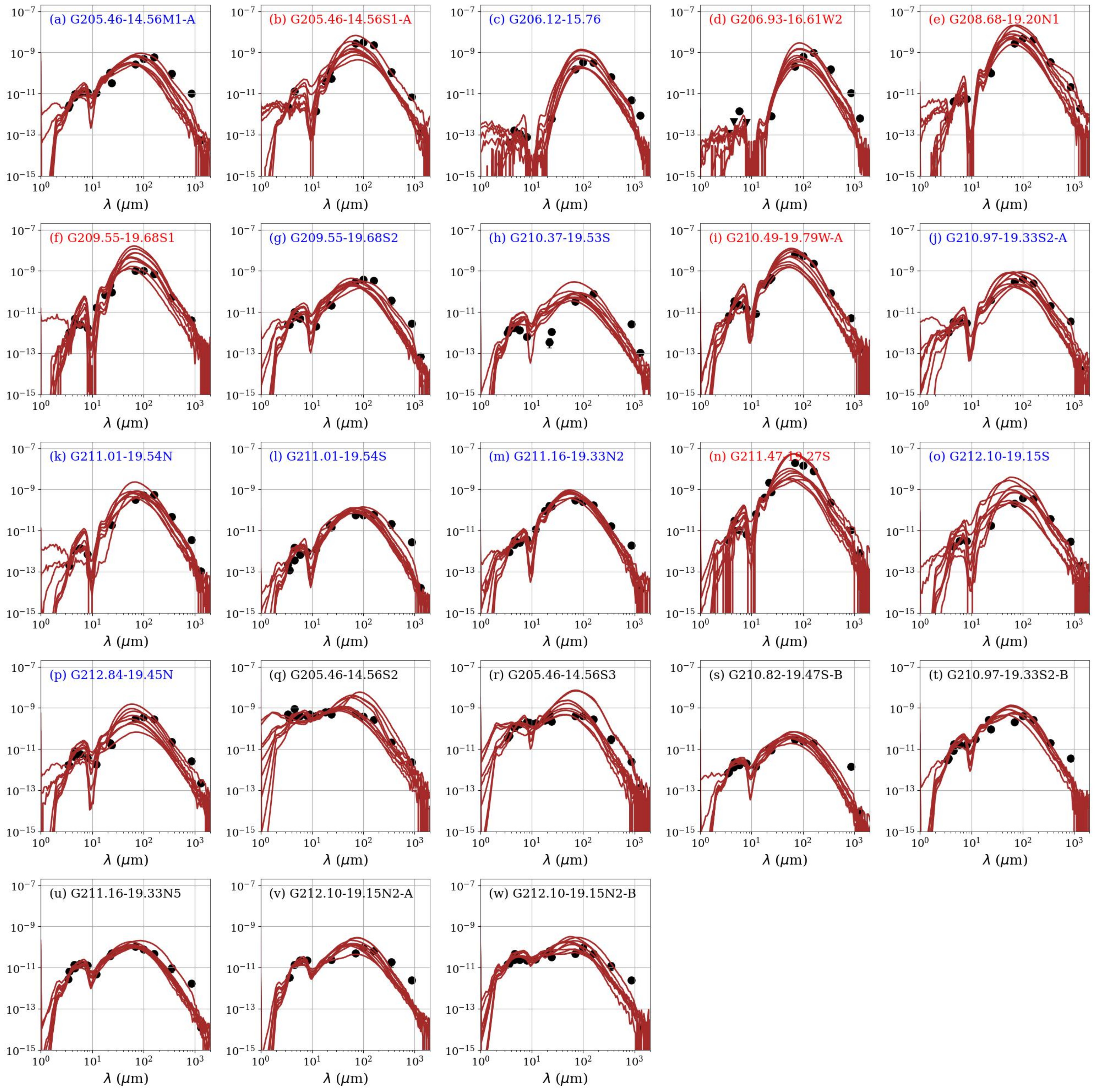}
\caption{\label{fig:2023:sed} 
The SEDs of the input data points and the nine best-fit YSO models of the sample. 
The triangle represents the upper limit of the observation. 
\editbf{The source names of the Class~0 YSOs with hot--corino signatures, Class~0 YSOs without hot--corino signatures, and Class~I YSOs are labeled in red, blue, and black, respectively.
\blue The uncertainties of the measurements are shown by the error bars around each data point. In most cases, though, the uncertainties are too small to be visible.}
}
\end{figure*}

\begin{figure*}[htb!]
    \gridline{\fig{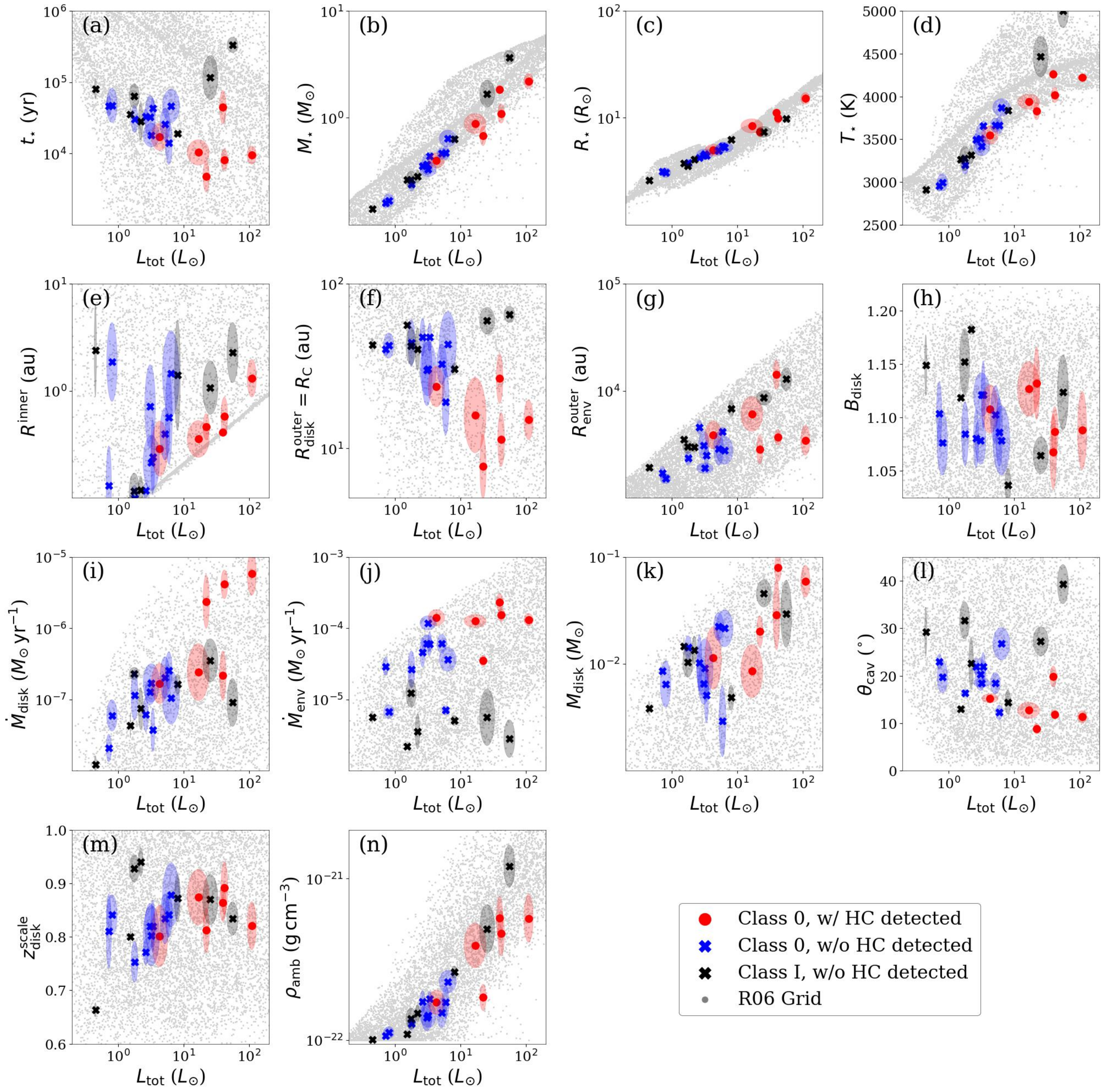}{0.95\textwidth}{}}
\caption{\label{fig:2023:YSOPar_complete} 
Weighted average and standard deviation of each YSO model parameter versus total luminosity (\Ltot). 
The x-axes of all sub-figures are total luminosities (\Ltot). 
The text of each sub-figure shows the label of the y-axis. 
The \tstar\ is the evolutionary age of the protostellar system. 
The \Mstar, \Rstar, and \Tstar\ are the mass, the radius, and the temperature of the central protostar, respectively. 
The \RMIN\ is the inner radius of both envelope and disk.
The \RMAXdisk\ and \RMAXenv\ are the outer radius of the disk and envelope, respectively. 
The \Bdisk\ is the disk-flaring factor. 
The \MDOTdisk and \MDOTenv\ are the disk accretion rate and the envelope infall rate, respectively.
The \Mdisk\ is the total mass of the disk.
The \THETAcav\ is the opening angle of the cavity. 
The \ZFAC\ is the factor to the disk scale height under hydrostatic equilibrium. 
The \RHOamb\ is the ambient density. 
The \RHOenvC\ is the characteristic envelope density. 
The red circles represent the sources with hot--corino signature \citet{2020Hsu_ALMASOP,2022Hsu_ALMASOP}. 
The blue and black crosses represent the Class~0 and Class~I protostars where the hot corinos are not detected, respectively. 
The grey dots represent all the YSO models in the R06 grid. 
\hfill\break
\hfill\break
\hfill\break
\hfill\break
}
\end{figure*}

Figure~\ref{fig:2023:sed} shows the SED data points and the nine best-fit YSO models in the R06 grid of our sample. 
Table \ref{tab:2023:sed_info} shows representative wavelengths and apertures of the archival SED data points.
Table \ref{tab:2023:YSOModel_name} shows the names of the top nine best-fit models in the R06 grid. 
Table \ref{tab:2023:YSOModel_par} shows the weighted average and standard deviation of the YSO parameters. 
Figure~\ref{fig:2023:YSOPar_complete} shows the distributions of the YSO parameters for the ALMASOP sample as well as those in the R06 grid. 
Table \ref{tab:2023:YSOModel_sparx} shows the YSO model parameters for the \texttt{SPARX} simulation. 

\begin{deluxetable*}{l|c|c|c|c|c|c|c}
\tablecaption{The information of the observation for SED. \label{tab:2023:sed_info}}
\tabletypesize{\small}
\tablehead{
\colhead{Observatory} & \colhead{WISE}  & \colhead{\Spitzer} & \colhead{AKARI PSC} & \colhead{\Spitzer} & \colhead{\Herschel} &  \colhead{APEX} & \colhead{ALMA}
}
\startdata
Filter or Band & W1, W2, W3, W4 & IRAC & IRC S9W, L18W & MIPS & PACS & SABOCA, LABOCA & Band 6 \\
Wavelength (\micron) & 3.4, 4.6, 12, 22 & 3.6, 4.5, 5.8, & 9, 18 & 24 & 70, 100, 160 & 350, 870 & 1300 \\
Aperture radius & 6\farcs{1}, 6\farcs{4}, 6\farcs{5}, 12\farcs{0} & 2\farcs{4} & 5\farcs{5}, 5\farcs{7} & 6\farcs{0} & 9\farcs{6}, 9\farcs{6}, 12\farcs{8} & 3\farcs{7}, 8\farcs{5} & $\sim$0\farcs{45} \\
\enddata
\tablecomments{
The SED flux data points were collected from \citet{2020Dutta_ALMASOP}.
}
\tablerefs{
WISE: \citet{2010Wright_SED_WISE}
\Spitzer:~\citet{2004Werner_Spitzer}; 
IRAC:~\citet{2004Fazio_IRAC}; 
MIPS:~\citet{2004Rieke_MIPS}; 
AKARI PSC: \citet{2010Ishihara_SED_AKARI_IRC};
\Herschel:~\citet{2010Pilbratt_Herschel};
PACS:~\citet{2010Poglitsch_PACS}; 
APEX:~\citet{2006Gusten_APEX}; 
SABOCA:~\citet{2010Siringo_APEX350_SABOCA}; 
LABOCA:~\citet{2009Siringo_APEX870_LABOCA}; 
}
\end{deluxetable*}

\clearpage
\startlongtable
\begin{deluxetable}{lrrr}
\setlength{\tabcolsep}{3pt}
\tabletypesize{\small}
\tablecaption{\label{tab:2023:YSOModel_name} YSO model names in R06 grid.}
\tablehead{\colhead{Model name in R06 grid} & \colhead{$N_\mathrm{data}$} & \colhead{$A_\mathrm{V}$} & \colhead{$\chi^2/N_\mathrm{data}$}
}
\startdata
\multicolumn{4}{c}{G205.46-14.56M1-A (HOPS--317)} \\
\hline
3006805\_10 & 15 & 23.135 & 78.583 \\
3006939\_10 & 15 & 23.516 & 82.092 \\
3012536\_10 & 15 & 19.237 & 83.349 \\
3005566\_10 & 15 & 22.333 & 85.900 \\
3009879\_10 & 15 & 23.034 & 88.395 \\
3005643\_10 & 15 & 22.625 & 90.105 \\
3014595\_9 & 15 & 21.229 & 90.247 \\
3001845\_8 & 15 & 27.964 & 94.457 \\
3013127\_10 & 15 & 17.995 & 95.709 \\
\hline
\multicolumn{4}{c}{G205.46-14.56S1-A (HOPS--358)} \\
\hline
3014523\_9 & 12 & 3.613 & 208.878 \\
3011935\_8 & 12 & 5.935 & 221.710 \\
3002461\_10 & 12 & 23.007 & 245.541 \\
3009148\_10 & 12 & 30.000 & 248.360 \\
3010028\_7 & 12 & 0.000 & 255.945 \\
3005113\_5 & 12 & 1.461 & 256.502 \\
3010028\_6 & 12 & 0.000 & 262.822 \\
3008121\_7 & 12 & 2.148 & 266.323 \\
3008863\_9 & 12 & 5.414 & 270.784 \\
\hline
\multicolumn{4}{c}{G206.12-15.76 (HOPS--400)} \\
\hline
3017827\_6 & 11 & 20.574 & 229.908 \\
3019250\_6 & 11 & 0.000 & 234.913 \\
3009181\_5 & 11 & 0.000 & 246.033 \\
3009181\_4 & 11 & 0.000 & 246.353 \\
3019843\_5 & 11 & 0.000 & 248.696 \\
3018837\_6 & 11 & 5.321 & 252.159 \\
3006939\_6 & 11 & 0.000 & 253.697 \\
3006939\_8 & 11 & 15.450 & 265.850 \\
3019250\_7 & 11 & 6.619 & 267.953 \\
\hline
\multicolumn{4}{c}{G206.93-16.61W2 (HOPS--399)} \\
\hline
3002461\_9 & 8 & 15.752 & 252.666 \\
3005362\_8 & 8 & 0.000 & 284.334 \\
3019843\_5 & 8 & 0.000 & 320.805 \\
3012334\_9 & 8 & 30.000 & 329.824 \\
3009148\_8 & 8 & 0.000 & 330.642 \\
3009181\_5 & 8 & 0.000 & 331.898 \\
3019250\_8 & 8 & 16.745 & 367.835 \\
3009879\_7 & 8 & 30.000 & 368.282 \\
3010733\_8 & 8 & 0.000 & 374.607 \\
\hline
\multicolumn{4}{c}{G208.68-19.20N1 (HOPS--87)} \\
\hline
3016258\_9 & 13 & 8.095 & 54.252 \\
3014529\_9 & 13 & 5.296 & 59.694 \\
3006540\_8 & 13 & 2.223 & 63.229 \\
3005436\_8 & 13 & 6.909 & 66.938 \\
3005775\_8 & 13 & 9.710 & 76.891 \\
3002402\_8 & 13 & 0.169 & 78.863 \\
3014904\_7 & 13 & 5.035 & 82.825 \\
3010556\_9 & 13 & 5.908 & 89.031 \\
3008121\_8 & 13 & 8.064 & 103.819 \\
\hline
\multicolumn{4}{c}{G209.55-19.68S1 (HOPS--11)} \\
\hline
3011928\_4 & 16 & 10.960 & 122.152 \\
3011928\_5 & 16 & 11.881 & 130.989 \\
3004222\_3 & 16 & 7.772 & 145.699 \\
3009693\_2 & 16 & 1.848 & 157.422 \\
3009777\_4 & 16 & 10.125 & 170.472 \\
3008378\_4 & 16 & 11.437 & 171.059 \\
3005928\_4 & 16 & 11.379 & 172.466 \\
3009693\_3 & 16 & 1.974 & 176.950 \\
3002996\_5 & 16 & 11.828 & 177.034 \\
\hline
\multicolumn{4}{c}{G209.55-19.68S2 (HOPS--10)} \\
\hline
3009892\_9 & 14 & 16.250 & 44.997 \\
3009892\_10 & 14 & 27.630 & 77.512 \\
3007523\_8 & 14 & 11.352 & 93.191 \\
3004378\_9 & 14 & 17.203 & 113.368 \\
3014404\_7 & 14 & 14.263 & 117.119 \\
3014404\_6 & 14 & 8.636 & 122.589 \\
3012738\_9 & 14 & 14.629 & 122.823 \\
3017475\_10 & 14 & 27.317 & 134.346 \\
3014404\_8 & 14 & 19.390 & 139.609 \\
\hline
\multicolumn{4}{c}{G210.37-19.53S (HOPS--164)} \\
\hline
3016014\_8 & 13 & 0.000 & 225.875 \\
3016088\_7 & 13 & 0.000 & 339.335 \\
3017966\_8 & 13 & 0.000 & 342.525 \\
3016014\_9 & 13 & 18.473 & 356.746 \\
3016405\_7 & 13 & 0.585 & 368.181 \\
3016088\_8 & 13 & 8.136 & 369.887 \\
3001740\_7 & 13 & 0.000 & 428.333 \\
3017966\_9 & 13 & 16.040 & 431.541 \\
3016405\_6 & 13 & 0.000 & 436.016 \\
\hline
\multicolumn{4}{c}{G210.49-19.79W-A (HOPS--168)} \\
\hline
3001830\_7 & 16 & 11.145 & 153.828 \\
3011394\_9 & 16 & 16.180 & 172.130 \\
3011804\_9 & 16 & 0.000 & 175.677 \\
3018997\_7 & 16 & 10.516 & 180.504 \\
3009709\_7 & 16 & 2.878 & 186.461 \\
3018997\_8 & 16 & 21.418 & 191.588 \\
3017681\_9 & 16 & 7.739 & 196.365 \\
3004153\_9 & 16 & 30.000 & 199.019 \\
3004153\_8 & 16 & 23.076 & 204.329 \\
\hline
\multicolumn{4}{c}{G210.97-19.33S2-A (HOPS--377)} \\
\hline
3013814\_8 & 11 & 5.853 & 20.312 \\
3000179\_7 & 11 & 5.800 & 23.044 \\
3006477\_5 & 11 & 2.726 & 24.096 \\
3014595\_8 & 11 & 8.858 & 24.317 \\
3016693\_10 & 11 & 30.000 & 24.697 \\
3015398\_8 & 11 & 4.719 & 27.010 \\
3012983\_7 & 11 & 4.184 & 29.173 \\
3006477\_6 & 11 & 8.451 & 29.361 \\
3010511\_9 & 11 & 0.000 & 29.877 \\
\hline
\multicolumn{4}{c}{G211.01-19.54N (HOPS--153)} \\
\hline
3014886\_8 & 11 & 0.000 & 96.148 \\
3013814\_8 & 11 & 21.044 & 106.305 \\
3009879\_8 & 11 & 9.614 & 107.687 \\
3013407\_7 & 11 & 0.000 & 111.200 \\
3007761\_6 & 11 & 10.438 & 119.725 \\
3014595\_7 & 11 & 9.052 & 135.005 \\
3009785\_5 & 11 & 19.017 & 140.722 \\
3007761\_5 & 11 & 7.335 & 141.676 \\
3001459\_5 & 11 & 18.543 & 144.308 \\
\hline
\multicolumn{4}{c}{G211.01-19.54S (HOPS--152)} \\
\hline
3016405\_10 & 15 & 29.014 & 85.718 \\
3016405\_9 & 15 & 22.442 & 89.371 \\
3019917\_5 & 15 & 26.139 & 98.265 \\
3019266\_9 & 15 & 14.416 & 101.512 \\
3016405\_8 & 15 & 15.040 & 102.875 \\
3016436\_6 & 15 & 19.492 & 103.737 \\
3000014\_3 & 15 & 19.884 & 104.242 \\
3016933\_4 & 15 & 11.905 & 105.633 \\
3016436\_7 & 15 & 25.796 & 106.186 \\
\hline
\multicolumn{4}{c}{G211.16-19.33N2 (HOPS--133)} \\
\hline
3013740\_2 & 16 & 15.108 & 34.785 \\
3014342\_4 & 16 & 13.264 & 37.383 \\
3013740\_1 & 16 & 13.474 & 45.458 \\
3011658\_7 & 16 & 15.492 & 45.862 \\
3019064\_2 & 16 & 15.789 & 46.484 \\
3008896\_2 & 16 & 14.151 & 47.783 \\
3004836\_2 & 16 & 18.687 & 48.488 \\
3014342\_3 & 16 & 11.539 & 49.737 \\
3012810\_2 & 16 & 15.469 & 50.911 \\
\hline
\multicolumn{4}{c}{G211.47-19.27S (HOPS--288)} \\
\hline
3013227\_3 & 16 & 2.480 & 527.642 \\
3001398\_5 & 16 & 9.019 & 556.939 \\
3001398\_4 & 16 & 7.123 & 564.907 \\
3003853\_5 & 16 & 2.751 & 582.541 \\
3003853\_6 & 16 & 5.320 & 588.719 \\
3002254\_7 & 16 & 0.920 & 597.693 \\
3001398\_6 & 16 & 11.424 & 600.840 \\
3015663\_7 & 16 & 2.601 & 606.336 \\
3018117\_3 & 16 & 6.943 & 612.036 \\
\hline
\multicolumn{4}{c}{G212.10-19.15S (HOPS--247)} \\
\hline
3005643\_6 & 10 & 1.048 & 235.502 \\
3003151\_4 & 10 & 30.000 & 261.213 \\
3017475\_5 & 10 & 16.717 & 289.215 \\
3004609\_5 & 10 & 30.000 & 306.891 \\
3009907\_4 & 10 & 7.759 & 389.509 \\
3003151\_3 & 10 & 25.165 & 408.486 \\
3004356\_6 & 10 & 6.092 & 415.491 \\
3014900\_1 & 10 & 29.053 & 421.456 \\
3007124\_3 & 10 & 0.000 & 453.359 \\
\hline
\multicolumn{4}{c}{G212.84-19.45N (HOPS--224)} \\
\hline
3019250\_9 & 15 & 0.000 & 248.868 \\
3003541\_9 & 15 & 0.461 & 335.326 \\
3008016\_10 & 15 & 27.766 & 358.632 \\
3019250\_10 & 15 & 24.251 & 365.757 \\
3008016\_8 & 15 & 25.439 & 372.744 \\
3005566\_9 & 15 & 4.101 & 394.948 \\
3002333\_9 & 15 & 0.000 & 411.531 \\
3020156\_8 & 15 & 0.000 & 424.541 \\
3010316\_10 & 15 & 26.172 & 435.676 \\
\hline
\multicolumn{4}{c}{G205.46-14.56S2 (HOPS--385)} \\
\hline
3016440\_10 & 17 & 4.909 & 75.117 \\
3011896\_6 & 17 & 8.022 & 96.039 \\
3011896\_5 & 17 & 6.546 & 105.905 \\
3011528\_9 & 17 & 7.285 & 115.080 \\
3012552\_10 & 17 & 2.238 & 121.903 \\
3009117\_10 & 17 & 16.597 & 122.467 \\
3011528\_10 & 17 & 4.301 & 123.620 \\
3016440\_9 & 17 & 5.757 & 129.535 \\
3011528\_7 & 17 & 9.356 & 131.427 \\
\hline
\multicolumn{4}{c}{G205.46-14.56S3 (HOPS--315)} \\
\hline
3008309\_7 & 17 & 13.150 & 226.404 \\
3011896\_9 & 17 & 24.156 & 238.779 \\
3008309\_6 & 17 & 10.996 & 246.721 \\
3019313\_9 & 17 & 12.236 & 252.863 \\
3005398\_10 & 17 & 13.213 & 255.839 \\
3005398\_9 & 17 & 10.780 & 274.785 \\
3012951\_10 & 17 & 18.821 & 284.815 \\
3019313\_8 & 17 & 11.665 & 285.554 \\
3012951\_8 & 17 & 20.191 & 305.153 \\
\hline
\multicolumn{4}{c}{G210.82-19.47S-B (HOPS--156)} \\
\hline
3010723\_9 & 14 & 21.197 & 35.742 \\
3010723\_10 & 14 & 20.958 & 35.882 \\
3010723\_8 & 14 & 20.030 & 37.207 \\
3010723\_7 & 14 & 17.999 & 40.553 \\
3018870\_8 & 14 & 20.266 & 42.793 \\
3018870\_9 & 14 & 20.773 & 43.170 \\
3019589\_3 & 14 & 8.568 & 45.000 \\
3010723\_6 & 14 & 15.452 & 45.383 \\
3009216\_6 & 14 & 9.514 & 46.723 \\
\hline
\multicolumn{4}{c}{G210.97-19.33S2-B (HOPS--144)} \\
\hline
3013740\_5 & 15 & 12.112 & 74.878 \\
3011479\_7 & 15 & 25.758 & 79.555 \\
3004010\_4 & 15 & 13.951 & 83.067 \\
3013740\_4 & 15 & 9.617 & 83.537 \\
3011479\_6 & 15 & 24.852 & 83.943 \\
3019417\_3 & 15 & 14.649 & 84.484 \\
3011479\_8 & 15 & 26.378 & 84.711 \\
3013740\_7 & 15 & 17.472 & 85.651 \\
3011479\_5 & 15 & 23.674 & 85.834 \\
\hline
\multicolumn{4}{c}{G211.16-19.33N5 (HOPS--135)} \\
\hline
3014344\_7 & 15 & 18.646 & 29.519 \\
3014344\_6 & 15 & 17.417 & 31.366 \\
3008016\_9 & 15 & 20.244 & 31.843 \\
3008016\_10 & 15 & 21.310 & 32.069 \\
3014344\_8 & 15 & 19.848 & 32.703 \\
3014344\_5 & 15 & 16.185 & 34.242 \\
3008016\_8 & 15 & 18.865 & 34.478 \\
3014344\_9 & 15 & 21.111 & 34.554 \\
3014344\_4 & 15 & 14.497 & 36.609 \\
\hline
\multicolumn{4}{c}{G212.10-19.15N2-A (HOPS--263)} \\
\hline
3007105\_9 & 11 & 21.206 & 24.748 \\
3007105\_10 & 11 & 21.678 & 26.044 \\
3013344\_6 & 11 & 25.332 & 28.741 \\
3013344\_7 & 11 & 28.806 & 30.606 \\
3013909\_10 & 11 & 24.551 & 31.183 \\
3005143\_10 & 11 & 25.532 & 33.147 \\
3013344\_5 & 11 & 20.258 & 34.300 \\
3013909\_9 & 11 & 22.049 & 34.627 \\
3007105\_8 & 11 & 16.563 & 36.950 \\
\hline
\multicolumn{4}{c}{G212.10-19.15N2-B (HOPS--262)} \\
\hline
3002125\_6 & 16 & 9.734 & 67.505 \\
3002125\_7 & 16 & 11.270 & 68.280 \\
3009559\_7 & 16 & 10.327 & 69.411 \\
3013344\_6 & 16 & 11.133 & 69.458 \\
3009559\_8 & 16 & 12.033 & 70.556 \\
3013344\_7 & 16 & 14.650 & 71.304 \\
3009783\_7 & 16 & 9.412 & 71.585 \\
3002125\_8 & 16 & 12.553 & 71.788 \\
3011716\_8 & 16 & 14.274 & 72.791 \\
\enddata
\tablecomments{For each source, nine best-fit YSO models are presented in the order of the goodness of the model. 
\editbf{
$N_\mathrm{data}$ is the number of photometric measurement for fitting. 
$A_\mathrm{V}$ is the additional extinction. 
$\chi^2$ is the goodness of fitting defined by \citet{2007Robitaille_sedfitter} and shown in Equation \ref{eq:chi2}. 
\blue $\chi^2/N_\mathrm{data}$ is the $\chi^2$-per-data point value defined by \citet{2007Robitaille_sedfitter}, indicating the goodness of the fitting. 
}
}
\tablereferences{R06 grid: \citet{2006Robitaille_grid}}
\end{deluxetable}

\clearpage
\begin{deluxetable*}{lrrrrrrrrrrrrrrrrr}
\rotate
\setlength{\tabcolsep}{3pt}
\renewcommand{\arraystretch}{1}
\tabletypesize{\scriptsize}
\tablecaption{\label{tab:2023:YSOModel_par} The weighted average and standard deviation of the YSO model parameters. }
\tablehead{
\colhead{Source} & \colhead{$L_\mathrm{tot}$} & \colhead{$t_{\star}$} & \colhead{$M_{\star}$} & \colhead{$R_{\star}$} & \colhead{$T_{\star}$} & \colhead{$R^\mathrm{inner}$} & \colhead{$R^\mathrm{outer}_\mathrm{disk}$} & \colhead{$\dot{M}_\mathrm{disk}$} & \colhead{$M_\mathrm{disk}$} & \colhead{$z^\mathrm{scale}_\mathrm{disk}$} & \colhead{$B_\mathrm{disk}$} & \colhead{$R^\mathrm{outer}_\mathrm{env}$} & \colhead{$\dot{M}_\mathrm{env}$} & \colhead{$\theta_\mathrm{cav}$} & \colhead{$\rho_\mathrm{cav}$} & \colhead{$\rho_\mathrm{amb}$} & \colhead{$\varphi$}\\
\colhead{} & \colhead{$L_\odot$} & \colhead{yr} & \colhead{\Msun} & \colhead{\Rsun} & \colhead{K} & \colhead{au} & \colhead{au}  & \colhead{\Msun\ yr$^{-1}$} & \colhead{} & \colhead{\Msun} & \colhead{} & \colhead{au} & \colhead{\Msun} & \colhead{$^\circ$} & \colhead{g cm$^{-3}$} & \colhead{g cm$^{-3}$} & \colhead{$^{\circ}$}\\
\colhead{} & \colhead{($\log_{10}$)} & \colhead{($\log_{10}$)} & \colhead{($\log_{10}$)} & \colhead{($\log_{10}$)} & \colhead{} &  \colhead{($\log_{10}$)} & \colhead{($\log_{10}$)} & \colhead{($\log_{10}$)} & \colhead{($\log_{10}$)} & \colhead{($\log_{10}$)} & \colhead{($\log_{10}$)} & \colhead{($\log_{10}$)} & \colhead{($\log_{10}$)} & \colhead{} & \colhead{($\log_{10}$)} & \colhead{($\log_{10}$)} & \colhead{}
}
\startdata
G205.46--14.56M1--A & 0.5$\pm$0.2 & 4.5$\pm$0.6 & -0.4$\pm$0.2 & 0.6$\pm$0.1 & 3509$\pm$327 & -0.1$\pm$0.8 & 1.5$\pm$0.4 & -6.9$\pm$0.4 & -2.2$\pm$0.4 & 0.8$\pm$0.1 & 1.08$\pm$0.04 & 3.5$\pm$0.3 & -4.2$\pm$0.2 & 20.3$\pm$6.8 & -19.8$\pm$0.5 & -21.9$\pm$0.2 & 22.0$\pm$7.8 \\
G205.46--14.56S1--A & 1.2$\pm$0.3 & 4.0$\pm$0.4 & -0.1$\pm$0.2 & 0.9$\pm$0.1 & 3940$\pm$175 & -0.5$\pm$0.3 & 1.2$\pm$0.5 & -6.6$\pm$0.8 & -2.1$\pm$0.5 & 0.9$\pm$0.1 & 1.13$\pm$0.05 & 3.8$\pm$0.3 & -3.9$\pm$0.2 & 12.7$\pm$3.4 & -19.7$\pm$0.2 & -21.4$\pm$0.3 & 39.6$\pm$15.1 \\
G206.12--15.76 & 0.5$\pm$0.2 & 4.3$\pm$0.3 & -0.5$\pm$0.1 & 0.7$\pm$0.1 & 3421$\pm$230 & -0.7$\pm$0.5 & 1.5$\pm$0.3 & -6.8$\pm$0.5 & -2.0$\pm$0.6 & 0.8$\pm$0.1 & 1.12$\pm$0.04 & 3.3$\pm$0.1 & -3.9$\pm$0.2 & 18.5$\pm$1.9 & -19.8$\pm$0.4 & -21.8$\pm$0.2 & 57.2$\pm$7.6 \\
G206.93--16.61W2 & 0.6$\pm$0.2 & 4.2$\pm$0.4 & -0.4$\pm$0.1 & 0.7$\pm$0.1 & 3548$\pm$214 & -0.5$\pm$0.5 & 1.4$\pm$0.2 & -6.8$\pm$0.6 & -1.9$\pm$0.6 & 0.8$\pm$0.1 & 1.11$\pm$0.05 & 3.6$\pm$0.3 & -3.9$\pm$0.2 & 15.2$\pm$1.9 & -19.8$\pm$0.3 & -21.8$\pm$0.2 & 44.6$\pm$11.2 \\
G208.68--19.20N1 & 1.6$\pm$0.1 & 3.9$\pm$0.3 & 0.0$\pm$0.2 & 1.0$\pm$0.1 & 4017$\pm$181 & -0.2$\pm$0.4 & 1.1$\pm$0.3 & -5.4$\pm$0.4 & -1.1$\pm$0.3 & 0.9$\pm$0.1 & 1.09$\pm$0.05 & 3.6$\pm$0.1 & -3.8$\pm$0.2 & 11.9$\pm$2.0 & -19.7$\pm$0.3 & -21.3$\pm$0.3 & 38.6$\pm$5.7 \\
G209.55--19.68S1 & 1.3$\pm$0.1 & 3.7$\pm$0.4 & -0.2$\pm$0.2 & 0.9$\pm$0.1 & 3828$\pm$202 & -0.3$\pm$0.2 & 0.9$\pm$0.4 & -5.6$\pm$0.7 & -1.7$\pm$0.3 & 0.8$\pm$0.1 & 1.13$\pm$0.07 & 3.5$\pm$0.3 & -4.5$\pm$0.2 & 8.9$\pm$2.3 & -19.5$\pm$0.2 & -21.7$\pm$0.2 & 70.7$\pm$5.6 \\
G209.55--19.68S2 & 0.4$\pm$0.1 & 4.5$\pm$0.2 & -0.4$\pm$0.1 & 0.6$\pm$0.0 & 3498$\pm$135 & -0.9$\pm$0.1 & 1.7$\pm$0.2 & -7.2$\pm$0.5 & -2.0$\pm$0.5 & 0.8$\pm$0.0 & 1.08$\pm$0.05 & 3.7$\pm$0.1 & -4.3$\pm$0.4 & 22.0$\pm$3.5 & -19.7$\pm$0.3 & -21.8$\pm$0.1 & 34.4$\pm$11.5 \\
G210.37--19.53S & -0.1$\pm$0.1 & 4.7$\pm$0.3 & -0.8$\pm$0.1 & 0.5$\pm$0.1 & 2953$\pm$120 & -0.9$\pm$0.7 & 1.6$\pm$0.1 & -7.7$\pm$0.4 & -2.1$\pm$0.1 & 0.8$\pm$0.1 & 1.10$\pm$0.06 & 3.2$\pm$0.1 & -4.5$\pm$0.3 & 23.0$\pm$3.1 & -19.9$\pm$0.2 & -22.0$\pm$0.0 & 43.4$\pm$7.5 \\
G210.49--19.79W--A & 1.6$\pm$0.1 & 4.6$\pm$0.5 & 0.3$\pm$0.1 & 1.1$\pm$0.1 & 4261$\pm$100 & -0.4$\pm$0.0 & 1.4$\pm$0.3 & -6.7$\pm$0.7 & -1.5$\pm$0.7 & 0.9$\pm$0.1 & 1.07$\pm$0.07 & 4.2$\pm$0.3 & -3.6$\pm$0.2 & 19.8$\pm$4.4 & -19.9$\pm$0.5 & -21.2$\pm$0.3 & 40.0$\pm$7.9 \\
G210.97--19.33S2--A & 0.5$\pm$0.1 & 4.6$\pm$0.3 & -0.4$\pm$0.1 & 0.6$\pm$0.0 & 3655$\pm$122 & -0.6$\pm$0.7 & 1.7$\pm$0.2 & -7.4$\pm$0.6 & -2.3$\pm$0.5 & 0.8$\pm$0.1 & 1.12$\pm$0.06 & 3.4$\pm$0.3 & -4.2$\pm$0.3 & 21.9$\pm$6.1 & -19.9$\pm$0.3 & -21.7$\pm$0.1 & 43.6$\pm$12.6 \\
G211.01--19.54N & 0.7$\pm$0.2 & 4.4$\pm$0.5 & -0.3$\pm$0.1 & 0.7$\pm$0.1 & 3665$\pm$165 & -0.4$\pm$0.5 & 1.5$\pm$0.2 & -6.7$\pm$0.6 & -1.7$\pm$0.3 & 0.8$\pm$0.1 & 1.10$\pm$0.05 & 3.5$\pm$0.3 & -4.2$\pm$0.3 & 18.4$\pm$4.5 & -19.9$\pm$0.2 & -21.8$\pm$0.2 & 51.0$\pm$9.1 \\
G211.01--19.54S & -0.1$\pm$0.2 & 4.7$\pm$0.5 & -0.8$\pm$0.1 & 0.5$\pm$0.1 & 2994$\pm$205 & 0.3$\pm$0.7 & 1.6$\pm$0.2 & -7.2$\pm$0.4 & -2.2$\pm$0.4 & 0.8$\pm$0.1 & 1.08$\pm$0.06 & 3.2$\pm$0.1 & -5.2$\pm$0.2 & 19.8$\pm$6.7 & -20.1$\pm$0.4 & -22.0$\pm$0.1 & 47.6$\pm$18.5 \\
G211.16--19.33N2 & 0.8$\pm$0.1 & 4.2$\pm$0.6 & -0.3$\pm$0.1 & 0.7$\pm$0.1 & 3664$\pm$197 & -0.3$\pm$0.9 & 1.3$\pm$0.4 & -6.6$\pm$0.5 & -2.5$\pm$0.7 & 0.8$\pm$0.1 & 1.09$\pm$0.06 & 3.6$\pm$0.1 & -5.2$\pm$0.1 & 12.3$\pm$3.1 & -19.6$\pm$0.4 & -21.8$\pm$0.2 & 76.4$\pm$10.5 \\
G211.47--19.27S & 2.0$\pm$0.1 & 4.0$\pm$0.3 & 0.3$\pm$0.1 & 1.2$\pm$0.1 & 4221$\pm$57 & 0.1$\pm$0.4 & 1.2$\pm$0.2 & -5.2$\pm$0.6 & -1.2$\pm$0.3 & 0.8$\pm$0.1 & 1.09$\pm$0.07 & 3.5$\pm$0.3 & -3.9$\pm$0.2 & 11.4$\pm$2.6 & -19.6$\pm$0.2 & -21.2$\pm$0.3 & 62.4$\pm$9.4 \\
G212.10--19.15S & 0.8$\pm$0.3 & 4.7$\pm$0.5 & -0.2$\pm$0.2 & 0.7$\pm$0.1 & 3871$\pm$206 & 0.2$\pm$0.8 & 1.6$\pm$0.4 & -7.0$\pm$0.7 & -1.7$\pm$0.3 & 0.9$\pm$0.1 & 1.08$\pm$0.06 & 3.4$\pm$0.3 & -4.4$\pm$0.3 & 26.7$\pm$6.8 & -20.0$\pm$0.3 & -21.6$\pm$0.1 & 67.3$\pm$8.9 \\
G212.84--19.45N & 0.3$\pm$0.1 & 4.5$\pm$0.4 & -0.6$\pm$0.1 & 0.6$\pm$0.1 & 3198$\pm$216 & -1.0$\pm$0.2 & 1.6$\pm$0.3 & -6.9$\pm$0.7 & -1.8$\pm$0.2 & 0.8$\pm$0.1 & 1.08$\pm$0.06 & 3.4$\pm$0.1 & -4.6$\pm$0.5 & 16.3$\pm$1.4 & -20.1$\pm$0.3 & -21.9$\pm$0.1 & 29.5$\pm$8.5 \\
\hline
G205.46--14.56S2 & 1.7$\pm$0.2 & 5.5$\pm$0.2 & 0.6$\pm$0.1 & 1.0$\pm$0.0 & 5000$\pm$392 & 0.4$\pm$0.6 & 1.8$\pm$0.1 & -7.0$\pm$0.7 & -1.5$\pm$0.6 & 0.8$\pm$0.1 & 1.12$\pm$0.07 & 4.1$\pm$0.2 & -5.5$\pm$0.5 & 39.3$\pm$8.3 & -20.3$\pm$0.1 & -20.9$\pm$0.3 & 34.0$\pm$17.4 \\
G205.46--14.56S3 & 1.4$\pm$0.2 & 5.1$\pm$0.6 & 0.2$\pm$0.3 & 0.9$\pm$0.1 & 4467$\pm$479 & 0.0$\pm$0.5 & 1.8$\pm$0.2 & -6.5$\pm$0.6 & -1.3$\pm$0.3 & 0.9$\pm$0.1 & 1.06$\pm$0.03 & 3.9$\pm$0.1 & -5.2$\pm$0.8 & 27.3$\pm$6.1 & -20.3$\pm$0.5 & -21.3$\pm$0.3 & 35.9$\pm$12.4 \\
G210.82--19.47S--B & -0.3$\pm$0.0 & 4.9$\pm$0.3 & -0.9$\pm$0.1 & 0.4$\pm$0.0 & 2907$\pm$130 & 0.4$\pm$0.9 & 1.6$\pm$0.1 & -7.9$\pm$0.2 & -2.4$\pm$0.1 & 0.7$\pm$0.0 & 1.15$\pm$0.06 & 3.3$\pm$0.1 & -5.3$\pm$0.2 & 29.2$\pm$10.4 & -20.0$\pm$0.2 & -22.0$\pm$0.0 & 43.6$\pm$16.0 \\
G210.97--19.33S2--B & 0.9$\pm$0.1 & 4.3$\pm$0.1 & -0.2$\pm$0.1 & 0.8$\pm$0.0 & 3839$\pm$73 & 0.1$\pm$1.0 & 1.5$\pm$0.1 & -6.8$\pm$0.3 & -2.3$\pm$0.2 & 0.9$\pm$0.1 & 1.04$\pm$0.03 & 3.8$\pm$0.2 & -5.3$\pm$0.1 & 14.4$\pm$5.1 & -19.7$\pm$0.1 & -21.6$\pm$0.1 & 59.8$\pm$10.6 \\
G211.16--19.33N5 & 0.2$\pm$0.0 & 4.6$\pm$0.2 & -0.6$\pm$0.1 & 0.6$\pm$0.0 & 3265$\pm$109 & -1.0$\pm$0.1 & 1.8$\pm$0.1 & -7.4$\pm$0.3 & -1.8$\pm$0.2 & 0.8$\pm$0.0 & 1.12$\pm$0.01 & 3.5$\pm$0.0 & -5.7$\pm$0.1 & 12.9$\pm$1.5 & -20.0$\pm$0.3 & -22.0$\pm$0.0 & 44.6$\pm$15.4 \\
G212.10--19.15N2--A & 0.2$\pm$0.2 & 4.8$\pm$0.3 & -0.6$\pm$0.2 & 0.5$\pm$0.0 & 3277$\pm$287 & -0.9$\pm$0.3 & 1.6$\pm$0.2 & -6.6$\pm$0.2 & -2.0$\pm$0.3 & 0.9$\pm$0.1 & 1.15$\pm$0.06 & 3.5$\pm$0.1 & -4.9$\pm$0.4 & 31.7$\pm$7.8 & -20.1$\pm$0.3 & -21.9$\pm$0.1 & 36.1$\pm$16.1 \\
G212.10--19.15N2--B & 0.3$\pm$0.1 & 4.4$\pm$0.5 & -0.5$\pm$0.1 & 0.6$\pm$0.1 & 3318$\pm$211 & -0.9$\pm$0.3 & 1.6$\pm$0.3 & -7.1$\pm$0.6 & -1.9$\pm$0.3 & 0.9$\pm$0.1 & 1.18$\pm$0.01 & 3.5$\pm$0.1 & -5.4$\pm$0.5 & 22.6$\pm$10.9 & -19.9$\pm$0.4 & -21.8$\pm$0.1 & 48.5$\pm$5.6 \\
\enddata
\tablecomments{
The $L_\mathrm{tot}$ is the total luminosity.
The \tstar\ is the evolutionary age. 
The $M_{\star}$, $R_{\star}$ and $T_{\star}$ are the mass, the radius and the temperature of central protostar, respectively.
The $M_\mathrm{disk}$ is the total mass of disk.
The $\dot{M}_\mathrm{disk}$ and $\dot{M}_\mathrm{env}$ are the mass accretion rate of disk and the in--fall rate of envelope, respectively.
The $R^{\mathrm{inner}}$ is the inner radius of both envelope and disk.
The $R^\mathrm{outer}_\mathrm{disk}$ and $R^\mathrm{outer}_\mathrm{env}$ are respectively the outer radius of disk and envelope.
The $\theta_\mathrm{cav}$ is the opening angle of cavity. 
The $\rho_\mathrm{cav}$ and the $\rho_\mathrm{amb}$ are respectively cavity density and ambient density.
The $\varphi$ is the viewing angle. 
}
\end{deluxetable*}

\clearpage
\begin{deluxetable}{lrrrr}
\tablecaption{\label{tab:2023:YSOModel_sparx} The YSO model parameters for \texttt{SPARX} simulation. }
\tablehead{
\colhead{} & \colhead{} & \colhead{G208N1} & \colhead{G210WA} & \colhead{G211S}
}
\startdata
$M_{\star}$ & $M_{\odot}$ & 1.586 & 2.446 & 1.939 \\
$R_{\star}$ & $R_{\odot}$ & 10.8 & 12.7 & 16.2 \\
$T_{\star}$ & $\mathrm{K}$ & 4202 & 4357 & 4131 \\
$R^\mathrm{inner}$ & $\mathrm{au}$ & 0.38 & 0.48 & 1.87 \\
$R^\mathrm{outer}_\mathrm{env}$ & $\mathrm{au}$ & 2219 & 3507 & 3108 \\
$R^\mathrm{outer}_\mathrm{disk}=R_\mathrm{C}$ & $\mathrm{au}$ & 16.0 & 34.2 & 8.5 \\
$\dot{M}_\mathrm{env}$ & $M_{\odot}\,\mathrm{yr}^{-1}$ & 3.1$\times$10$^{-4}$ & 3.0$\times$10$^{-4}$ & 9.7$\times$10$^{-5}$ \\
$\dot{M}_\mathrm{disk}$ & $M_{\odot}\,\mathrm{yr}^{-1}$ & 2.1$\times$10$^{-6}$ & 3.6$\times$10$^{-7}$ & 5.6$\times$10$^{-6}$ \\
$M_\mathrm{disk}$ & $M_{\odot}$ & 0.139 & 0.080 & 0.027 \\
$z^\mathrm{scale}_\mathrm{disk}$ &  & 0.87 & 0.73 & 1.00 \\
$B_\mathrm{disk}$ &  & 1.087 & 1.098 & 1.216 \\
$\theta_\mathrm{cav}$ & $^{\circ}$ & 13.3 & 30.4 & 7.1 \\
$\varphi$ & $^{\circ}$ & 31.8 & 49.5 & 75.5 \\
$\rho_\mathrm{cav}$ & $\mathrm{g}\,\mathrm{cm}^{-3}$ & 1.2$\times$10$^{-20}$ & 4.8$\times$10$^{-21}$ & 2.8$\times$10$^{-20}$ \\
$\rho_\mathrm{amb}$ & $\mathrm{g}\,\mathrm{cm}^{-3}$ & 9.3$\times$10$^{-22}$ & 1.2$\times$10$^{-21}$ & 4.2$\times$10$^{-22}$ \\
\enddata
\tablecomments{
See Table \ref{tab:2023:YSOModel_par} for the meaning of the symbols.} 
\end{deluxetable}

\clearpage
\section{Methanol Emission Simulation\label{appx:sparx}}
\resetapptablenumbers

Table \ref{tab:2023:sparx_top9} shows the \texttt{SPARX} simulation results of the nine best-matching YSO models. 

\begin{deluxetable*}{lrrrrr}
\tabletypesize{\small}
\setlength{\tabcolsep}{10pt}
\tablecaption{\label{tab:2023:sparx_top9} The \texttt{SPARX} simulation results of the nine best-matching YSO models. }
\tablehead{
\colhead{Model Name} & \colhead{FWHM$_\mathrm{maj}$} & \colhead{FWHM$_\mathrm{min}$} & \colhead{PA} & \colhead{$\overline{\mathrm{FWHM}}$} & \colhead{Peak}\\
\colhead{} & \colhead{} & \colhead{} & \colhead{} & \colhead{} & \colhead{(mJy beam$^{-1}$ km s$^{-1}$)}
}
\startdata
\multicolumn{6}{c}{G208.68-19.20N1} \\
\hline
3016258\_9 & 0$\farcs{48}$ & 0$\farcs{46}$ & 149$^\circ$ & 0$\farcs{47}$ & 583 \\
3014529\_9 & 0$\farcs{51}$ & 0$\farcs{47}$ & 147$^\circ$ & 0$\farcs{49}$ & 799 \\
3006540\_8 & 0$\farcs{52}$ & 0$\farcs{46}$ & 160$^\circ$ & 0$\farcs{49}$ & 750 \\
3005436\_8 & 0$\farcs{58}$ & 0$\farcs{48}$ & 174$^\circ$ & 0$\farcs{53}$ & 474 \\
3005775\_8 & 0$\farcs{51}$ & 0$\farcs{47}$ & 158$^\circ$ & 0$\farcs{49}$ & 683 \\
3002402\_8 & 0$\farcs{54}$ & 0$\farcs{47}$ & 178$^\circ$ & 0$\farcs{51}$ & 211 \\
3014904\_7 & 0$\farcs{50}$ & 0$\farcs{45}$ & 155$^\circ$ & 0$\farcs{48}$ & 351 \\
3010556\_9 & 0$\farcs{54}$ & 0$\farcs{48}$ & 165$^\circ$ & 0$\farcs{51}$ & 1057 \\
3008121\_8 & 0$\farcs{53}$ & 0$\farcs{47}$ & 169$^\circ$ & 0$\farcs{50}$ & 711 \\
\hline
\multicolumn{6}{c}{G210.49-19.79W-A} \\
\hline
3001830\_7 & 0$\farcs{74}$ & 0$\farcs{69}$ & 85$^\circ$ & 0$\farcs{71}$ & 250 \\
3011394\_9 & 0$\farcs{74}$ & 0$\farcs{70}$ & 90$^\circ$ & 0$\farcs{72}$ & 85 \\
3011804\_9 & 0$\farcs{72}$ & 0$\farcs{69}$ & 120$^\circ$ & 0$\farcs{71}$ & 270 \\
3018997\_7 & 0$\farcs{80}$ & 0$\farcs{67}$ & 84$^\circ$ & 0$\farcs{73}$ & 106 \\
3009709\_7 & 0$\farcs{82}$ & 0$\farcs{79}$ & 90$^\circ$ & 0$\farcs{81}$ & 171 \\
3018997\_8 & 0$\farcs{80}$ & 0$\farcs{79}$ & 38$^\circ$ & 0$\farcs{80}$ & 95 \\
3017681\_9 & 0$\farcs{77}$ & 0$\farcs{66}$ & 90$^\circ$ & 0$\farcs{71}$ & 89 \\
3004153\_9 & 0$\farcs{67}$ & 0$\farcs{67}$ & 32$^\circ$ & 0$\farcs{67}$ & 94 \\
3004153\_8 & 0$\farcs{67}$ & 0$\farcs{65}$ & 77$^\circ$ & 0$\farcs{66}$ & 108 \\
\hline
\multicolumn{6}{c}{G211.47-19.27S} \\
\hline
3013227\_3 & 0$\farcs{72}$ & 0$\farcs{51}$ & 72$^\circ$ & 0$\farcs{61}$ & 2013 \\
3001398\_5 & 0$\farcs{73}$ & 0$\farcs{53}$ & 74$^\circ$ & 0$\farcs{62}$ & 1819 \\
3001398\_4 & 0$\farcs{77}$ & 0$\farcs{54}$ & 72$^\circ$ & 0$\farcs{65}$ & 1819 \\
3003853\_5 & 0$\farcs{68}$ & 0$\farcs{49}$ & 77$^\circ$ & 0$\farcs{58}$ & 1106 \\
3003853\_6 & 0$\farcs{66}$ & 0$\farcs{49}$ & 78$^\circ$ & 0$\farcs{57}$ & 1060 \\
3002254\_7 & 0$\farcs{78}$ & 0$\farcs{56}$ & 74$^\circ$ & 0$\farcs{66}$ & 1949 \\
3001398\_6 & 0$\farcs{71}$ & 0$\farcs{52}$ & 75$^\circ$ & 0$\farcs{61}$ & 1723 \\
3015663\_7 & 0$\farcs{83}$ & 0$\farcs{58}$ & 76$^\circ$ & 0$\farcs{69}$ & 2141 \\
3018117\_3 & 0$\farcs{73}$ & 0$\farcs{52}$ & 71$^\circ$ & 0$\farcs{61}$ & 1704 \\
\enddata
\tablecomments{
The FWHM$_\mathrm{maj}$, FWHM$_\mathrm{min}$, and PA are the FWHM along the major axis, FWHM along the minor axis, and the position angle exported by the 2D Gaussian fitting. 
The \FWHM\ is the geometric mean defined as  (FWHM$_\mathrm{maj}\times$FWHM$_\mathrm{min}$)$^{1/2}$.
The transition is the CH$_3$OH-46 transition. 
The methanol abundance model in \texttt{SPARX} is the ``envelope+disk'' model. 
}
\end{deluxetable*}


\bibliography{REFERENCE.bib}{}

\begin{thebibliography}{}
\expandafter\ifx\csname natexlab\endcsname\relax\def\natexlab#1{#1}\fi
\providecommand{\url}[1]{\href{#1}{#1}}
\providecommand{\dodoi}[1]{doi:~\href{http://doi.org/#1}{\nolinkurl{#1}}}
\providecommand{\doeprint}[1]{\href{http://ascl.net/#1}{\nolinkurl{http://ascl.net/#1}}}
\providecommand{\doarXiv}[1]{\href{https://arxiv.org/abs/#1}{\nolinkurl{https://arxiv.org/abs/#1}}}

\bibitem[{Arce \& Sargent(2006)}]{2006Arce_cavity_angle}
Arce, H.~G., \& Sargent, A.~I. 2006, \apj, 646, 1070, \dodoi{10.1086/505104}

\bibitem[{Artur de~la Villarmois {et~al.}(2018)Artur de~la Villarmois,
  Kristensen, J{\o}rgensen, Bergin, Brinch, Frimann, Harsono, Sakai, \&
  Yamamoto}]{2018Artur_IRS67_disk}
Artur de~la Villarmois, E., Kristensen, L.~E., J{\o}rgensen, J.~K., {et~al.}
  2018, \aap, 614, A26, \dodoi{10.1051/0004-6361/201731603}

\bibitem[{{Astropy Collaboration} {et~al.}(2013){Astropy Collaboration},
  {Robitaille}, {Tollerud}, {Greenfield}, {Droettboom}, {Bray}, {Aldcroft},
  {Davis}, {Ginsburg}, {Price-Whelan}, {Kerzendorf}, {Conley}, {Crighton},
  {Barbary}, {Muna}, {Ferguson}, {Grollier}, {Parikh}, {Nair}, {Unther},
  {Deil}, {Woillez}, {Conseil}, {Kramer}, {Turner}, {Singer}, {Fox}, {Weaver},
  {Zabalza}, {Edwards}, {Azalee Bostroem}, {Burke}, {Casey}, {Crawford},
  {Dencheva}, {Ely}, {Jenness}, {Labrie}, {Lim}, {Pierfederici}, {Pontzen},
  {Ptak}, {Refsdal}, {Servillat}, \& {Streicher}}]{astropy:2013}
{Astropy Collaboration}, {Robitaille}, T.~P., {Tollerud}, E.~J., {et~al.} 2013,
  \aap, 558, A33, \dodoi{10.1051/0004-6361/201322068}

\bibitem[{Belloche {et~al.}(2020)Belloche, Maury, Maret, Anderl, Bacmann,
  André, Bontemps, Cabrit, Codella, Gaudel, Gueth, Lefèvre, Lefloch, Podio,
  \& Testi}]{2020Belloche_COM_CALYPSO}
Belloche, A., Maury, A.~J., Maret, S., {et~al.} 2020, \aap, 635, A198,
  \dodoi{10.1051/0004-6361/201937352}

\bibitem[{Bergner {et~al.}(2019)Bergner, Martín-Doménech, Öberg,
  J{\o}rgensen, Artur de~la Villarmois, \& Brinch}]{2019Bergner_Ser-emb_COM}
Bergner, J.~B., Martín-Doménech, R., Öberg, K.~I., {et~al.} 2019, ACS Earth
  and Space Chemistry, 3, 1564, \dodoi{10.1021/acsearthspacechem.9b00059}

\bibitem[{Bergner {et~al.}(2020)Bergner, Öberg, Bergin, Andrews, Blake,
  Carpenter, Cleeves, Guzmán, Huang, Jørgensen, Qi, Schwarz, Williams, \&
  Wilner}]{2020Bergner_Ser-emb-17_chem}
Bergner, J.~B., Öberg, K.~I., Bergin, E.~A., {et~al.} 2020, \apj, 898, 97,
  \dodoi{10.3847/1538-4357/ab9e71}

\bibitem[{Bianchi {et~al.}(2022)Bianchi, López-Sepulcre, Ceccarelli, Codella,
  Podio, Bouvier, \& Enrique-Romero}]{2022Bianchi_SVS13A_COM_binary}
Bianchi, E., López-Sepulcre, A., Ceccarelli, C., {et~al.} 2022, \apj, 928, L3,
  \dodoi{10.3847/2041-8213/ac5a56}

\bibitem[{Bianchi {et~al.}(2020)Bianchi, Chandler, Ceccarelli, Codella, Sakai,
  López-Sepulcre, Maud, Moellenbrock, Svoboda, Watanabe, Sakai, Ménard,
  Aikawa, Alves, Balucani, Bouvier, Caselli, Caux, Charnley, Choudhury,
  De~Simone, Dulieu, Durán, Evans, Favre, Fedele, Feng, Fontani, Francis,
  Hama, Hanawa, Herbst, Hirota, Imai, Isella, Jiménez-Serra, Johnstone,
  Kahane, Lefloch, Loinard, Maureira, Mercimek, Miotello, Mori, Nakatani,
  Nomura, Oba, Ohashi, Okoda, Ospina-Zamudio, Oya, Pineda, Podio, Rimola, Cox,
  Shirley, Taquet, Testi, Vastel, Viti, Watanabe, Witzel, Xue, Zhang, Zhao, \&
  Yamamoto}]{2020Bianchi_L1551-IRS5}
Bianchi, E., Chandler, C.~J., Ceccarelli, C., {et~al.} 2020, \mnras, 498, L87,
  \dodoi{10.1093/mnrasl/slaa130}

\bibitem[{Boogert {et~al.}(2015)Boogert, Gerakines, \&
  Whittet}]{2015Boogert_ice_review}
Boogert, A. C.~A., Gerakines, P.~A., \& Whittet, D. C.~B. 2015, \araa, 53, 541,
  \dodoi{10.1146/annurev-astro-082214-122348}

\bibitem[{Booth {et~al.}(2021)Booth, Walsh, Terwisscha~van Scheltinga, van
  Dishoeck, Ilee, Hogerheijde, Kama, \& Nomura}]{2021Booth_CH3OH_HD100546}
Booth, A.~S., Walsh, C., Terwisscha~van Scheltinga, J., {et~al.} 2021, NatAs,
  5, 684, \dodoi{10.1038/s41550-021-01352-w}

\bibitem[{Bouvier {et~al.}(2022)Bouvier, Ceccarelli, López-Sepulcre, Sakai,
  Yamamoto, \& Yang}]{2022Bouvier_ORANGES}
Bouvier, M., Ceccarelli, C., López-Sepulcre, A., {et~al.} 2022, \apj, 929, 10,
  \dodoi{10.3847/1538-4357/ac5904}

\bibitem[{Brunken {et~al.}(2022)Brunken, Booth, Leemker, Nazari, van~der Marel,
  \& van Dishoeck}]{2022Brunken_CH3OCH3}
Brunken, N. G.~C., Booth, A.~S., Leemker, M., {et~al.} 2022, \aap, 659, A29,
  \dodoi{10.1051/0004-6361/202142981}

\bibitem[{Cassen \& Moosman(1981)}]{1981Cassen_CMU}
Cassen, P., \& Moosman, A. 1981, Icarus, 48, 353,
  \dodoi{10.1016/0019-1035(81)90051-8}

\bibitem[{Cazaux {et~al.}(2003)Cazaux, Tielens, Ceccarelli, Castets, Wakelam,
  Caux, Parise, \& Teyssier}]{2003Cazaux_IRAS16293-2422}
Cazaux, S., Tielens, A. G. G.~M., Ceccarelli, C., {et~al.} 2003, \apj, 593,
  L51, \dodoi{10.1086/378038}

\bibitem[{Ceccarelli(2004)}]{2004Ceccarelli_HotCorino}
Ceccarelli, C. 2004, in Star Formation in the Interstellar Medium: In Honor of
  David Hollenbach, Chris McKee and Frank Shu, ASP Conference Proceeding, ed.
  D.~Johnstone, F.~Adams, D.~Lin, D.~Neufeld, \& E.~Ostriker, Vol. 323, San
  Francisco, Astronomical Society of the Pacific (ASP Conference Proceedings)

\bibitem[{Charnley {et~al.}(1992)Charnley, Tielens, \&
  Millar}]{1992Charnley_hot_core_orion}
Charnley, S.~B., Tielens, A. G. G.~M., \& Millar, T.~J. 1992, \apj, 399, L71,
  \dodoi{10.1086/186609}

\bibitem[{Codella {et~al.}(2016)Codella, Ceccarelli, Cabrit, Gueth, Podio,
  Bachiller, Fontani, Gusdorf, Lefloch, Leurini, \&
  Tafalla}]{2016Codella_HH212}
Codella, C., Ceccarelli, C., Cabrit, S., {et~al.} 2016, \aap, 586, L3,
  \dodoi{10.1051/0004-6361/201527424}

\bibitem[{De~Simone {et~al.}(2020)De~Simone, Ceccarelli, Codella, Svoboda,
  Chandler, Bouvier, Yamamoto, Sakai, Caselli, Favre, Loinard, Lefloch, Liu,
  López-Sepulcre, Pineda, Taquet, \& Testi}]{2020DeSimone_dust_opacity}
De~Simone, M., Ceccarelli, C., Codella, C., {et~al.} 2020, \apj, 896, L3,
  \dodoi{10.3847/2041-8213/ab8d41}

\bibitem[{Di~Francesco {et~al.}(2008)Di~Francesco, Johnstone, Kirk, MacKenzie,
  \& Ledwosinska}]{2008Francesco_JCMTS_submm}
Di~Francesco, J., Johnstone, D., Kirk, H., MacKenzie, T., \& Ledwosinska, E.
  2008, \apjs, 175, 277, \dodoi{10.1086/523645}

\bibitem[{Drozdovskaya {et~al.}(2015)Drozdovskaya, Walsh, Visser, Harsono, \&
  van Dishoeck}]{2015Drozdovskaya_cavity_COMs}
Drozdovskaya, M.~N., Walsh, C., Visser, R., Harsono, D., \& van Dishoeck, E.~F.
  2015, \mnras, 451, 3836, \dodoi{10.1093/mnras/stv1177}

\bibitem[{Dutta {et~al.}(2020)Dutta, Lee, Liu, Hirano, Liu, Tatematsu, Kim,
  Shang, Sahu, Kim, Moraghan, Jhan, Hsu, Evans, Johnstone, Ward-Thompson, Kuan,
  Lee, Lee, Traficante, Juvela, Vastel, Zhang, Sanhueza, Soam, Kwon, Bronfman,
  Eden, Goldsmith, He, Wu, Pelkonen, Qin, Li, \& Li}]{2020Dutta_ALMASOP}
Dutta, S., Lee, C.-F., Liu, T., {et~al.} 2020, \apjs, 251, 20,
  \dodoi{10.3847/1538-4365/abba26}

\bibitem[{Eden {et~al.}(2019)Eden, Liu, Kim, Juvela, Liu, Tatematsu, Francesco,
  Wang, Wu, Thompson, Fuller, Li, Ristorcelli, Kang, Hirano, Johnstone, Lin,
  He, Koch, Sanhueza, Qin, Zhang, Goldsmith, Evans, Yuan, Zhang, White, Choi,
  Lee, Toth, Mairs, Yi, Tang, Soam, Peretto, Samal, Fich, Parsons, Malinen,
  Bendo, Rivera-Ingraham, Liu, Wouterloot, Li, Qian, Rawlings, Rawlings, Feng,
  Wang, Li, Liu, Luo, Marston, Pattle, Pelkonen, Rigby, Zahorecz, Zhang,
  Bőgner, Aikawa, Akhter, Alina, Bell, Bernard, Blain, Bronfman, Byun,
  Chapman, Chen, Chen, Chen, Chen, Chen, Chrysostomou, Chu, Chung, Cornu,
  Cosentino, Cunningham, Demyk, Drabek-Maunder, Doi, Eswaraiah, Falgarone,
  Fehér, Fraser, Friberg, Garay, Ge, Gear, Greaves, Guan, Harvey-Smith,
  Hasegawa, He, Henkel, Hirota, Holland, Hughes, Jarken,
  {et~al.}}]{2019Eden_SCOPE}
Eden, D.~J., Liu, T., Kim, K.-T., {et~al.} 2019, \mnras, 485, 2895,
  \dodoi{10.1093/mnras/stz574}

\bibitem[{Enoch {et~al.}(2011)Enoch, Corder, Duchêne, Bock, Bolatto,
  Culverhouse, Kwon, Lamb, Leitch, Marrone, Muchovej, Pérez, Scott, Teuben,
  Wright, \& Zauderer}]{2011Enoch_Ser-emb_bol}
Enoch, M.~L., Corder, S., Duchêne, G., {et~al.} 2011, \apjs, 195, 21,
  \dodoi{10.1088/0067-0049/195/2/21}

\bibitem[{Evans {et~al.}(2022)Evans, Yang, Green, Zhao, Di~Francesco, Lee,
  J{\o}rgensen, Choi, Myers, \& Mardones}]{2022Evans_B335_modeling}
Evans, Neal~J., I., Yang, Y.-L., Green, J.~D., {et~al.} 2022, arXiv e-prints,
  arXiv:2212.03746, \dodoi{10.48550/arXiv.2212.03746}

\bibitem[{Fazio {et~al.}(2004)Fazio, Hora, Allen, Ashby, Barmby, Deutsch,
  Huang, Kleiner, Marengo, Megeath, Melnick, Pahre, Patten, Polizotti, Smith,
  Taylor, Wang, Willner, Hoffmann, Pipher, Forrest, McMurty, McCreight,
  McKelvey, McMurray, Koch, Moseley, Arendt, Mentzell, Marx, Losch, Mayman,
  Eichhorn, Krebs, Jhabvala, Gezari, Fixsen, Flores, Shakoorzadeh, Jungo,
  Hakun, Workman, Karpati, Kichak, Whitley, Mann, Tollestrup, Eisenhardt,
  Stern, Gorjian, Bhattacharya, Carey, Nelson, Glaccum, Lacy, Lowrance, Laine,
  Reach, Stauffer, Surace, Wilson, Wright, Hoffman, Domingo, \&
  Cohen}]{2004Fazio_IRAC}
Fazio, G.~G., Hora, J.~L., Allen, L.~E., {et~al.} 2004, \apjs, 154, 10,
  \dodoi{10.1086/422843}

\bibitem[{Ferreira {et~al.}(2006)Ferreira, Dougados, \&
  Cabrit}]{2006Ferreira_diskwind}
Ferreira, J., Dougados, C., \& Cabrit, S. 2006, \aap, 453, 785,
  \dodoi{10.1051/0004-6361:20054231}

\bibitem[{{Fiorellino} {et~al.}(2023){Fiorellino}, {Tychoniec}, {Cruz-S{\'a}enz
  de Miera}, {Antoniucci}, {K{\'o}sp{\'a}l}, {Manara}, {Nisini}, \&
  {Rosotti}}]{2023Fiorellino_MDOTdisk_ClassI}
{Fiorellino}, E., {Tychoniec}, {\L}., {Cruz-S{\'a}enz de Miera}, F., {et~al.}
  2023, \apj, 944, 135, \dodoi{10.3847/1538-4357/aca320}

\bibitem[{Fiorellino {et~al.}(2021)Fiorellino, Manara, Nisini, Ramsay,
  Antoniucci, Giannini, Biazzo, Alcalà, \&
  Fedele}]{2021Fiorellino_MDOTdisk_ClassI}
Fiorellino, E., Manara, C.~F., Nisini, B., {et~al.} 2021, \aap, 650, A43,
  \dodoi{10.1051/0004-6361/202039264}

\bibitem[{Froebrich {et~al.}(2006)Froebrich, Schmeja, Smith, \&
  Klessen}]{2006Froebrich_Class0_age}
Froebrich, D., Schmeja, S., Smith, M.~D., \& Klessen, R.~S. 2006, \mnras, 368,
  435, \dodoi{10.1111/j.1365-2966.2006.10124.x}

\bibitem[{Furlan {et~al.}(2016)Furlan, Fischer, Ali, Stutz, Stanke, Tobin,
  Megeath, Osorio, Hartmann, Calvet, Poteet, Booker, Manoj, Watson, \&
  Allen}]{2016Furlan_HOPS_SED}
Furlan, E., Fischer, W.~J., Ali, B., {et~al.} 2016, \apjs, 224, 5,
  \dodoi{10.3847/0067-0049/224/1/5}

\bibitem[{Garrod \& Herbst(2006)}]{2006Garrod_3phase}
Garrod, R.~T., \& Herbst, E. 2006, \aap, 457, 927,
  \dodoi{10.1051/0004-6361:20065560}

\bibitem[{Garrod {et~al.}(2008)Garrod, Widicus~Weaver, \&
  Herbst}]{2008Garrod_3phase}
Garrod, R.~T., Widicus~Weaver, S.~L., \& Herbst, E. 2008, \apj, 682, 283,
  \dodoi{10.1086/588035}

\bibitem[{{G{\"u}sten} {et~al.}(2006){G{\"u}sten}, {Nyman}, {Schilke},
  {Menten}, {Cesarsky}, \& {Booth}}]{2006Gusten_APEX}
{G{\"u}sten}, R., {Nyman}, L.~{\r{A}}., {Schilke}, P., {et~al.} 2006, \aap,
  454, L13, \dodoi{10.1051/0004-6361:20065420}

\bibitem[{Herbst \& van Dishoeck(2009)}]{2009Herbst_COM_review}
Herbst, E., \& van Dishoeck, E.~F. 2009, \araa, 47, 427,
  \dodoi{10.1146/annurev-astro-082708-101654}

\bibitem[{Hincelin {et~al.}(2013)Hincelin, Wakelam, Commerçon, Hersant, \&
  Guilloteau}]{2013Hincelin_chemical_ISM_disk}
Hincelin, U., Wakelam, V., Commerçon, B., Hersant, F., \& Guilloteau, S. 2013,
  \apj, 775, 44, \dodoi{10.1088/0004-637X/775/1/44}

\bibitem[{Hirano {et~al.}(1988)Hirano, Kameya, Nakayama, \&
  Takakubo}]{1988Hirano_B335_incl}
Hirano, N., Kameya, O., Nakayama, M., \& Takakubo, K. 1988, \apj, 327, L69,
  \dodoi{10.1086/185142}

\bibitem[{Hsu {et~al.}(2020)Hsu, Liu, Liu, Sahu, Hirano, Lee, Tatematsu, Kim,
  Juvela, Sanhueza, He, Johnstone, Qin, Bronfman, Chen, Dutta, Eden, Jhan, Kim,
  Kuan, Kwon, Lee, Lee, Moraghan, Rawlings, Shang, Soam, Thompson, Traficante,
  Wu, Yang, \& Zhang}]{2020Hsu_ALMASOP}
Hsu, S.-Y., Liu, S.-Y., Liu, T., {et~al.} 2020, \apj, 898, 107,
  \dodoi{10.3847/1538-4357/ab9f3a}

\bibitem[{Hsu {et~al.}(2022)Hsu, Liu, Liu, Sahu, Lee, Tatematsu, Kim, Hirano,
  Yang, Johnstone, Liu, Juvela, Bronfman, Chen, Dutta, Eden, Jhan, Kuan, Lee,
  Lee, Li, Liu, Qin, Sanhueza, Shang, Soam, Traficante, \&
  Zhou}]{2022Hsu_ALMASOP}
---. 2022, \apj, 927, 218, \dodoi{10.3847/1538-4357/ac49e0}

\bibitem[{Imai {et~al.}(2016)}]{2016Imai_B335}
Imai, M., {et~al.} 2016, \apj, 830, L37, \dodoi{10.3847/2041-8205/830/2/l37}

\bibitem[{Ishihara {et~al.}(2010)Ishihara, Onaka, Kataza, Salama, Alfageme,
  Cassatella, Cox, García-Lario, Stephenson, Cohen, Fujishiro, Fujiwara,
  Hasegawa, Ita, Kim, Matsuhara, Murakami, Müller, Nakagawa, Ohyama, Oyabu,
  Pyo, Sakon, Shibai, Takita, Tanabé, Uemizu, Ueno, Usui, Wada, Watarai,
  Yamamura, \& Yamauchi}]{2010Ishihara_SED_AKARI_IRC}
Ishihara, D., Onaka, T., Kataza, H., {et~al.} 2010, \aap, 514, A1,
  \dodoi{10.1051/0004-6361/200913811}

\bibitem[{Jacobsen {et~al.}(2019)Jacobsen, J{\o}rgensen, Di~Francesco, Evans,
  Choi, \& Lee}]{2019Jacobsen_L483_COM}
Jacobsen, S.~K., J{\o}rgensen, J.~K., Di~Francesco, J., {et~al.} 2019, \aap,
  629, A29, \dodoi{10.1051/0004-6361/201833214}

\bibitem[{J{\o}rgensen {et~al.}(2009)J{\o}rgensen, van Dishoeck, Visser,
  Bourke, Wilner, Lommen, Hogerheijde, \&
  Myers}]{2009Jorgensen_PROSAC_II_Class0toI}
J{\o}rgensen, J.~K., van Dishoeck, E.~F., Visser, R., {et~al.} 2009, \aap, 507,
  861, \dodoi{10.1051/0004-6361/200912325}

\bibitem[{Keto \& Zhang(2010)}]{2010Keto_SPARX_velocity}
Keto, E., \& Zhang, Q. 2010, \mnras, 406, 102,
  \dodoi{10.1111/j.1365-2966.2010.16672.x}

\bibitem[{Kim {et~al.}(2020)Kim, Tatematsu, Liu, Yi, He, Hirano, Liu, Choi,
  Sanhueza, Tóth, Evans~Ii, Feng, Juvela, Kim, Vastel, Lee,
  Nguyễn~Lu’o’, Kang, Ristorcelli, Fehér, Wu, Ohashi, Wang, Kandori,
  Hirota, Sakai, Lu, Thompson, Fuller, Li, Shinnaga, \&
  Kim}]{2020Kim_ALMASOP_Nobeyama}
Kim, G., Tatematsu, K., Liu, T., {et~al.} 2020, \apjs, 249, 33,
  \dodoi{10.3847/1538-4365/aba746}

\bibitem[{Kounkel {et~al.}(2018)Kounkel, Covey, Suárez, Román-Zúñiga,
  Hernandez, Stassun, Jaehnig, Feigelson, Peña~Ramírez, Roman-Lopes, Da~Rio,
  Stringfellow, Kim, Borissova, Fernández-Trincado, Burgasser,
  García-Hernández, Zamora, Pan, \& Nitschelm}]{2018Kounkel_Orion_distance}
Kounkel, M., Covey, K., Suárez, G., {et~al.} 2018, \aj, 156, 84,
  \dodoi{10.3847/1538-3881/aad1f1}

\bibitem[{Lawrence {et~al.}(2007)Lawrence, Warren, Almaini, Edge, Hambly,
  Jameson, Lucas, Casali, Adamson, Dye, Emerson, Foucaud, Hewett, Hirst,
  Hodgkin, Irwin, Lodieu, McMahon, Simpson, Smail, Mortlock, \&
  Folger}]{2007Lawrence_SED_UKIDSS}
Lawrence, A., Warren, S.~J., Almaini, O., {et~al.} 2007, \mnras, 379, 1599,
  \dodoi{10.1111/j.1365-2966.2007.12040.x}

\bibitem[{Lee {et~al.}(2019{\natexlab{a}})Lee, Codella, Li, \&
  Liu}]{2019Lee_HH212_COM_atm}
Lee, C.-F., Codella, C., Li, Z.-Y., \& Liu, S.-Y. 2019{\natexlab{a}}, \apj,
  876, 63, \dodoi{10.3847/1538-4357/ab15db}

\bibitem[{Lee {et~al.}(2017)Lee, Li, Ho, Hirano, Zhang, \&
  Shang}]{2017Lee_HH212}
Lee, C.-F., Li, Z.-Y., Ho, P. T.~P., {et~al.} 2017, \apj, 843, 27,
  \dodoi{10.3847/1538-4357/aa7757}

\bibitem[{Lee {et~al.}(2019{\natexlab{b}})Lee, Lee, Baek, Aikawa, Cieza, Yoon,
  Herczeg, Johnstone, \& Casassus}]{2019Lee_snowline_outburst}
Lee, J.-E., Lee, S., Baek, G., {et~al.} 2019{\natexlab{b}}, NatAs, 3, 314,
  \dodoi{10.1038/s41550-018-0680-0}

\bibitem[{Lefèvre {et~al.}(2017)Lefèvre, Cabrit, Maury, Gueth, Tabone, Podio,
  Belloche, Codella, Maret, Anderl, André, \&
  Hennebelle}]{2017Lefevre_SVS13A_HotCorino}
Lefèvre, C., Cabrit, S., Maury, A.~J., {et~al.} 2017, \aap, 604, L1,
  \dodoi{10.1051/0004-6361/201730766}

\bibitem[{Lindberg {et~al.}(2014)Lindberg, J{\o}rgensen, Brinch, Haugbølle,
  Bergin, Harsono, Persson, Visser, \& Yamamoto}]{2014Lindberg_IRS7B_disk}
Lindberg, J.~E., J{\o}rgensen, J.~K., Brinch, C., {et~al.} 2014, \aap, 566,
  A74, \dodoi{10.1051/0004-6361/201322651}

\bibitem[{Liu {et~al.}(2018)Liu, Kim, Juvela, Wang, Tatematsu, Di~Francesco,
  Liu, Wu, Thompson, Fuller, Eden, Li, Ristorcelli, Kang, Lin, Johnstone, He,
  Koch, Sanhueza, Qin, Zhang, Hirano, Goldsmith, Evans, White, Choi, Lee, Toth,
  Mairs, Yi, Tang, Soam, Peretto, Samal, Fich, Parsons, Yuan, Zhang, Malinen,
  Bendo, Rivera-Ingraham, Liu, Wouterloot, Li, Qian, Rawlings, Rawlings, Feng,
  Aikawa, Akhter, Alina, Bell, Bernard, Blain, Bőgner, Bronfman, Byun,
  Chapman, Chen, Chen, Chen, Chen, Chen, Chrysostomou, Cosentino, Cunningham,
  Demyk, Drabek-Maunder, Doi, Eswaraiah, Falgarone, Fehér, Fraser, Friberg,
  Garay, Ge, Gear, Greaves, Guan, Harvey-Smith, Hasegawa, Hatchell, He, Henkel,
  Hirota, Holland, Hughes, Jarken, Ji, Jimenez-Serra, Kang, Kawabata, Kim, Kim,
  Kim, Kim, Koo, Kwon, Kuan, Lacaille, {et~al.}}]{2018Liu_TOP-SCOPE}
Liu, T., Kim, K.-T., Juvela, M., {et~al.} 2018, \apjs, 234, 28,
  \dodoi{10.3847/1538-4365/aaa3dd}

\bibitem[{{McMullin} {et~al.}(2007){McMullin}, Waters, Schiebel, Young, \&
  Golap}]{2007McMullin_CASA}
{McMullin}, J.~P., Waters, B., Schiebel, D., Young, W., \& Golap, K. 2007, in
  Astronomical Data Analysis Software and Systems XVI, ed. R.~A. Shaw, F.~Hill,
  \& D.~J. Bell, Vol. 376, San Francisco, Astronomical Society of the Pacific
  (ASP Conference Series)

\bibitem[{Megeath {et~al.}(2012)Megeath, Gutermuth, Muzerolle, Kryukova,
  Flaherty, Hora, Allen, Hartmann, Myers, Pipher, Stauffer, Young, \&
  Fazio}]{2012Megeath_MGM2012}
Megeath, S.~T., Gutermuth, R., Muzerolle, J., {et~al.} 2012, \aj, 144, 192,
  \dodoi{10.1088/0004-6256/144/6/192}

\bibitem[{Mendoza {et~al.}(2004)Mendoza, Cantó, \&
  Raga}]{2004Mendoza_YSO_velocity}
Mendoza, S., Cantó, J., \& Raga, A.~C. 2004, \rmxaa, 40, 147.
\newblock \url{https://ui.adsabs.harvard.edu/abs/2004RMxAA..40..147M}

\bibitem[{Mercimek {et~al.}(2022)Mercimek, Codella, Podio, Bianchi, Chahine,
  Bouvier, López-Sepulcre, Neri, \&
  Ceccarelli}]{2022Mercimek_ClassI_COMSurvey}
Mercimek, S., Codella, C., Podio, L., {et~al.} 2022, \aap, 659, A67,
  \dodoi{10.1051/0004-6361/202141790}

\bibitem[{Millar \& Hatchell(1998)}]{1998Millar_model_core}
Millar, T.~J., \& Hatchell, J. 1998, Faraday Discussions, 109, 15,
  \dodoi{10.1039/a800127h}

\bibitem[{Murillo {et~al.}(2015)Murillo, Bruderer, van Dishoeck, Walsh,
  Harsono, Lai, \& Fuchs}]{2015Murillo_disk-shadowing}
Murillo, N.~M., Bruderer, S., van Dishoeck, E.~F., {et~al.} 2015, \aap, 579,
  A114, \dodoi{10.1051/0004-6361/201425118}

\bibitem[{Nazari {et~al.}(2022)Nazari, Tabone, Rosotti, van Gelder, Meshaka, \&
  van Dishoeck}]{2022Nazari_CH3OH_YSOmodel}
Nazari, P., Tabone, B., Rosotti, G.~P., {et~al.} 2022, \aap, 663, A58,
  \dodoi{10.1051/0004-6361/202142777}

\bibitem[{Nomura {et~al.}(2009)Nomura, Aikawa, Nakagawa, \&
  Millar}]{2009Nomura_disk_accretion}
Nomura, H., Aikawa, Y., Nakagawa, Y., \& Millar, T.~J. 2009, \aap, 495, 183,
  \dodoi{10.1051/0004-6361:200810206}

\bibitem[{Okoda {et~al.}(2022)Okoda, Oya, Imai, Sakai, Watanabe,
  López-Sepulcre, Saigo, \& Yamamoto}]{2022Okoda_B335_few_au}
Okoda, Y., Oya, Y., Imai, M., {et~al.} 2022, \apj, 935, 136,
  \dodoi{10.3847/1538-4357/ac7ff4}

\bibitem[{Olguin {et~al.}(2022)Olguin, Sanhueza, Ginsburg, Chen, Zhang, Li, Lu,
  \& Sakai}]{2022Olguin_G335MM1ALMA1_CH3OH}
Olguin, F.~A., Sanhueza, P., Ginsburg, A., {et~al.} 2022, \apj, 929, 68,
  \dodoi{10.3847/1538-4357/ac5bd8}

\bibitem[{Osorio {et~al.}(2003)Osorio, D'Alessio, Muzerolle, Calvet, \&
  Hartmann}]{2003Osorio_L1551IRS5_SEDFitting}
Osorio, M., D'Alessio, P., Muzerolle, J., Calvet, N., \& Hartmann, L. 2003,
  \apj, 586, 1148, \dodoi{10.1086/367695}

\bibitem[{Oya {et~al.}(2016)Oya, Sakai, López-Sepulcre, Watanabe, Ceccarelli,
  Lefloch, Favre, \& Yamamoto}]{2016Oya_IRAS16293-2422A_envelope}
Oya, Y., Sakai, N., López-Sepulcre, A., {et~al.} 2016, \apj, 824, 88,
  \dodoi{10.3847/0004-637X/824/2/88}

\bibitem[{Oya {et~al.}(2018)Oya, Sakai, Watanabe, López-Sepulcre, Ceccarelli,
  Lefloch, \& Yamamoto}]{2018Oya_L483_incl}
Oya, Y., Sakai, N., Watanabe, Y., {et~al.} 2018, \apj, 863, 72,
  \dodoi{10.3847/1538-4357/aacf42}

\bibitem[{Oya \& Yamamoto(2020)}]{2020Oya_IRAS16293-2422-A_few_au}
Oya, Y., \& Yamamoto, S. 2020, \apj, 904, 185, \dodoi{10.3847/1538-4357/abbe14}

\bibitem[{Oya {et~al.}(2017)Oya, Sakai, Watanabe, Higuchi, Hirota,
  Lopez-Sepulcre, Sakai, Aikawa, Ceccarelli, Lefloch, Caux, Vastel, Kahane, \&
  Yamamoto}]{2017Oya_L483_HCC_WCCC}
Oya, Y., Sakai, N., Watanabe, Y., {et~al.} 2017, \apj, 837, 174, \dodoi{ARTN
  174 10.3847/1538-4357/aa6300}

\bibitem[{Oya {et~al.}(2019)Oya, López-Sepulcre, Sakai, Watanabe, Higuchi,
  Hirota, Aikawa, Sakai, Ceccarelli, Lefloch, Caux, Vastel, Kahane, \&
  Yamamoto}]{2019Oya_Elias29_sulfur}
Oya, Y., López-Sepulcre, A., Sakai, N., {et~al.} 2019, \apj, 881, 112,
  \dodoi{10.3847/1538-4357/ab2b97}

\bibitem[{Pilbratt {et~al.}(2010)Pilbratt, Riedinger, Passvogel, Crone, Doyle,
  Gageur, Heras, Jewell, Metcalfe, Ott, \& Schmidt}]{2010Pilbratt_Herschel}
Pilbratt, G.~L., Riedinger, J.~R., Passvogel, T., {et~al.} 2010, \aap, 518, L1,
  \dodoi{10.1051/0004-6361/201014759}

\bibitem[{Planck {et~al.}(2016)Planck, Ade, Aghanim, Arnaud, Ashdown, Aumont,
  Baccigalupi, Banday, Barreiro, Bartolo, Battaner, Benabed, Benoît,
  Benoit-Lévy, Bernard, Bersanelli, Bielewicz, Bonaldi, Bonavera, Bond,
  Borrill, Bouchet, Boulanger, Bucher, Burigana, Butler, Calabrese, Catalano,
  Chamballu, Chiang, Christensen, Clements, Colombi, Colombo, Combet, Couchot,
  Coulais, Crill, Curto, Cuttaia, Danese, Davies, Davis, de~Bernardis, de~Rosa,
  de~Zotti, Delabrouille, Désert, Dickinson, Diego, Dole, Donzelli, Doré,
  Douspis, Ducout, Dupac, Efstathiou, Elsner, Enßlin, Eriksen, Falgarone,
  Fergusson, Finelli, Forni, Frailis, Fraisse, Franceschi, Frejsel, Galeotta,
  Galli, Ganga, Giard, Giraud-Héraud, Gjerløw, González-Nuevo, Górski,
  Gratton, Gregorio, Gruppuso, Gudmundsson, Hansen, Hanson, Harrison, Helou,
  Henrot-Versillé, Hernández-Monteagudo, Herranz, Hildebrandt, Hivon, Hobson,
  Holmes, Hornstrup, Hovest, Huffenberger, Hurier, Jaffe, Jaffe, Jones, Juvela,
  Keihänen, {et~al.}}]{2016Planck_PGCC}
Planck, C., Ade, P. A.~R., Aghanim, N., {et~al.} 2016, \aap, 594, A28,
  \dodoi{10.1051/0004-6361/201525819}

\bibitem[{Poglitsch {et~al.}(2010)Poglitsch, Waelkens, Geis, Feuchtgruber,
  Vandenbussche, Rodriguez, Krause, Renotte, van Hoof, Saraceno, Cepa,
  Kerschbaum, Agnèse, Ali, Altieri, Andreani, Augueres, Balog, Barl, Bauer,
  Belbachir, Benedettini, Billot, Boulade, Bischof, Blommaert, Callut, Cara,
  Cerulli, Cesarsky, Contursi, Creten, De~Meester, Doublier, Doumayrou, Duband,
  Exter, Genzel, Gillis, Grözinger, Henning, Herreros, Huygen, Inguscio,
  Jakob, Jamar, Jean, de~Jong, Katterloher, Kiss, Klaas, Lemke, Lutz, Madden,
  Marquet, Martignac, Mazy, Merken, Montfort, Morbidelli, Müller, Nielbock,
  Okumura, Orfei, Ottensamer, Pezzuto, Popesso, Putzeys, Regibo, Reveret,
  Royer, Sauvage, Schreiber, Stegmaier, Schmitt, Schubert, Sturm, Thiel,
  Tofani, Vavrek, Wetzstein, Wieprecht, \& Wiezorrek}]{2010Poglitsch_PACS}
Poglitsch, A., Waelkens, C., Geis, N., {et~al.} 2010, \aap, 518, L2,
  \dodoi{10.1051/0004-6361/201014535}

\bibitem[{{Price-Whelan} {et~al.}(2018){Price-Whelan}, {Sip{\H{o}}cz},
  {G{\"u}nther}, {Lim}, {Crawford}, {Conseil}, {Shupe}, {Craig}, {Dencheva},
  {Ginsburg}, {VanderPlas}, {Bradley}, {P{\'e}rez-Su{\'a}rez}, {de Val-Borro},
  {Paper Contributors}, {Aldcroft}, {Cruz}, {Robitaille}, {Tollerud},
  {Coordination Committee}, {Ardelean}, {Babej}, {Bach}, {Bachetti}, {Bakanov},
  {Bamford}, {Barentsen}, {Barmby}, {Baumbach}, {Berry}, {Biscani}, {Boquien},
  {Bostroem}, {Bouma}, {Brammer}, {Bray}, {Breytenbach}, {Buddelmeijer},
  {Burke}, {Calderone}, {Cano Rodr{\'\i}guez}, {Cara}, {Cardoso}, {Cheedella},
  {Copin}, {Corrales}, {Crichton}, {D{\textquoteright}Avella}, {Deil},
  {Depagne}, {Dietrich}, {Donath}, {Droettboom}, {Earl}, {Erben}, {Fabbro},
  {Ferreira}, {Finethy}, {Fox}, {Garrison}, {Gibbons}, {Goldstein}, {Gommers},
  {Greco}, {Greenfield}, {Groener}, {Grollier}, {Hagen}, {Hirst}, {Homeier},
  {Horton}, {Hosseinzadeh}, {Hu}, {Hunkeler}, {Ivezi{\'c}}, {Jain}, {Jenness},
  {Kanarek}, {Kendrew}, {Kern}, {Kerzendorf}, {Khvalko}, {King}, {Kirkby},
  {Kulkarni}, {Kumar}, {Lee}, {Lenz}, {Littlefair}, {Ma}, {Macleod},
  {Mastropietro}, {McCully}, {Montagnac}, {Morris}, {Mueller}, {Mumford},
  {Muna}, {Murphy}, {Nelson}, {Nguyen}, {Ninan}, {N{\"o}the}, {Ogaz}, {Oh},
  {Parejko}, {Parley}, {Pascual}, {Patil}, {Patil}, {Plunkett}, {Prochaska},
  {Rastogi}, {Reddy Janga}, {Sabater}, {Sakurikar}, {Seifert}, {Sherbert},
  {Sherwood-Taylor}, {Shih}, {Sick}, {Silbiger}, {Singanamalla}, {Singer},
  {Sladen}, {Sooley}, {Sornarajah}, {Streicher}, {Teuben}, {Thomas},
  {Tremblay}, {Turner}, {Terr{\'o}n}, {van Kerkwijk}, {de la Vega}, {Watkins},
  {Weaver}, {Whitmore}, {Woillez}, {Zabalza}, \& {Contributors}}]{astropy:2018}
{Price-Whelan}, A.~M., {Sip{\H{o}}cz}, B.~M., {G{\"u}nther}, H.~M., {et~al.}
  2018, \aj, 156, 123, \dodoi{10.3847/1538-3881/aabc4f}

\bibitem[{Pringle(1981)}]{1981Pringle_disk_model}
Pringle, J.~E. 1981, \araa, 19, 137,
  \dodoi{10.1146/annurev.aa.19.090181.001033}

\bibitem[{Rieke {et~al.}(2004)Rieke, Young, Engelbracht, Kelly, Low, Haller,
  Beeman, Gordon, Stansberry, Misselt, Cadien, Morrison, Rivlis, Latter,
  Noriega-Crespo, Padgett, Stapelfeldt, Hines, Egami, Muzerolle,
  Alonso-Herrero, Blaylock, Dole, Hinz, Le~Floc'h, Papovich, Pérez-González,
  Smith, Su, Bennett, Frayer, Henderson, Lu, Masci, Pesenson, Rebull, Rho,
  Keene, Stolovy, Wachter, Wheaton, Werner, \& Richards}]{2004Rieke_MIPS}
Rieke, G.~H., Young, E.~T., Engelbracht, C.~W., {et~al.} 2004, \apjs, 154, 25,
  \dodoi{10.1086/422717}

\bibitem[{Robitaille {et~al.}(2007)Robitaille, Whitney, Indebetouw, \&
  Wood}]{2007Robitaille_sedfitter}
Robitaille, T.~P., Whitney, B.~A., Indebetouw, R., \& Wood, K. 2007, \apjs,
  169, 328, \dodoi{10.1086/512039}

\bibitem[{Robitaille {et~al.}(2006)Robitaille, Whitney, Indebetouw, Wood, \&
  Denzmore}]{2006Robitaille_grid}
Robitaille, T.~P., Whitney, B.~A., Indebetouw, R., Wood, K., \& Denzmore, P.
  2006, \apjs, 167, 256, \dodoi{10.1086/508424}

\bibitem[{Sahu {et~al.}(2019)Sahu, Liu, Su, Li, Lee, Hirano, \&
  Takakuwa}]{2019Sahu_IRAS4A1_hot_corino_atmosphere}
Sahu, D., Liu, S.~Y., Su, Y.~N., {et~al.} 2019, \apj, 872, 196, \dodoi{ARTN 196
  10.3847/1538-4357/aaffda}

\bibitem[{Sakai {et~al.}(2008)Sakai, Sakai, Hirota, \&
  Yamamoto}]{2008Sakai_L1527_WCCC}
Sakai, N., Sakai, T., Hirota, T., \& Yamamoto, S. 2008, \apj, 672, 371,
  \dodoi{Doi 10.1086/523635}

\bibitem[{Seale \& Looney(2008)}]{2008Seale_YSO_cavity}
Seale, J.~P., \& Looney, L.~W. 2008, \apj, 675, 427, \dodoi{10.1086/526766}

\bibitem[{Shirley {et~al.}(2000)Shirley, Evans, Rawlings, \&
  Gregersen}]{2000Shirley_L483_Tbol}
Shirley, Y.~L., Evans, Neal~J., I., Rawlings, J. M.~C., \& Gregersen, E.~M.
  2000, \apjs, 131, 249, \dodoi{10.1086/317358}

\bibitem[{Siringo {et~al.}(2009)Siringo, Kreysa, Kovács, Schuller, Weiß,
  Esch, Gemünd, Jethava, Lundershausen, Colin, Güsten, Menten, Beelen,
  Bertoldi, Beeman, \& Haller}]{2009Siringo_APEX870_LABOCA}
Siringo, G., Kreysa, E., Kovács, A., {et~al.} 2009, \aap, 497, 945,
  \dodoi{10.1051/0004-6361/200811454}

\bibitem[{Siringo {et~al.}(2010)Siringo, Kreysa, De~Breuck, Kovacs, Lundgren,
  Schuller, Stanke, Weiss, Guesten, Jethava, May, Menten, Meyer, Starkloff, \&
  Zakosarenko}]{2010Siringo_APEX350_SABOCA}
Siringo, G., Kreysa, E., De~Breuck, C., {et~al.} 2010, The Messenger, 139, 20.
\newblock \url{https://ui.adsabs.harvard.edu/abs/2010Msngr.139...20S}

\bibitem[{Stutz {et~al.}(2013)Stutz, Tobin, Stanke, Megeath, Fischer,
  Robitaille, Henning, Ali, di~Francesco, Furlan, Hartmann, Osorio, Wilson,
  Allen, Krause, \& Manoj}]{2013Stutz_HOPS_APEX}
Stutz, A.~M., Tobin, J.~J., Stanke, T., {et~al.} 2013, \apj, 767, 36,
  \dodoi{10.1088/0004-637X/767/1/36}

\bibitem[{Tabone {et~al.}(2017)Tabone, Cabrit, Bianchi, Ferreira, Pineau~des
  Forêts, Codella, Gusdorf, Gueth, Podio, \&
  Chapillon}]{2021Tabone_HH212_cavity}
Tabone, B., Cabrit, S., Bianchi, E., {et~al.} 2017, \aap, 607, L6,
  \dodoi{10.1051/0004-6361/201731691}

\bibitem[{Tatematsu {et~al.}(2017)Tatematsu, Liu, Ohashi, Sanhueza,
  Nguyen~Lu'o'ng, Hirota, Liu, Hirano, Choi, Kang, Thompson, Fuller, Wu, Li,
  Di~Francesco, Kim, Wang, Ristorcelli, Juvela, Shinnaga, Cunningham, Saito,
  Lee, Tóth, He, Sakai, Kim, Collaboration, \&
  Collaboration}]{2017Tatematsu_PGCC_N2Dp}
Tatematsu, K., Liu, T., Ohashi, S., {et~al.} 2017, \apjs, 228, 12,
  \dodoi{10.3847/1538-4365/228/2/12}

\bibitem[{Tatematsu {et~al.}(2020)Tatematsu, Liu, Kim, Yi, Lee, Hirano, Liu,
  Ohashi, Sanhueza, Francesco, Evans~Ii, Fuller, Kandori, Choi, Kang, Feng,
  Hirota, Sakai, Lu, Lu’o’ng, Thompson, Wu, Li, Kim, Wang, Ristorcelli,
  Juvela, \& Tóth}]{2020Tatematsu_ALMASOP}
Tatematsu, K., Liu, T., Kim, G., {et~al.} 2020, \apj, 895, 119,
  \dodoi{10.3847/1538-4357/ab8d3e}

\bibitem[{Tobin {et~al.}(2015)Tobin, Stutz, Megeath, Fischer, Henning, Ragan,
  Ali, Stanke, Manoj, Calvet, \& Hartmann}]{2015Tobin_SED_HOPS}
Tobin, J.~J., Stutz, A.~M., Megeath, S.~T., {et~al.} 2015, \apj, 798, 128,
  \dodoi{10.1088/0004-637X/798/2/128}

\bibitem[{Tobin {et~al.}(2020)Tobin, Sheehan, Megeath, Díaz-Rodríguez,
  Offner, Murillo, van~'t Hoff, van Dishoeck, Osorio, Anglada, Furlan, Stutz,
  Reynolds, Karnath, Fischer, Persson, Looney, Li, Stephens, Chandler, Cox,
  Dunham, Tychoniec, Kama, Kratter, Kounkel, Mazur, Maud, Patel, Perez,
  Sadavoy, Segura-Cox, Sharma, Stephenson, Watson, \&
  Wyrowski}]{2020Tobin_VANDAM-II}
Tobin, J.~J., Sheehan, P.~D., Megeath, S.~T., {et~al.} 2020, \apj, 890, 130,
  \dodoi{10.3847/1538-4357/ab6f64}

\bibitem[{Ulrich(1976)}]{1976Ulrich_CMU}
Ulrich, R.~K. 1976, \apj, 210, 377, \dodoi{10.1086/154840}

\bibitem[{van~der Marel {et~al.}(2021)van~der Marel, Booth, Leemker, van
  Dishoeck, \& Ohashi}]{2021vanderMarel_CH3OH_OphIRS48}
van~der Marel, N., Booth, A.~S., Leemker, M., van Dishoeck, E.~F., \& Ohashi,
  S. 2021, \aap, 651, L5, \dodoi{10.1051/0004-6361/202141051}

\bibitem[{van Gelder {et~al.}(2020)van Gelder, Tabone, Tychoniec, van Dishoeck,
  Beuther, Boogert, Caratti~o Garatti, Klaassen, Linnartz, Müller, \&
  Taquet}]{2020vanGelder_COM}
van Gelder, M.~L., Tabone, B., Tychoniec, {\L}., {et~al.} 2020, \aap, 639, A87,
  \dodoi{10.1051/0004-6361/202037758}

\bibitem[{van Gelder {et~al.}(2022)van Gelder, Nazari, Tabone, Ahmadi, van
  Dishoeck, Beltrán, Fuller, Sakai, Sánchez-Monge, Schilke, Yang, \&
  Zhang}]{2022vanGelder_MCH3OH}
van Gelder, M.~L., Nazari, P., Tabone, B., {et~al.} 2022, \aap, 662, A67,
  \dodoi{10.1051/0004-6361/202142769}

\bibitem[{van~'t Hoff {et~al.}(2018)van~'t Hoff, Persson, Harsono, Taquet,
  Jørgensen, Visser, Bergin, \& van Dishoeck}]{2018vantHoff_snowline}
van~'t Hoff, M. L.~R., Persson, M.~V., Harsono, D., {et~al.} 2018, \aap, 613,
  A29, \dodoi{10.1051/0004-6361/201731656}

\bibitem[{van't Hoff {et~al.}(2020)van't Hoff, Harsono, Tobin, Bosman, van
  Dishoeck, J{\o}rgensen, Miotello, Murillo, \&
  Walsh}]{2020vantHoff_temperature_disk_Taurus}
van't Hoff, M. L.~R., Harsono, D., Tobin, J.~J., {et~al.} 2020, \apj, 901, 166,
  \dodoi{10.3847/1538-4357/abb1a2}

\bibitem[{Vastel {et~al.}(2022)Vastel, Alves, Ceccarelli, Bouvier,
  Jiménez-Serra, Sakai, Caselli, Evans, Fontani, Le~Gal, Chandler, Svoboda,
  Maud, Codella, Sakai, López-Sepulcre, Moellenbrock, Aikawa, Balucani,
  Bianchi, Busquet, Caux, Charnley, Cuello, De~Simone, Dulieu, Durân, Fedele,
  Feng, Francis, Hama, Hanawa, Herbst, Hirota, Imai, Isella, Johnstone,
  Lefloch, Loinard, Maureira, Murillo, Mercimek, Mori, Menard, Miotello,
  Nakatani, Nomura, Oba, Ohashi, Okoda, Ospina-Zamudio, Oya, Pineda, Podio,
  Rimola, Segura~Cox, Shirley, Testi, Viti, Watanabe, Watanabe, Witzel, Xue,
  Zhang, Zhao, \& Yamamoto}]{2022Vastel_BHB2007-11_COM_outflow}
Vastel, C., Alves, F., Ceccarelli, C., {et~al.} 2022, \aap, 664, A171,
  \dodoi{10.1051/0004-6361/202243414}

\bibitem[{Watson(2020)}]{2020Watson_B335_distance}
Watson, D.~M. 2020, RNAAS, 4, 88, \dodoi{10.3847/2515-5172/ab9df4}

\bibitem[{Werner {et~al.}(2004)Werner, Roellig, Low, Rieke, Rieke, Hoffmann,
  Young, Houck, Brandl, Fazio, Hora, Gehrz, Helou, Soifer, Stauffer, Keene,
  Eisenhardt, Gallagher, Gautier, Irace, Lawrence, Simmons, Van~Cleve, Jura,
  Wright, \& Cruikshank}]{2004Werner_Spitzer}
Werner, M.~W., Roellig, T.~L., Low, F.~J., {et~al.} 2004, \apjs, 154, 1,
  \dodoi{10.1086/422992}

\bibitem[{Whitney {et~al.}(2003{\natexlab{a}})Whitney, Wood, Bjorkman, \&
  Cohen}]{2003Whitney_SEDII}
Whitney, B.~A., Wood, K., Bjorkman, J.~E., \& Cohen, M. 2003{\natexlab{a}},
  \apj, 598, 1079, \dodoi{10.1086/379068}

\bibitem[{Whitney {et~al.}(2003{\natexlab{b}})Whitney, Wood, Bjorkman, \&
  Wolff}]{2003Whitney_SED}
Whitney, B.~A., Wood, K., Bjorkman, J.~E., \& Wolff, M.~J. 2003{\natexlab{b}},
  \apj, 591, 1049, \dodoi{10.1086/375415}

\bibitem[{Wright {et~al.}(2010)Wright, Eisenhardt, Mainzer, Ressler, Cutri,
  Jarrett, Kirkpatrick, Padgett, McMillan, Skrutskie, Stanford, Cohen, Walker,
  Mather, Leisawitz, Gautier, McLean, Benford, Lonsdale, Blain, Mendez, Irace,
  Duval, Liu, Royer, Heinrichsen, Howard, Shannon, Kendall, Walsh, Larsen,
  Cardon, Schick, Schwalm, Abid, Fabinsky, Naes, \& Tsai}]{2010Wright_SED_WISE}
Wright, E.~L., Eisenhardt, P. R.~M., Mainzer, A.~K., {et~al.} 2010, \aj, 140,
  1868, \dodoi{10.1088/0004-6256/140/6/1868}

\bibitem[{Wu {et~al.}(2009)Wu, Takakuwa, \& Lim}]{2009Wu_L1551-IRS5_outflow}
Wu, P.-F., Takakuwa, S., \& Lim, J. 2009, \apj, 698, 184,
  \dodoi{10.1088/0004-637X/698/1/184}

\bibitem[{Yamamura(2010)}]{2010Yamamura_SED_AKARI_FIS}
Yamamura, I. 2010, in 38th COSPAR Scientific Assembly, Vol.~38, 2.
\newblock \url{https://ui.adsabs.harvard.edu/abs/2010cosp...38.2496Y}

\bibitem[{Yang {et~al.}(2021)Yang, Sakai, Zhang, Murillo, Zhang, Higuchi, Zeng,
  López-Sepulcre, Yamamoto, Lefloch, Bouvier, Ceccarelli, Hirota, Imai, Oya,
  Sakai, \& Watanabe}]{2021Yang_PEACHES_COM}
Yang, Y.-L., Sakai, N., Zhang, Y., {et~al.} 2021, \apj, 910, 20,
  \dodoi{10.3847/1538-4357/abdfd6}

\bibitem[{Yen {et~al.}(2017)Yen, Koch, Takakuwa, Krasnopolsky, Ohashi, \&
  Aso}]{2017Yen_MDOTdisk_Class0_to_I}
Yen, H.-W., Koch, P.~M., Takakuwa, S., {et~al.} 2017, \apj, 834, 178,
  \dodoi{10.3847/1538-4357/834/2/178}

\bibitem[{Yen {et~al.}(2015)Yen, Takakuwa, Koch, Aso, Koyamatsu, Krasnopolsky,
  \& Ohashi}]{2015Yen_B335_nodisk}
Yen, H.-W., Takakuwa, S., Koch, P.~M., {et~al.} 2015, \apj, 812, 129,
  \dodoi{10.1088/0004-637X/812/2/129}

\bibitem[{Yen {et~al.}(2010)Yen, Takakuwa, \& Ohashi}]{2010Yen_B335_incl}
Yen, H.-W., Takakuwa, S., \& Ohashi, N. 2010, \apj, 710, 1786,
  \dodoi{10.1088/0004-637X/710/2/1786}

\bibitem[{Yi {et~al.}(2018)Yi, Lee, Liu, Kim, Choi, Eden, Evans, Di~Francesco,
  Fuller, Hirano, Juvela, Kang, Kim, Koch, Lee, Li, Liu, Liu, Liu, Rawlings,
  Ristorcelli, Sanhueza, Soam, Tatematsu, Thompson, Toth, Wang, White, Wu,
  Yang, Collaboration, \& Collaboration}]{2018Yi_PGCC_Orion}
Yi, H.-W., Lee, J.-E., Liu, T., {et~al.} 2018, \apjs, 236, 51,
  \dodoi{10.3847/1538-4365/aac2e0}

\bibitem[{Yoneda {et~al.}(2016)Yoneda, Tsukamoto, Furuya, \&
  Aikawa}]{2016Yoneda_chemistry_disk}
Yoneda, H., Tsukamoto, Y., Furuya, K., \& Aikawa, Y. 2016, \apj, 833, 105,
  \dodoi{10.3847/1538-4357/833/1/105}

\end{thebibliography}
\bibliographystyle{aasjournal}




\end{CJK*}
\end{document}